\documentclass[pra,showpacs,twocolumn,superscriptaddress,floatfix]{revtex4}

\newcommand{\ket}[1]{|{#1}\rangle}
\newcommand{\bra}[1]{\langle{#1}|}

\newcommand{\Melement}[3]{\left\langle #1 \right| #3 \left| #2 \right\rangle } 

\usepackage{amsfonts}
\usepackage{amsmath}
\usepackage{graphicx}
\usepackage{amssymb}
\usepackage{color}
\usepackage{subfigure}
\usepackage[perpage]{footmisc}
\usepackage{mathtools}

\begin{document}
\title{Black-hole lasing in coherently coupled two-component atomic condensates}

\author{Salvatore Butera}
\author{Patrik \"Ohberg}
\affiliation{SUPA, Institute of Photonics and Quantum Sciences, Heriot-Watt University, Edinburgh EH14 4AS, United Kingdom}
\author{Iacopo Carusotto}
\affiliation{INO-CNR BEC Center and Dipartimento di Fisica, Universit\`a di Trento, I-38123 Povo, Italy}
\begin{abstract}
We theoretically study the black-hole lasing phenomenon in a flowing one-dimensional, coherently coupled two component atomic Bose-Einstein condensate whose constituent atoms interact via a spin-dependent s-wave contact interaction. We show by a numerical analysis the onset of the dynamical instability in the spin branch of the excitations, once a finite supersonic region is created in this branch. We study both a spatially homogeneous geometry and a harmonically trapped condensate. Experimental advantages of the two-component configuration are pointed out, with an eye towards studies of back-reaction phenomena.
\end{abstract}
\maketitle


\section{Introduction \label{sec:intro}}
Quantum field theories on curved spacetime (QFT on CS) represent a first step towards including the effects of gravity into the dynamics of quantum matter fields \citep{birrellBook,fullingBook}. The approach is similar to the well known semi classical treatment of the interaction of an atom with an electromagnetic field, where light is considered as a classical field governed by the Maxwell equations, while the internal structure of the atom is quantized. In the QFT on CS case, this approach treats the interaction between spacetime and matter fields by considering the former as a classical field, whose dynamics is described by the Einstein's theory of general relativity, while the latter as quantized fields. The predictions of QFT on CS did open the door to a plethora of new intriguing effects which were unexpected in a Minkowskian background. A prominent example is the evaporation of Black-Holes (BH) in the form of a thermal Hawking radiation \citep{hawking1974,hawking1975}.

Hawking radiation consists of the emission of thermal particles from an astrophysical BH, characterized by a temperature whose value is inversely proportional to the BH mass. The typical value of this temperature, for ordinary BH is of the order of less than micro-kelvin, which makes its detection extremely difficult and very unlikely with state-of-the-art astrophysical observational techniques. Because of this reason then, the direct observation of Hawking radiation from astrophysical objects cannot provide so far a useful test for the validity of QFT on CS.\\
However, Hawking radiation is not a phenomenon relevant only to gravitational systems, but is instead a purely kinematical effect of quantum fields living on a curved spacetime and experiencing an event horizon. As pointed out by Unruh \citep{Unruh1981}, the same physics appears for example in hydrodynamical systems in which the inhomogeneity of the background flow results in a curvature of the effective spacetime seen by the phononic (long wave) excitations \cite{Barcel2005,Visser2002,Visser1998,faccioBook,Balbinot2005}. The analogy with gravitational systems is even more interesting in the regions where the background flow becomes supersonic. These supersonic regions, enclosed by (sonic) horizons defined as the surfaces at which the velocity of the background flow equals the local speed of sound, represent the analogue of the interior region of an ordinary BH. The sound waves are in fact dragged in by the supersonic flow, without having any chance to escape. The analogy between gravitation and hydrodynamics, and in particular between astrophysical and sonic Black-Holes, is a promising tool in order to reveal the existence of the Hawking radiation, and more generally to test the predictions of QFT on CS.

In this respect, superfluids and in particular Bose-Einstein condensates (BEC) are useful systems because of the high level of controllability reached in state-of-the-art cold atoms experiments \cite{Garay2000,Garay2001,Balbinot2013,Carusotto2008,Carusotto2010,Larre2012,Recati2009,Parentani2009}. The experimental realisation of sonic BH configurations in a BEC has been for example achieved \cite{Lahav2010}. Moreover, the knowledge of the microscopic theory describing a BEC, in the context of gravity analogues, allows for the investigation of effects related to the so-called trans-planckian problem, which is concerned with the physics at scales shorter than the Planck scale. In such analogue models, the equivalent of the Planck scale is represented by the so-called healing length for which the (long wave) hydrodynamic description of the excitations breaks down. This leads to superluminal (for the case of a BEC for example) or subluminal deviations from the linear dispersion relation. While the Hawking radiation is quantitatively robust as long as the BH temperature is far below the energy scale characteristic of this short distance physics \cite{Unruh1995,Unruh2005,Parentani1995,Corley1996}, its qualitative features persist even for extremely sharp horizons \cite{Carusotto2008,coutant2012black}. Pioneering experimental studies of this Hawking radiation have been recently reported \cite{Steinhauer2016}.

The physics is even richer when configurations characterized by an inner (white) and a outer (black) horizon are considered. For the case of superluminal bosonic field, in this configuration the Hawking emission between the two horizons is expected to be self-amplified \cite{Corley1999}. Correspondingly, the radiated flux of excitations is anticipated to grow exponentially in time, as a result of the emergence of a lasing cavity, in which the negative energy companions of the Hawking particles bounce back and forth between the two horizons stimulating further emissions \cite{Finazzi2010,Coutant2010,MichelParentani1,MichelParentani2,deNova2016}. A claim of experimental observation of this black-hole lasing mechanism has been reported in \cite{Steinhauer2014}.

In this paper we study the black-hole lasing phenomenon in a (pseudo)-spinorial condensate composed by two-level atoms, where the two levels are coherently coupled and the atoms are assumed to interact via a spin-dependent s-wave contact interaction. First steps in the study of Hawking physics in two-components condensates were reported in~\cite{larre}. Our goal here is to show the occurrence of a lasing effect in the ``spin'' modes rather than in the usual ``density'' modes. The former are fluctuations in the excess of atoms in one component with respect to the other, while the latter represent local oscillations in the density of the total number of particles. 

Several promising advantages of two-component atomic condensates over standard single-component ones are pointed out: on one hand, the dispersion relation of the spin excitations can be easily controlled in a spatial dependent way by shaping the local amplitude of the coherent coupling field, which allows to generate a wide variety of single- and multi-horizon configurations. On the other hand, using a two-photon Raman coupling allows to keep the mean-field dynamics simple and well-controlled without the need for a spatially dependent interaction parameters \cite{Carusotto2008}. Finally, our theoretical study focuses on the spin density in the atomic cloud as the main observable, for which advanced imaging techniques offer direct {\em in situ} experimental access~\cite{Carusotto2004,Seo:PRL2015}.

In the next Sec. \ref{sec:system} we describe in detail the physical system at hand and we introduce the aforementioned density and spin excitations. After having addressed a number of experimental considerations in Sec. \ref{sec:exp}, we show numerically, in Sec. \ref{sec:ring}, the appearance of the lasing mechanism in the spin modes of an effectively one-dimensional (1D) uniform condensate. As a first step we will study the problem of a wave packet of spin excitations propagating through the lasing region and will show how all the relevant physics emerges straightforwardly in this case, such as the mode conversion at the two horizons, the self-amplification of the radiation in the cavity, and the exponentially growing amplitude of the Hawking emission from the cavity. We will then consider the same system initially prepared in its ground state, and will discuss the dependence of the amplification rate of a initially imprinted random white noise, as a function of the strength of the coherent coupling. In Sec. \ref{sec:HO} we investigate the BH-lasing mechanism in the experimentally more feasible case of a 1D cloud confined by a harmonic potential and we highlight some observable consequences of the higher-order coupling of the spin modes to the density ones. Finally, we present in sec. \ref{sec:conclusions} our concluding remarks.

\section{Coherently coupled two-component condensates \label{sec:system}}
The properties of two-component BECs, which are coherently coupled and with the usual s-wave contact interaction, have been already addressed in the literature (see \cite{Abad2013} references therein). Within the meanfield approximation, the dynamics of the system is described by two coupled Gross-Pitaevskii (GP) equations for the order parameters $\psi_i$ (with $i=a,b$) of the two components, which read as
\begin{align}
	& i\hbar \frac{\partial\psi_a}{\partial t}=\left[-\frac{\hbar^2}{2m} \boldsymbol{\nabla}^2+V_a(\mathbf{r})+g_a|\psi_a|^2+g_{ab}|\psi_b|^2\right]\psi_a+\Omega \psi_b, \label{GP_a}\\
	& i\hbar \frac{\partial\psi_b}{\partial t}=\left[-\frac{\hbar^2}{2m} \boldsymbol{\nabla}^2+V_b(\mathbf{r})+g_b|\psi_b|^2+g_{ab}|\psi_a|^2\right]\psi_b+\Omega^* \psi_a. \label{GP_b}
\end{align}
In Eqs.\eqref{GP_a} and \eqref{GP_b} $\Omega$ is the coherent coupling strength, while $g_i$ (with $i=a,b,ab$) are the meanfield interaction parameters relative to the different scattering channels, which are related to the corresponding scattering lengths $a_i$ as $g_i=4\pi\hbar^2 a_i/m$, with $m$ the mass of the atomic species. Any external confining potentials, applied selectively to the two components, are taken into account by the terms $V_i$ $(i=a,b)$. 

\subsection{Thomas-Fermi approximation\label{subsec:ThomasFermi}}
Working in the Thomas-Fermi (TF) regime, in which the cloud is sufficiently large with a smooth density profile, the kinetic energy can be neglected, and the ground state is obtained by minimizing the energy density of the system which has the form:
\begin{multline}
	\epsilon=\frac{g_a n_a^2}{2}+\frac{g_b n_b^2}{2}+g_{ab}n_a\,n_b+2|\Omega|\cos\phi\sqrt{n_a n_b}\\
	+V_a n_a+V_b n_b-\mu(n_a+n_b)
\label{En_dens_GS}
\end{multline}
In Eq.\eqref{En_dens_GS} we introduced the phase $\phi\equiv\phi_{ab}+\phi_\Omega$, with $\phi_{ab}\equiv\phi_b-\phi_a$ and $\phi_\Omega$ the phase characterizing the coherent coupling $\Omega=|\Omega|\, e^{i\phi_\Omega}$. The minimum energy configuration is obtained for $\cos\phi=-1$, which implies $\phi_{ab}+\phi_\Omega=(2n+1) \pi$ with $n\in\mathbb{N}$. In the following we will choose $\Omega<0$, so that $\phi_{ab}=0$. Under this condition, the minimization of Eq.\eqref{En_dens_GS} leads to the equations defining the ground state in terms of the local density of particles $n=n_a+n_b$ and of the population difference $\delta n=n_a-n_b$. Considering the case of a symmetric contact interaction between particles in states $a$ and $b$, for which $g_a=g_b=g$, one obtains:
\begin{align}
	& \left(g-g_{ab}+\frac{|\Omega|}{\sqrt{n_a n_b}}\right)(n_a-n_b)=V_b-V_a, \label{GS_def_delta}\\
	& \left(g+g_{ab}-\frac{|\Omega|}{\sqrt{n_a n_b}}\right)(n_a+n_b)=2\mu -\left(V_b+V_a\right). \label{GS_def_n}
\end{align}
For the simplest case in which the trapping potential is the same for both the atomic states $V_a=V_b$, Eq.\eqref{GS_def_delta} and Eq.\eqref{GS_def_n} imply the existence of either a symmetric (GS) or a polarised (GA) ground state:
\begin{align}
	& \text{(GS)}\; \delta n=0, &\left(g_{ab}\leq \tilde{g}_{ab}\right) \label{GS1}\\
	& \text{(GA)}\; \delta n_\pm=\pm n\sqrt{1-\left(\frac{1-\tilde{g}_{ab}/g}{1-g_{ab}/g}\right)^2} , & \left(g_{ab}> \tilde{g}_{ab}.  \label{GS2}
\right)\end{align}
depending on the actual strength of the coupling constant $g_{ab}$ with respect to the critical value $\tilde{g}_{ab}=g+2|\Omega|/n$ at which a bifurcation in the ground state solution occurs. Eq.\eqref{GS_def_n} simply gives the profile of the particle density that, in the local density approximation, takes the form:
\begin{equation}
	n(\mathbf{r})=2\,\left(\frac{\mu+|\Omega|-V(\mathbf{r})}{g+g_{ab}}\right),
\label{n}
\end{equation}
where the value of the chemical potential is then determined by the normalization condition $N=\int{d^3 r \,n(\mathbf{r})}$, with $N$ the total number of particles.

Introducing an asymmetry in the external potentials $\left(V_a\neq V_b\right)$, and/or in the interaction between atoms, i.e. for $g_a\neq g_b$, a certain polarization is always present in the system, so that the singularities occurring at $g_{ab}=\tilde{g}_{ab}$ is smoothed out, and the bifurcation is removed. In the following we will work in the symmetric state, where the density and spin branches of the Bogoliubov excitations, introduced in the next section, decouple.

\begin{figure*}[!htbp]
\centering%
\subfigure[\label{fig:DispRelOUT_NONlasing}]
{\includegraphics[width=0.45\linewidth]{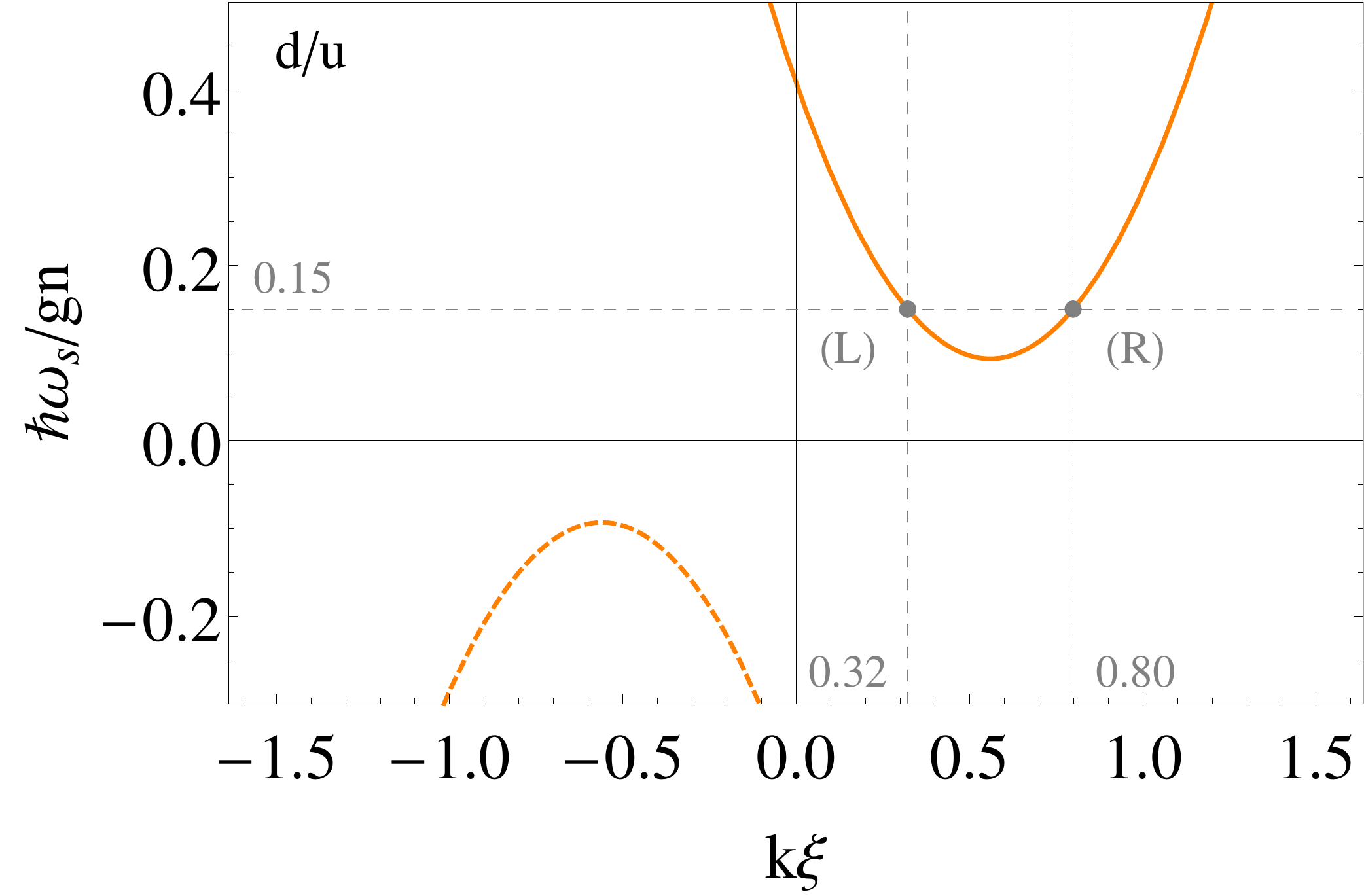}}\quad
\subfigure [\label{fig:DispRelIN_NONlasing}]
{\includegraphics[width=0.45\linewidth]{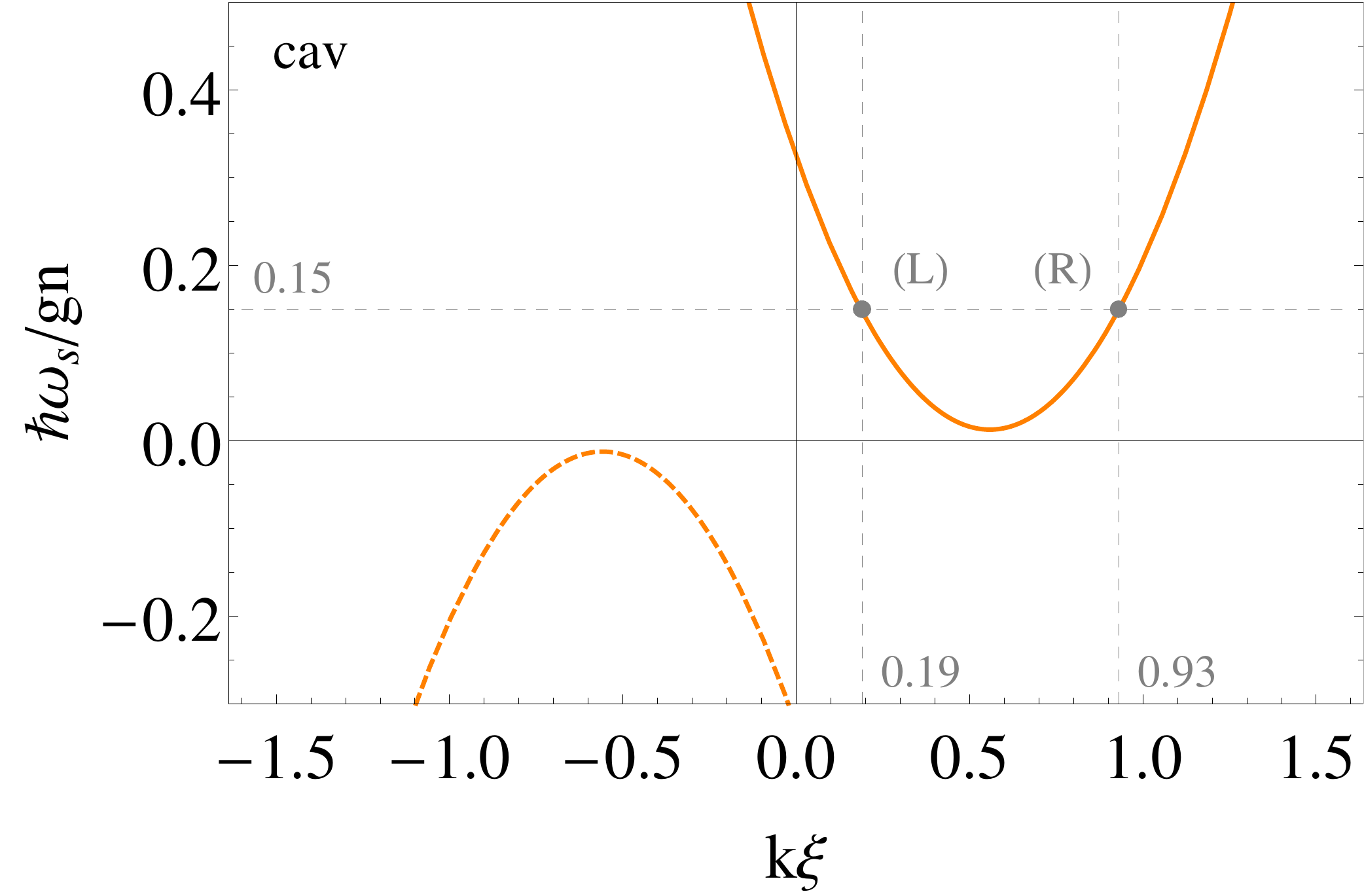}}
\caption{Spin-modes dispersion relations: non-lasing configuration. a) Downstream and upstream regions: $\Omega/gn=-0.16$, $g_{ab}/g=0.8$. b)  Cavity region: $\Omega/gn=-0.12$, $g_{ab}/g=0.8$.}
\label{fig:DispRel_Nonlasing}
\end{figure*}

\begin{figure*}[!htbp]
\centering%
\subfigure[\label{fig:DispRelOUT_lasing}]
{\includegraphics[width=0.45\linewidth]{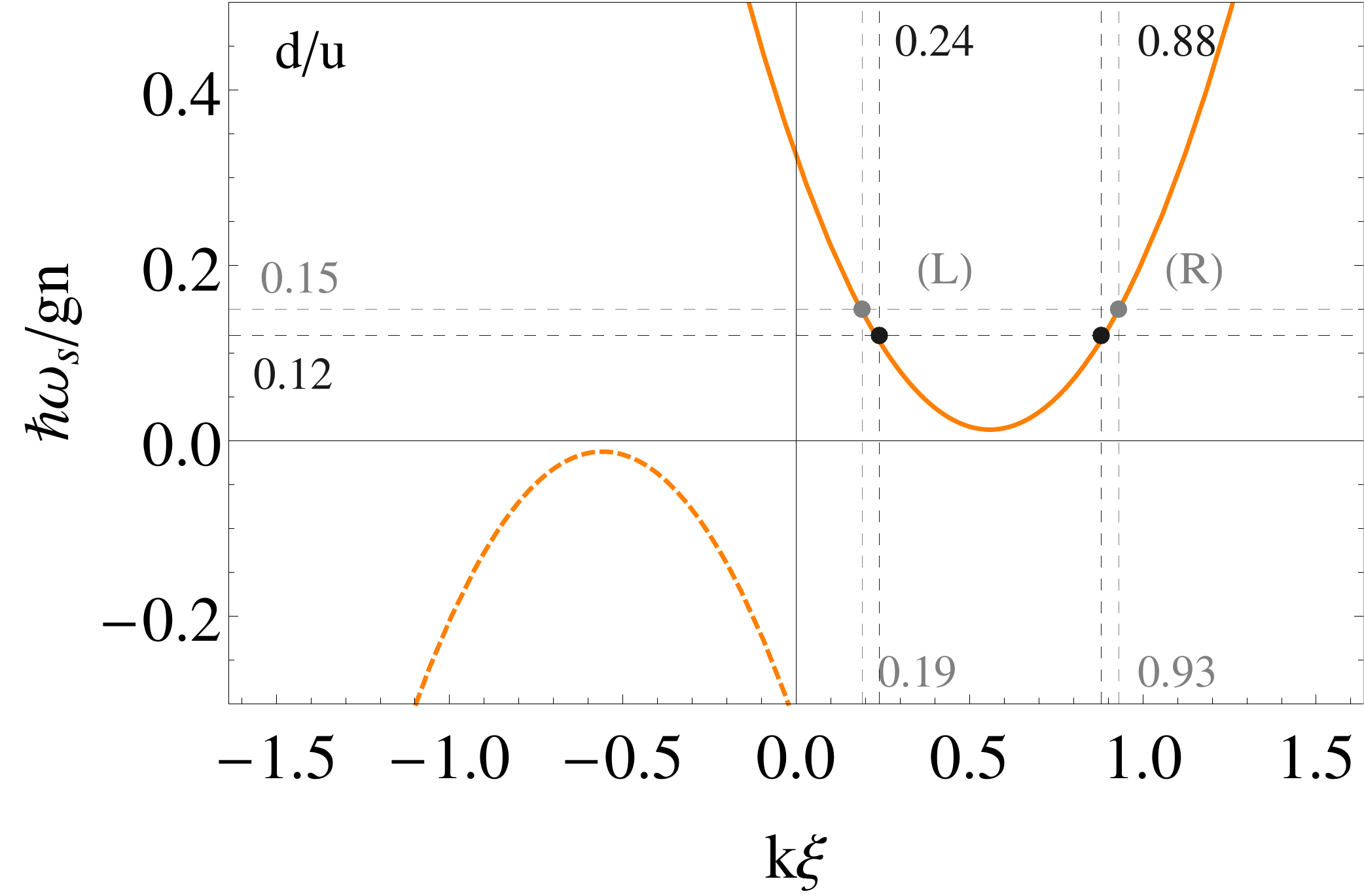}}\quad
\subfigure [\label{fig:DispRelIN_lasing}]
{\includegraphics[width=0.45\linewidth]{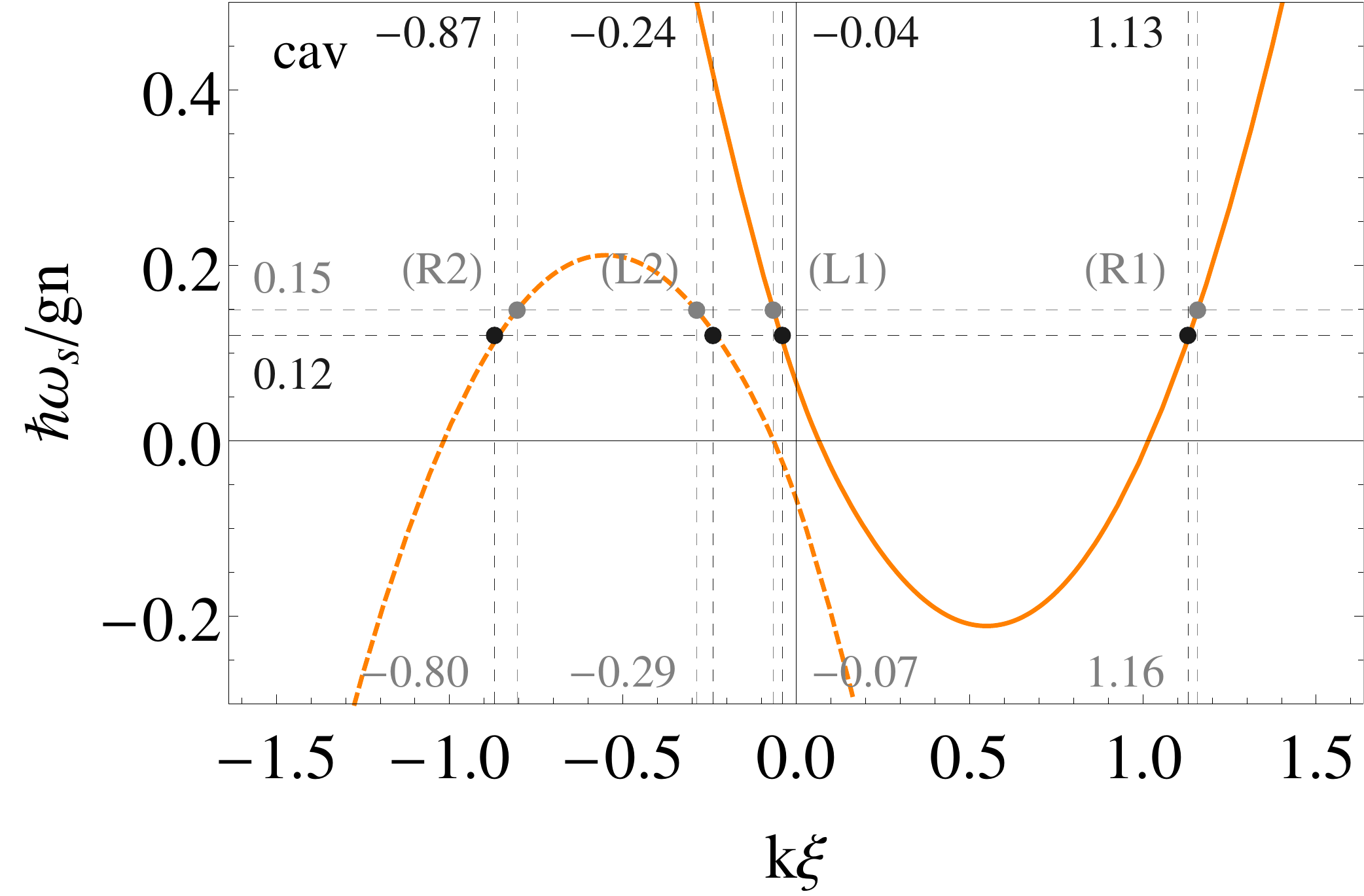}}
\caption{Spin-modes dispersion relations: lasing configuration. a) Downstream and upstream regions: $\Omega/gn=-0.12$, $g_{ab}/g=0.8$. b) Cavity region: $\Omega/gn=-0.01$, $g_{ab}/g=0.8$. }
\label{fig:DispRel_lasing}
\end{figure*} 

\subsection{Bogoliubov excitations\label{subsec:bogoliubov}}
\vspace{3mm} From now on we restrict our attention to excitations on top of a condensate in the symmetric ground state defined above. As discussed in detail in \cite{Abad2013} and references therein, in this case the spin and density branches are by symmetry decoupled and do not hybridize, which is a condition that we need for the following of this work. In the Bogoliubov formalism, the order parameters are perturbatively expanded as $\psi_i=\psi_{i0}+\psi_{i1}$, where $\psi_{i0}$ and $\psi_{i1}$  describe the condensed and the excited components respectively. By substituting the expression for the single mode
\begin{equation}
	\left( \begin{array}{c} \psi_{i1}\left(\mathbf{r}\right) \\ \psi_{i1}^*\left(\mathbf{r}\right) \end{array} \right) = \left( \begin{array}{c} u_i(\mathbf{r}) \\ v_i(\mathbf{r}) \end{array} \right) e^{-i\omega t}+\left( \begin{array}{c} v_i^*(\mathbf{r}) \\ u_i^*(\mathbf{r}) \end{array} \right) e^{i\omega t}\quad (i=a,b)
\label{Bogoliubov}
\end{equation}
into Eqs.\eqref{GP_a} and \eqref{GP_b}, retaining terms up to first order in $\psi_{i1}$, we obtain the equations for the Bogoliubov $u$'s and $v$'s functions, along with the spectrum of the excitations. 

For the case of a homogeneous system, $u_i(\mathbf{r})$ and $v_i(\mathbf{r})$ take the simple form of plane waves $u_i\left(\mathbf{r}\right)=u_i e^{i\mathbf{k}\cdot\mathbf{r}}$, $v_i\left(\mathbf{r}\right)=v_i e^{i\mathbf{k}\cdot\mathbf{r}}$ (with $u_i$ and $v_i$ constant coefficients), and the Bogolioubov-De Gennes equations take the form
\begin{equation}
	\mathcal{L}\left( \begin{array}{c} u_a \\ v_a\\ u_b\\ v_b \end{array} \right)=\hbar\omega \left( \begin{array}{c} u_a \\ v_a\\ u_b\\ v_b \end{array} \right),
\label{Bogoliubov_ab}
\end{equation}
where the Bogoliubov operator $\mathcal{L}$ is defined in the Appendix \ref{app:BogolOp}. A clearer picture of the physics at hand can be drawn by expressing the excited components in the density $\psi_{d}=(\psi_{a1}+\psi_{b1})/2$ and spin $\psi_{s}=(\psi_{a1}-\psi_{b1})/2$ basis. Within this formalism, because of the symmetry of the underlying condensate, the excitations decouple into two channels, that are exactly the density and the spin branches. In terms of $\psi_d$ and $\psi_s$, the full eigenvalue problem in eq.(\ref{Bogoliubov_ab}) diagonalizes into the two independent sets of equations:
\begin{align}
	\mathcal{L}_d\left( \begin{array}{c} u_d \\ v_d \end{array} \right)&=\hbar\omega \left( \begin{array}{c} u_d \\ v_d\end{array} \right),\label{Bogoliubov_d}\\
	\mathcal{L}_s\left( \begin{array}{c} u_s \\ v_s \end{array} \right)&=\hbar\omega \left( \begin{array}{c} u_s \\ v_s\end{array} \right),
\label{Bogoliubov_s}
\end{align}
where
\begin{equation}
	\mathcal{L}_d= \begin{pmatrix}
	\frac{\hbar^2 k^2}{2m}+\frac{n}{2}(g+g_{ab})& \frac{n}{2}(g+g_{ab})\\
	-\frac{n}{2}(g+g_{ab})& -(\frac{\hbar^2 k^2}{2m}+\frac{n}{2}(g+g_{ab}))
	\end{pmatrix},
\label{L_D}
\end{equation}
and
\begin{equation}
	\mathcal{L}_s= \begin{pmatrix}
	\frac{\hbar^2 k^2}{2m}+\frac{n}{2}(g-g_{ab})+2|\Omega|& \frac{n}{2}(g-g_{ab})\\
	-\frac{n}{2}(g-g_{ab}) & -(\frac{\hbar^2 k^2}{2m}+\frac{n}{2}(g-g_{ab})+2|\Omega|)
	\end{pmatrix},
\label{L_S}
\end{equation}
are the Bogoliubov operators for the density and the spin excitations respectively, and we defined $u_{d}=(u_a+u_b)/2$, $u_{s}=(u_a-u_b)/2$ and analogously $v_{d}$ and $v_{s}$. The spectra of the two branches are obtained by diagonalizing the operators in Eqs.\eqref{L_D} and \eqref{L_S}, giving:
\begin{align}
	(\hbar\omega_d)^2&=\frac{\hbar^2 k^2}{2m}\left(\frac{\hbar^2 k^2}{2m}+(g+g_{ab})n\right),\label{Spectrum_D}\\
	(\hbar\omega_s)^2&=\frac{\hbar^2 k^2}{2m}\left(\frac{\hbar^2 k^2}{2m}+(g-g_{ab})n+4|\Omega|\right)\label{Spectrum_S}\\
	&+2|\Omega|\left[(g-g_{ab})n+2|\Omega|\right].\nonumber
\end{align}
In the Bogoliubov formalism, the positive (negative) energy solutions correspond to modes with the positive (negative) norm $||\psi_{i1}||^2=\int d\mathbf{r}\left(|u_i|^2-|v_i|^2\right)$. A peculiar feature of the spin excitations is the gap that appears in the low energy limit, which makes not obvious the definition of an effective metric for these (massive) quasi-particles \cite{Visser2005}. While this property may appear as a serious drawback in view of using these systems as analog models, we shall see in the following that its detrimental effect on the possibility of achieving supersonic regimes is not that dramatic and, rather, it opens interesting new perspectives.

\subsection{Moving condensate}
\vspace{3mm}The spectra in Eqs.\eqref{Spectrum_D} and \eqref{Spectrum_S} are obtained in the reference frame in which the condensate is at rest, i.e. in what is called the reference frame \emph{co-moving} with the superfluid. In the laboratory reference frame, in which the condensate is moving with velocity $v$, the corresponding expressions are obtained by the usual Galilean transformation for the energy, that is by substituting $\hbar\omega_i\rightarrow\hbar\omega_i+\hbar k v$, with $\left(i=d,s\right)$ in Eqs.\eqref{Spectrum_D} and \eqref{Spectrum_S}. 

For the density modes, the supersonic regime is attained as usual once the background velocity of the system exceeds the sound speed $c_d$, locally defined as $c_d=\lim_{k\to 0} \omega_d(k)/|k|=\sqrt{\left(g+g_{ab}\right)n/2m}$. On the other hand, the definition of the supersonic regime for the spin excitations is slightly more involved since, in the low energy limit, the corresponding dispersion relation is not linear, and so a characteristic speed for these excitations cannot be defined in the usual sense. To overcome this ambiguity, we adopt a more general definition of the supersonic regime when there exists a spin mode with positive (negative) norm having a negative (positive) energy in the laboratory frame. 

As can be inferred from Figs. \ref{fig:DispRelOUT_NONlasing},\ref{fig:DispRelIN_NONlasing},\ref{fig:DispRelOUT_lasing} and \ref{fig:DispRelIN_lasing}, the number of real solutions for a fixed value of the energy strictly depends on the particular hydrodynamical regime: whenever the flow is subsonic, two positive (negative) norm solutions exist with wave-vectors $k=k_{R1}$ and $k=k_{L1}$ for each $\omega>\omega_t$ ($\omega<-\omega_t$) above a threshold value ($R$, $L$ in Fig. \ref{fig:DispRelOUT_NONlasing},\ref{fig:DispRelIN_NONlasing},\ref{fig:DispRelOUT_lasing}). Throughout the article, the different branches of the solutions will be labeled as ``$R$'' and ``$L$'', depending on whether their group velocity $v_g=d\omega/dk$ is positive (right-moving solutions) or negative (left-moving solutions). Since the dispersion relation is of fourth order in $k$, two other complex solutions exist, conjugated to each other, representing evanescent waves and so unable to sustain a propagating particle. 

If the flow is supersonic instead, the scenario is much richer since for $|\omega|<|\omega'_t|$ below a threshold value $\omega'_t$, two additional positive (negative) norm propagating solutions appear with negative (positive) frequency $\omega<0$ ($\omega>0$) ($R2$, $L2$ in Fig. \ref{fig:DispRelIN_lasing}). The presence of these extra solutions is exactly what makes possible the onset of the Hawking physics: pairs of particles with opposite energy can be in fact created, while conserving the total energy of the system. Thanks to the superluminal (quadratic for the Bogoliubov modes) dispersion relation. the BH-lasing effect can then be established once the system presents a finite supersonic region enclosed between a pair of white and black hole horizons and playing the role of a lasing cavity.

\begin{figure*}[!t]
\centering%
\subfigure[\label{fig:two_levels}]
{\includegraphics[height=2.5cm]{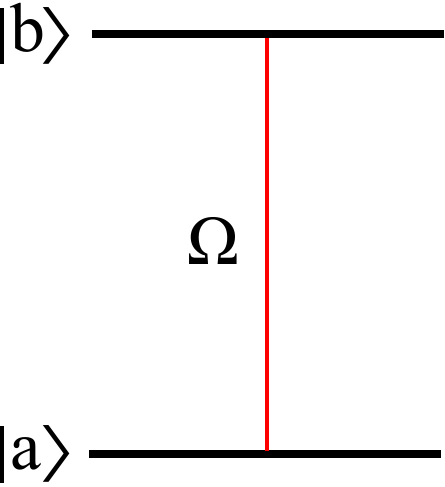}}\qquad
\subfigure[\label{fig:lambda_levels}]
{\includegraphics[height=2.5cm]{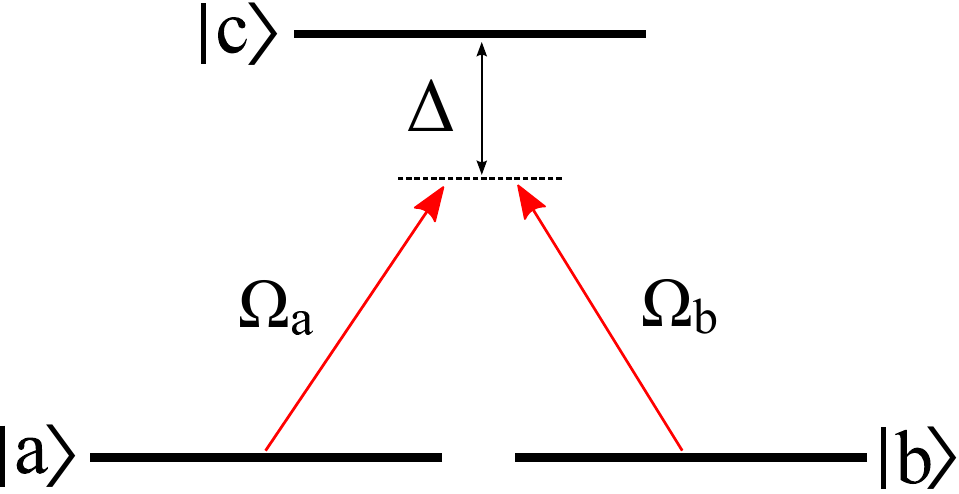}} \qquad
\subfigure[\label{fig:Exp_1}]
{\includegraphics[width=0.26\linewidth]{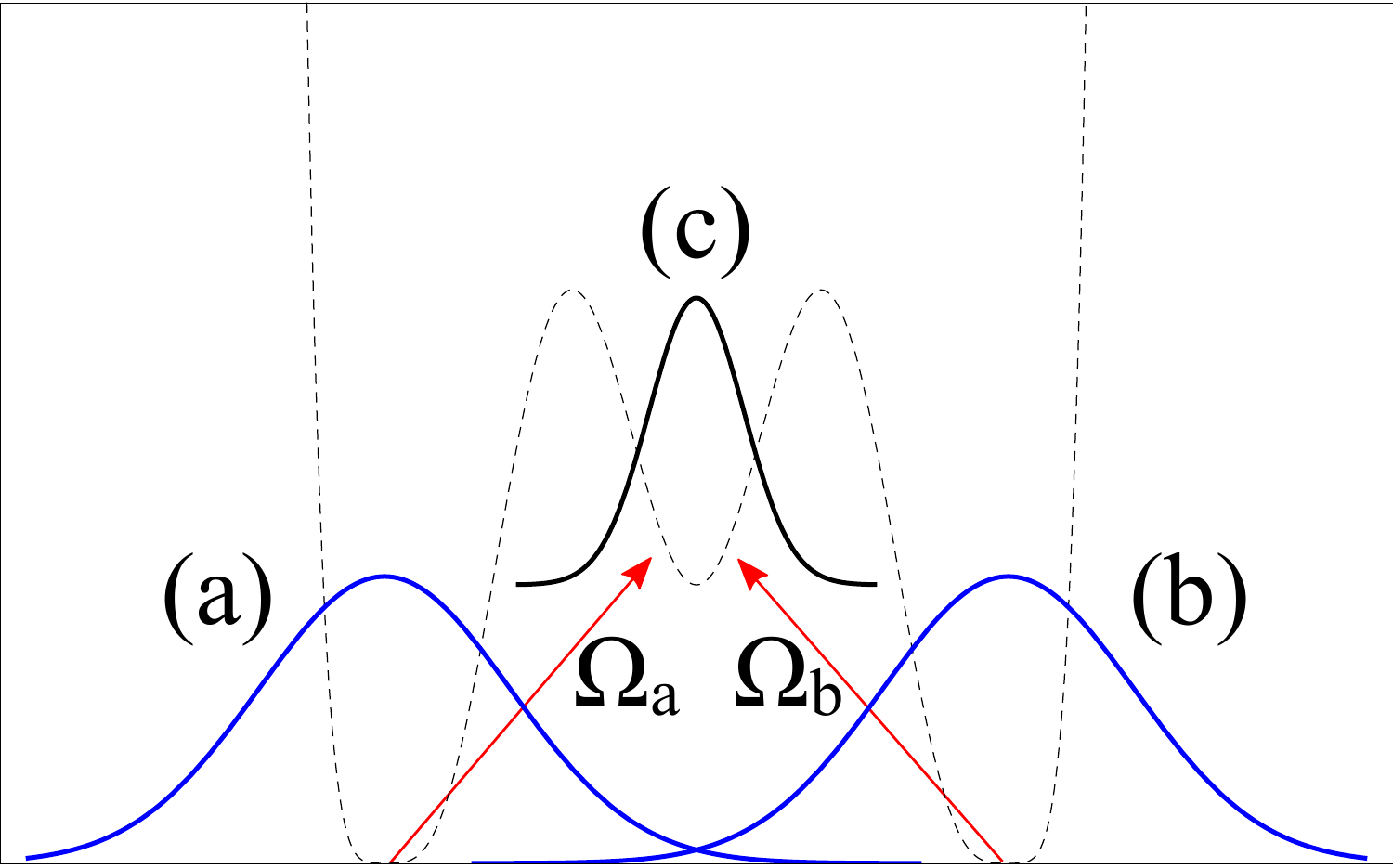}}\qquad
\subfigure [\label{fig:Exp_2}]
{\includegraphics[width=0.17\linewidth]{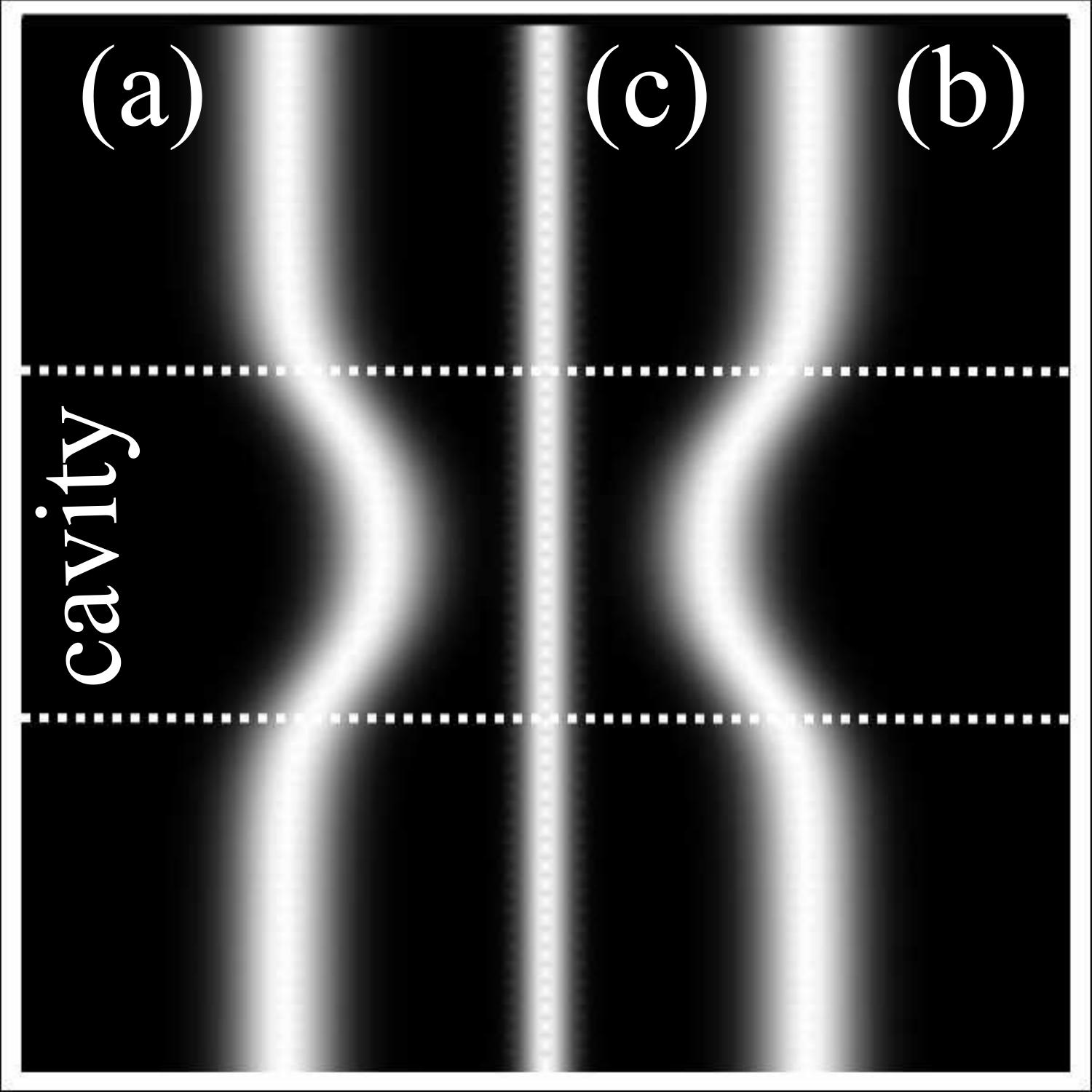}}
\caption{Different schemes for the coherent coupling of the atomic levels. a) Two-level model. b) $\Lambda$-structure model. c) Scheme of $\Lambda$-model implemented with an elongated atomic cloud in a three-tube inhomogeneous potential. d) Here, the hopping strength between  $a-b$ and $b-c$ tubes can be controlled via the distance between the clouds.}
\label{fig:Exp}
\end{figure*}

\section{Experimental considerations \label{sec:exp}}
\subsection{Two-level atom \label{sec:expA}}
The simplest atomic configuration which involves a coherent coupling in a spinorial system is represented by the two level scheme, where the coupling between the internal states $\left\{\ket{a},\ket{b}\right\}$ is given by a single resonant electromagnetic field electric-dipole coupled to the transition between them. In such a basis, and in the rotating wave approximation (RWA) in which the rapidly oscillating terms can be averaged to zero in favour of the resonant ones \cite{loudonBook,cohenBook}, the Hamiltonian $H$ modelling the internal dynamics of the atom can be written as
\begin{equation}
	H=H_0+U
\label{H_atom}
\end{equation}
where
\begin{align}
	H_0&=\omega_0\ket{b}\bra{b}\label{H_0}\\
	U&=\Omega e^{-i\omega_L t}\ket{b}\bra{a}+\Omega^* e^{i\omega_L t}\ket{a}\bra{b}\label{U_2_lev}
\end{align}
are respectively the Hamiltonian of the free atom (with the ground state energy set to zero) and the operator describing the dipole interaction $U=-\mathbf{d}\cdot\mathbf{E}$, with $\mathbf{d}$ the electric dipole moment operator of the atom and $\mathbf{E}$ the amplitude of the electric field. In Eq. \eqref{U_2_lev}, $\Omega$ is the Rabi frequency, which characterizes the strength of the light-matter interaction, and is defined in terms of the matrix element of the atomic electric dipole operator between the two internal states $\mathbf{d}_{ab}=\Melement{a}{b}{\mathbf{d}}$ with $\Omega=-\mathbf{d}_{ab}\cdot \mathbf{E}/2$, while $\omega_L$ is the frequency of the electric field. This type of coupling is the one appearing in the GP Eqs.\eqref{GP_a} and \eqref{GP_b}.\\
A lasing cavity can be established in the system by properly tuning the spatial profile of $g,\, g_{ab}$ in order to induce the lasing in the density modes~\cite{deNova2016}, or of $ |\Omega|$ in the case of the spin modes. The former can be done by exploiting for example Feshbach resonances in order to modify the scattering lengths of the atomic interactions. The latter by tuning the amplitude of the electric field of the laser beam which couples the internal states of the atoms.

Unfortunately, the local modification of these physical parameters induces an inhomogeneity in the effective chemical potential according to (\ref{n}) and, consequently, may be responsible for spurious excitations in the condensate via, e.g., a Bogoliubov-Cherenkov mechanism~\cite{Carusotto:PRL2006}. Unless this effect is compensated by a suitably tailored external potential~\cite{Carusotto2008}, the Hawking emission and the self-amplification mechanism of black hole lasing may be then obscured. The required compensation appears to be experimentally very challenging since it calls for a rather extreme fine and possibly even atom-number-dependent tuning of the experimental parameters. Possible solutions based on, e.g., the spontaneous horizon generation mechanism of \cite{Pavloff2012} seem restricted to black hole configurations and, in addition, involve a very complicate horizon formation hydrodynamics already at mean-field level, with unknown consequences on the quantum fluctuations.

\subsection{$\Lambda$-scheme}
\label{sec:lambda}
This difficulty can be overcome by replacing the simple two level configuration with a  Raman-coupled $\Lambda$ scheme of internal atomic levels as shown in Fig.\ref{fig:lambda_levels}. For simplicity, and without affecting the generality of the following arguments, we consider the case of a two-fold degenerate ground state spanned by the two $\left\{\ket{a},\ket{b}\right\}$ states. Each of them is coupled to the excited state $\ket{c}$ by a far off-resonant electromagnetic field of Rabi frequency $\Omega_{a,b}$ and detuned by $\Delta$ from atomic resonance. In these conditions, the operator in Eq. \eqref{U_2_lev}, which describes the internal coupling, can be straightforwardly generalized as
\begin{multline}
	U_{\Lambda}=e^{-i\omega_L t}\left(\Omega_a \ket{c}\bra{a}+\Omega_b \ket{c}\bra{b}\right)\\
	+e^{i\omega_L t}\left(\Omega_a^* \ket{a}\bra{c}+\Omega_b^* \ket{b}\bra{c}\right).
\label{U_lamda}
\end{multline}
The Schr\"odinger equation describing the full internal dynamics of the atoms, is given as
\begin{equation}
	\left[H_0+U_\Lambda\right] \ket{\psi}= E\ket{\psi}.
\label{H_lambda_full}
\end{equation}
where $\ket{\psi}$ is the internal state of the atom, which consist of a linear combination of the basis vectors $\left\{\ket{a},\ket{b},\ket{c}\right\}$, and $E$ its energy. By projecting Eq.\eqref{H_lambda_full} onto the excited state (via the projector $Q$) and onto the two-fold degenerate ground state manifold (via the projector $P$), we get
\begin{align}
	&H_0\left(Q\ket{\psi}\right)+Q U_\Lambda P\left(P\ket{\psi}\right)=E\left(Q\ket{\psi}\right),\label{P_proj}\\
	&H_0\left(P\ket{\psi}\right)+P U_\Lambda Q\left(Q\ket{\psi}\right)=E\left(P\ket{\psi}\right),\label{Q_proj}
\end{align}
We solve now for $\left(Q\ket{\psi}\right)$ from Eq.~\eqref{P_proj} and substitute the result into Eq.~\eqref{Q_proj}. We obtain the effective Sch\"odeinger equation for the ground manifold dynamics, which reads
\begin{equation}
	\left[H_0+U_{\text{eff}}\right]\left(P\ket{\psi}\right)=E\left(P\ket{\psi}\right),
\label{LambdaAtEffHam}
\end{equation}
with
\begin{equation}
	U_{\text{eff}}(E)=(PU_{\Lambda}Q)\,\frac{1}{E-H_0}\,(QU_{\Lambda}P),
\label{LambdaAtEffCoupl}
\end{equation}
the energy dependent effective coupling Hamiltonian. The Eq.~\eqref{LambdaAtEffCoupl} is still exact at this stage. Because of the far off-resonance nature of the coherent coupling to the $\ket{c}$ state, the interaction $U_\Lambda$ can be treated as a small perturbation with respect to the free Hamiltonian $H_0$. In other words this means considering $|\kappa_{a}|, |\kappa_{b}|\ll \Delta$. By evaluating Eq.~\eqref{LambdaAtEffCoupl} for $E=E_0$, with $E_0$ the unperturbed ground states energy, we finally get the effective coupling Hamiltonian up to second order in $U_\Lambda$ that, in the usual time independent form, reads
\begin{multline}
	U_{\text{eff}}(E_0)=\left[\frac{|\Omega_a|^2}{\Delta}\ket{a}\bra{a}+\frac{|\Omega_b|^2}{ {\Delta}}\ket{b}\bra{b}\right.\\
	\left.+\frac{\Omega_a^*\Omega_b}{\Delta}\ket{a}\bra{b}+\frac{\Omega_a\Omega_b^*}{ {\Delta}}\ket{b}\bra{a}\right].
\label{H_eff}
\end{multline}

This Hamiltonian Eq. \eqref{H_eff} describes the consequences on the ground state manifold of the coupling to the excited state $\ket{c}$. On one hand, it  shows the same type of non-diagonal terms as in Eq.\eqref{U_2_lev}, with the coherent coupling amplitude $\Omega$ replaced by the product $\Omega_a\Omega_b^*/\Delta$. On the other hand, new diagonal terms appear that were not present in the simple two-level model. The presence of these additional external potential terms to be included in $V_{a,b}$ represents the main advantage in considering the atomic $\Lambda$-configuration instead of the simpler two-level scheme.

This mechanism is most effective if we take Raman fields with real and opposite $\Omega_a=-\Omega_b$ amplitudes and a positive detuning $\Delta>0$. In this case, the effective coherent coupling reads $\Omega=\Omega_a\Omega_b/\Delta=-|\Omega_a|^2/\Delta<0$ and the extra potential term is $|\Omega_a|^2/\Delta>0$ for both $\ket{a,b}$ atomic states.
Inserting these results into the state equation Eq.\eqref{n}, it is immediate to see that the two contributions cancel and the density no longer depends on the coherent coupling: physically, this can be understood as a destructive interference effect canceling the transition amplitude to the excited state from the condensate mode, which acquires a dark state character. 

As a result, the flow of the condensate is not affected by the presence of the Raman fields, while the dispersion of its spin modes is. Arbitrary multi-horizon configurations can therefore be created by simply shaping the spatial profile of the Raman fields, with no need for any tricky compensation.

\label{atomic_models}

\subsection{Tailoring the cross-state scattering length \label{sec:expB}}
A further issue concerning the experimental observation of the BH-lasing in the spin modes is related to the fact that the coupling between the positive and negative norm solutions of the Bogoliubov spin excitations is proportional to the difference $g-g_{ab}$. This coupling is represented by the off-diagonal terms in the operator $\mathcal{L}_s$ in Eq. \eqref{Bogoliubov_s}. For typical atoms used in modern cold-atoms experiments, the values of the cross-state scattering length $a_{ab}$ are close to the values for the atoms in the same internal state (in $\text{Rb}^{87}$ for example $g_{ab}\approx 0.97\,g$ \cite{Oberthaler2015}). This would then strongly suppress the conversion, at the two horizons, of positive norm into negative norm modes and vice versa, practically inhibiting the lasing phenomenon. 

In order to attain higher values of $|g-g_{ab}|$, a variety of different techniques can be used. As a most direct idea, Feshbach resonances can be exploited for example in order to modulate the relative value of the cross-state vs. same-state scattering lengths. 

Another choice may be to use a one-dimensional atomic condensate split among three parallel tubes as sketched in Fig.\ref{fig:Exp_1}. In this way, contact interactions within the same tube are much stronger than the ones between different tubes which involve the superposition of the two condensate wave functions, giving a much suppressed $g_{ij}\simeq 0$ for any $i\neq j$. In this configuration the coherent coupling is provided by the hopping of particles from the clouds $a$ and $b$ into $c$, a process that could even be artificially assisted with external electromagnetic fields if necessary: A lasing region can be prepared by tuning in space the hopping strength, e.g. by opportunely tailoring the spatial profile of the fields and/or the geometry of the tube potential by varying the distance between the atomic guides $a$ and $b$ or the height of the potential which confines the cloud $c$.

\section{Homogeneous condensate\label{sec:ring}}
We start in this section the study of the BH-lasing in the spin branch of the excitations introduced in Sec. \ref{sec:system}, by considering first the simplest case of a longitudinally homogeneous 1D system. Effective quasi-1D systems can be experimentally realized by tightly confining the gases in the transverse direction, freezing-out the corresponding atomic degree of freedoms. The prefix ``quasi'' refers to the fact that, even if the kinematic of the system is effectively 1D, the interparticle interactions preserve a 3D character. Labelling with $\omega_y$ and $\omega_z$ the frequencies of the trapping potential in the transverse plane such a condition is achieved for temperatures $k_B T\ll \hbar\omega_y,\, \hbar\omega_z$.

We simulate an infinite homogeneous system by considering a ring geometry with periodic boundary conditions. An opportunely designed absorbing region where the excitations above the homogeneous and uniformly flowing condensate are suppressed (see Appendix \ref{app:absBC}) is inserted at a location opposite to the region of interest. This prescription avoids the occurrence of interference between excitations turning around the ring and allows to simulate a spatially unbounded system, where the excitations leaving the region around the cavity can escape to infinity.

As we have discussed in the previous Sec.\ref{sec:lambda} and Sec.\ref{sec:expB}, it is possible to selectively prepare a lasing cavity for the spin excitations without affecting the density degrees of freedom by modulating the spatial profile of $\Omega$ while keeping $g$ and $g_{ab}$ homogeneous throughout the system. In the following numerical simulations, we use the simplest possible configuration, that is a step-like profile, with the value of $\Omega$ chosen in such a way that the spin modes attain the supersonic regime in a finite region (the lasing cavity), while they are subsonic elsewhere. Of course, the external potential $V_{a,b}$ must include the compensating terms discussed in Sec.\ref{sec:lambda}.

\subsection{Wave packet propagation \label{sec:packet}}

We first analyse the propagation of a wave packet through the ring, in the absence and in presence of the lasing conditions, so to emphasize the dramatically different behaviour of the condensate in the two cases. From the experimental point of view, a spin excitation wavepacket can be created in the condensate by generalizing standard Bragg techniques~\cite{steinhauer:PRL2003} to the spinful case. A suitable tayloring of the polarization of the Bragg beams can be used to suppress the parasitic excitation of density excitations~\cite{carusotto2006} and the time-evolution of the spatial spin density profile can be detected with {\em in situ} techniques~\cite{Seo:PRL2015}. Even though our predictions can be in principle verified in single-shot experiments of the condensate dynamics, the deterministic nature of the Bragg scattering process allows to increase the signal-to-noise ratio by averaging over many shots of the same experiment.

When the lasing condition is not fulfilled, we recover the standard physics of a wave propagating in an inhomogeneous medium, incurring in the transmission and reflection phenomena in correspondence of the discontinuities of the coherent coupling amplitude. In the opposite case instead, the appearance inside the cavity of negative norm solutions with positive energy strongly enriches the physics at hand: As it is shown in the following, the availability of such modes enables the conversion between negative and positive norm waves at the two horizons, which is responsible for the black-hole lasing phenomenon. At late times, the dynamical instability due to the presence of these negative norm solutions starts dominating the scene as an exponential amplification of the excitation inside the cavity and of the emitted Hawking radiation.

The results we show in the following are expressed in the dimensionless unit $L'=L/\xi$, $t'=\hbar t/(2m\xi^2)$, $\hbar\omega'=\hbar\omega/gn$ and $n'=n\xi$ for lengths, time, energy and density respectively, where the healing length $\xi=\sqrt{\hbar^2/2mgn}$ and $g$ is hereafter the effective 1D meanfield coupling constant, defined in terms of the corresponding 3D value $g_{3D}$ and the effective transverse area $S_t$ of the 1D cloud as $g=g_{3D}/S_t$. We consider in the following simulations a condensate with $N=10^3$ atoms, leftward flowing through a ring of length $L'=1000$, with an initial phase profile $\theta(x')$ for both the order parameters which is linear and characterized by the winding number $w=\left(\theta(L')-\theta(0)\right)/2\pi=90$. For standard condensate clouds with $m\sim 10^{-25}\, \text{Kg}$ and $\xi\sim 1 \mu\text{m}$ this choice results in a flow velocity $v$ of the order $v\sim 1\,\text{mm/s}$. As discussed above, we can restrict ourselves to a constant value for the meanfield coupling strength, and in particular we take $g_{ab}/g=0.8$.

\begin{figure*}[htbp]
\centering%
\subfigure[\label{fig:F1}]
{\includegraphics[width=0.8\linewidth]{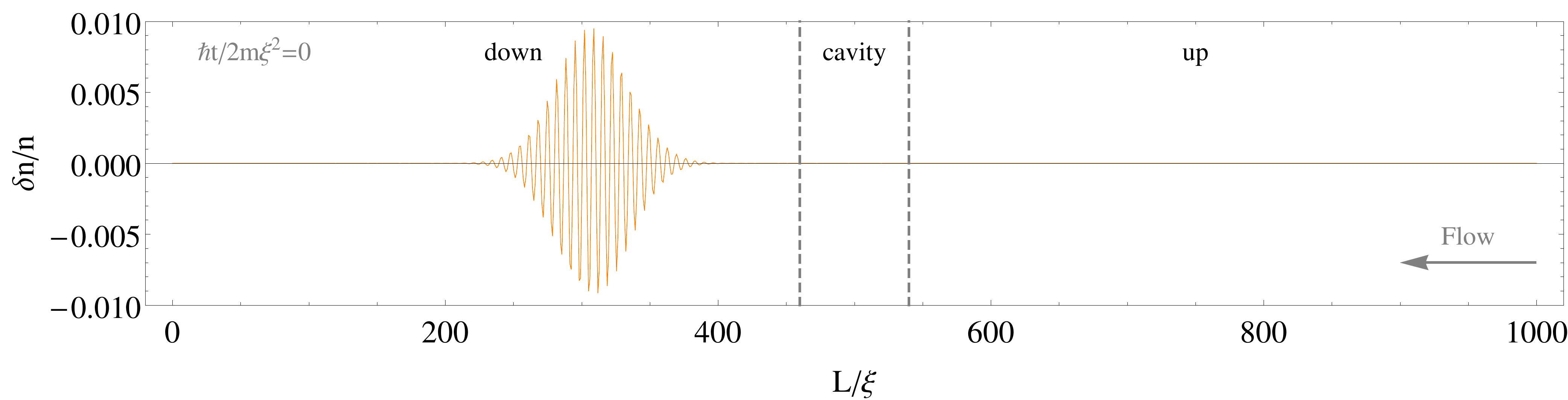}}\\
\subfigure [ \label{fig:F2}]
{\includegraphics[width=0.8\linewidth]{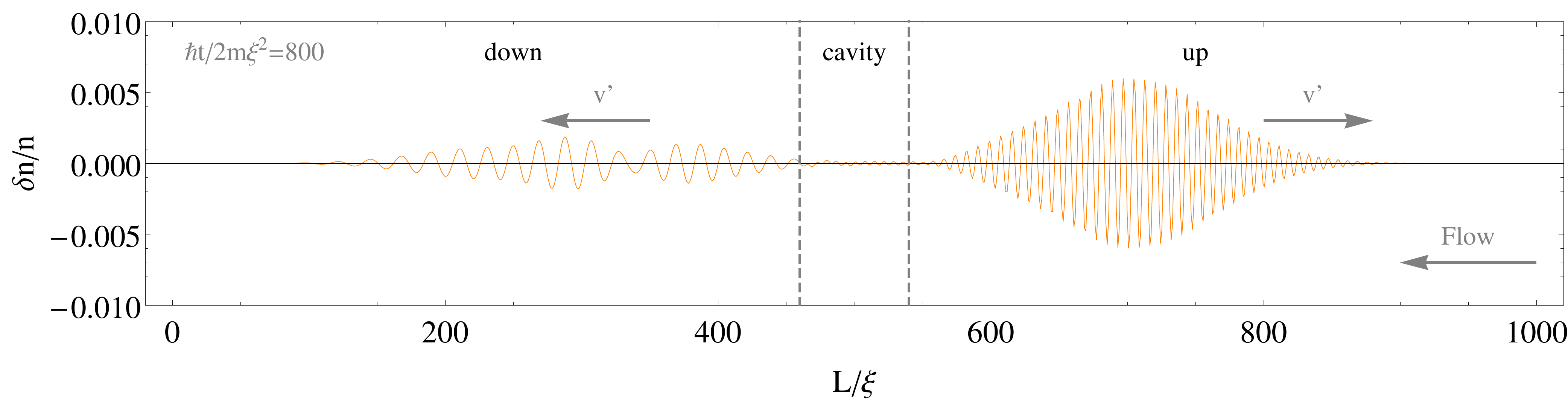}}\\
\subfigure [\label{fig:F3}]
{\includegraphics[width=0.8\linewidth]{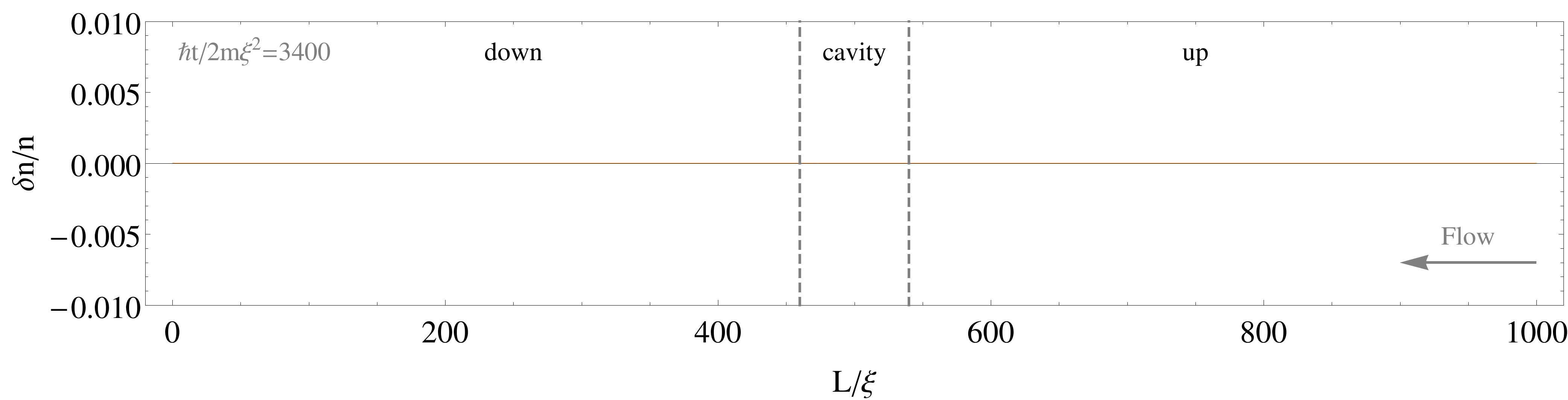}}\\
\subfigure [\label{fig:F4}]
{\includegraphics[width=0.8\linewidth]{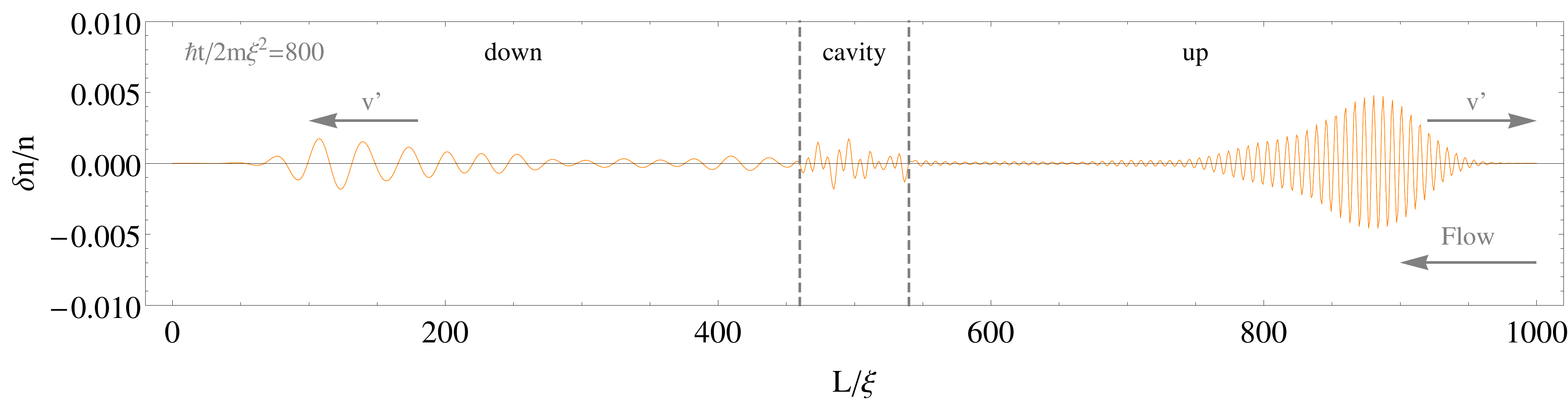}}\\
\subfigure [\label{fig:F5}]
{\includegraphics[width=0.8\linewidth]{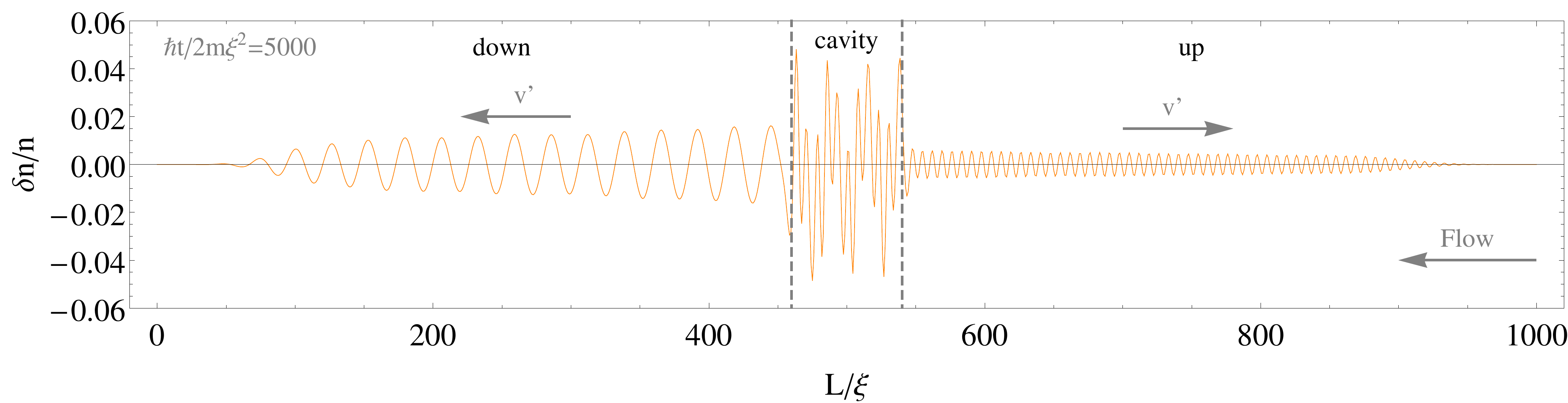}}
\caption{Time evolution of the wave-packet in real space. The dashed lines enclose the cavity region. a) Initial condition of the simulation. b-c) State of the spin density $\delta n$ of the system respectively at $\hbar t/2m \xi^2=800\text{ and }3400$, for the non-lasing conditions defined by the set of parameters given in the text, and corresponding to the dispersion plotted in Fig.\ref{fig:DispRel_Nonlasing}. d-e)  State of the spin density $\delta n$ of the system respectively at $\hbar t/2m \xi^2=800\text{ and }5000$, for the lasing condition defined by the set of parameters given in the text, and corresponding to the dispersion plotted in Fig.\ref{fig:DispRel_lasing}. The arrows indicate the group velocity $v'$ of the different emerging wavepackets.}
\label{fig:F}
\end{figure*}

\begin{figure*}[htbp]
\centering%
\subfigure[\label{fig:L1}]
{\includegraphics[width=0.32\linewidth]{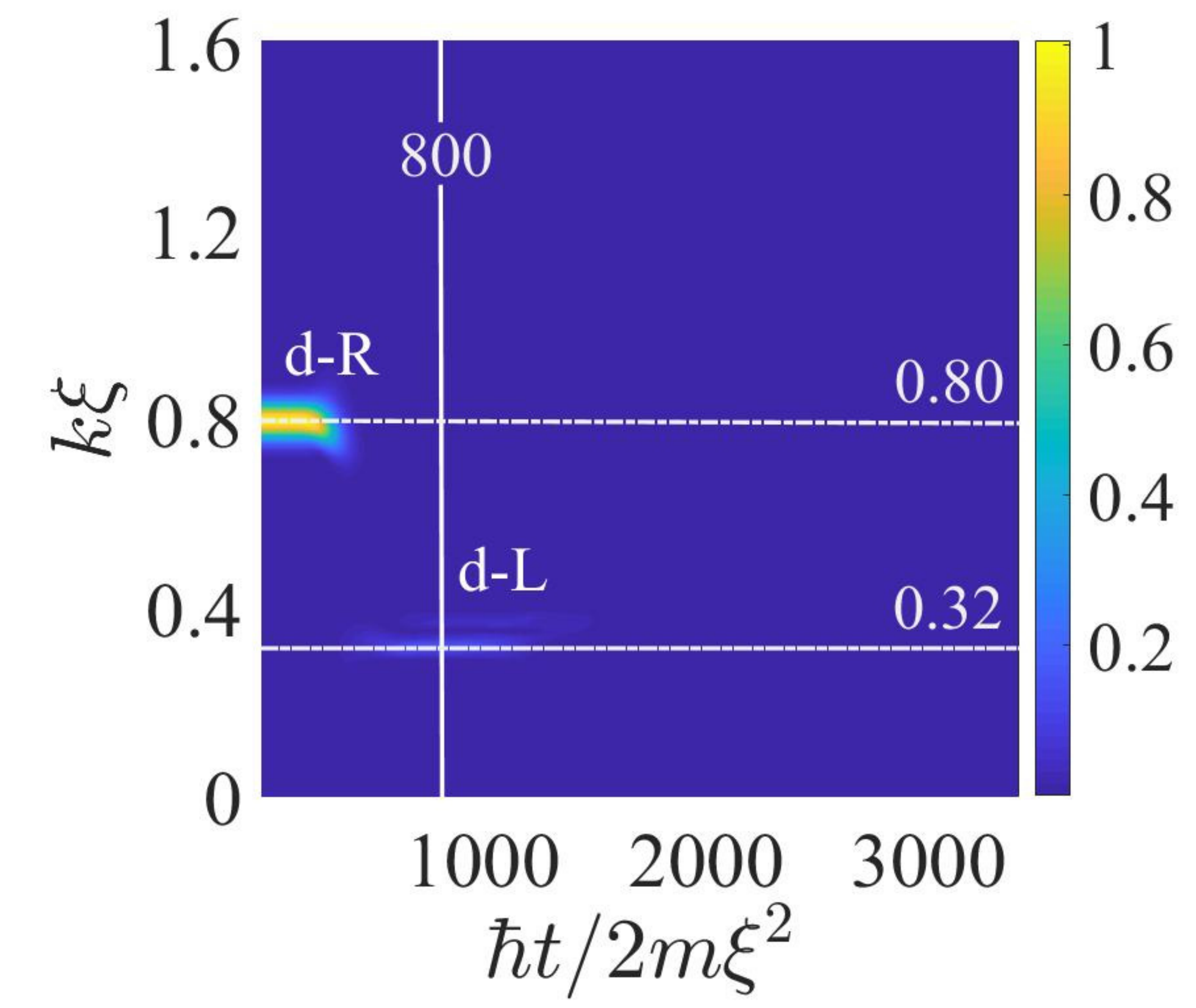}}
\subfigure [\label{fig:L2}]
{\includegraphics[width=0.32\linewidth]{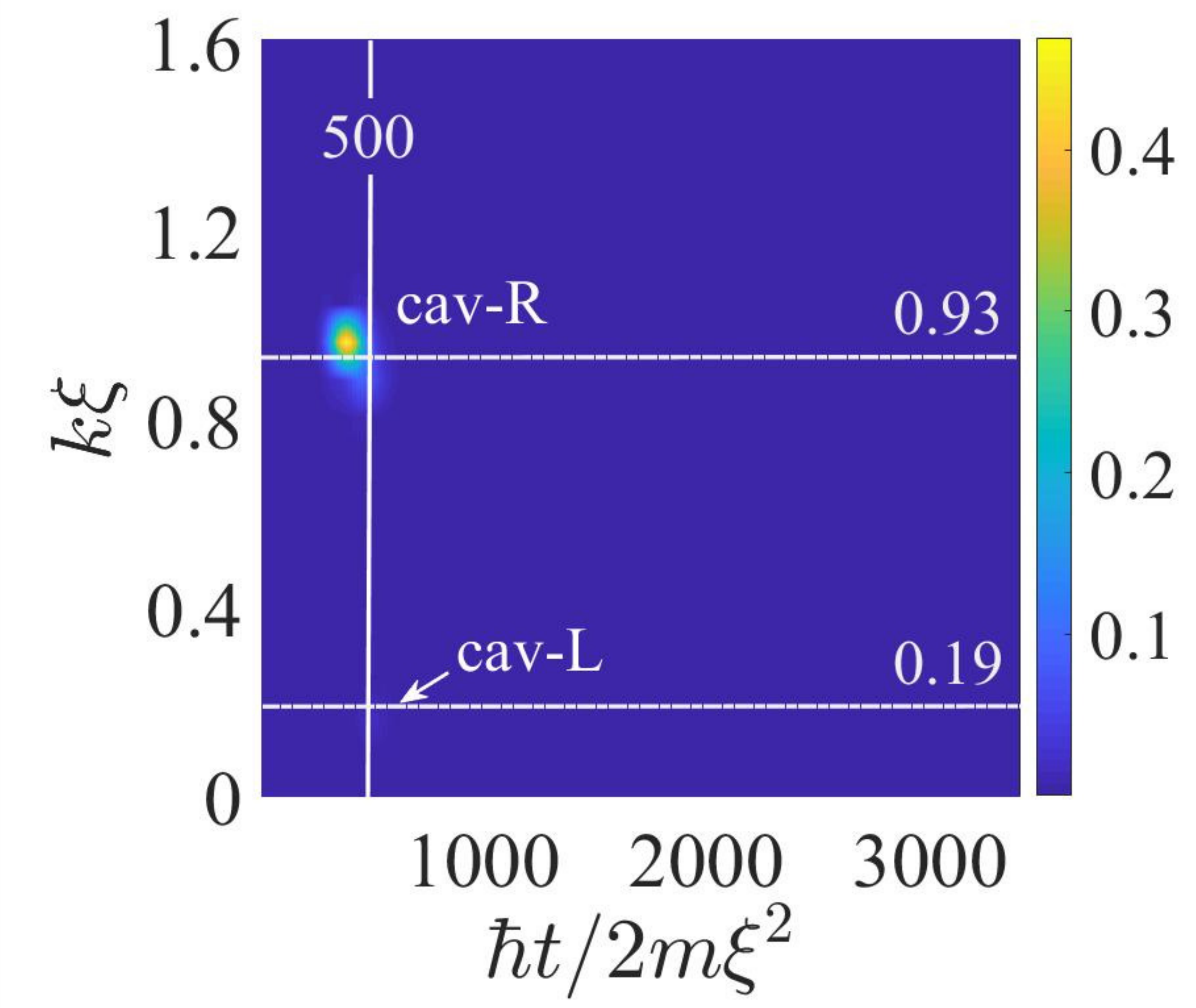}}
\subfigure [\label{fig:L3}]
{\includegraphics[width=0.32\linewidth]{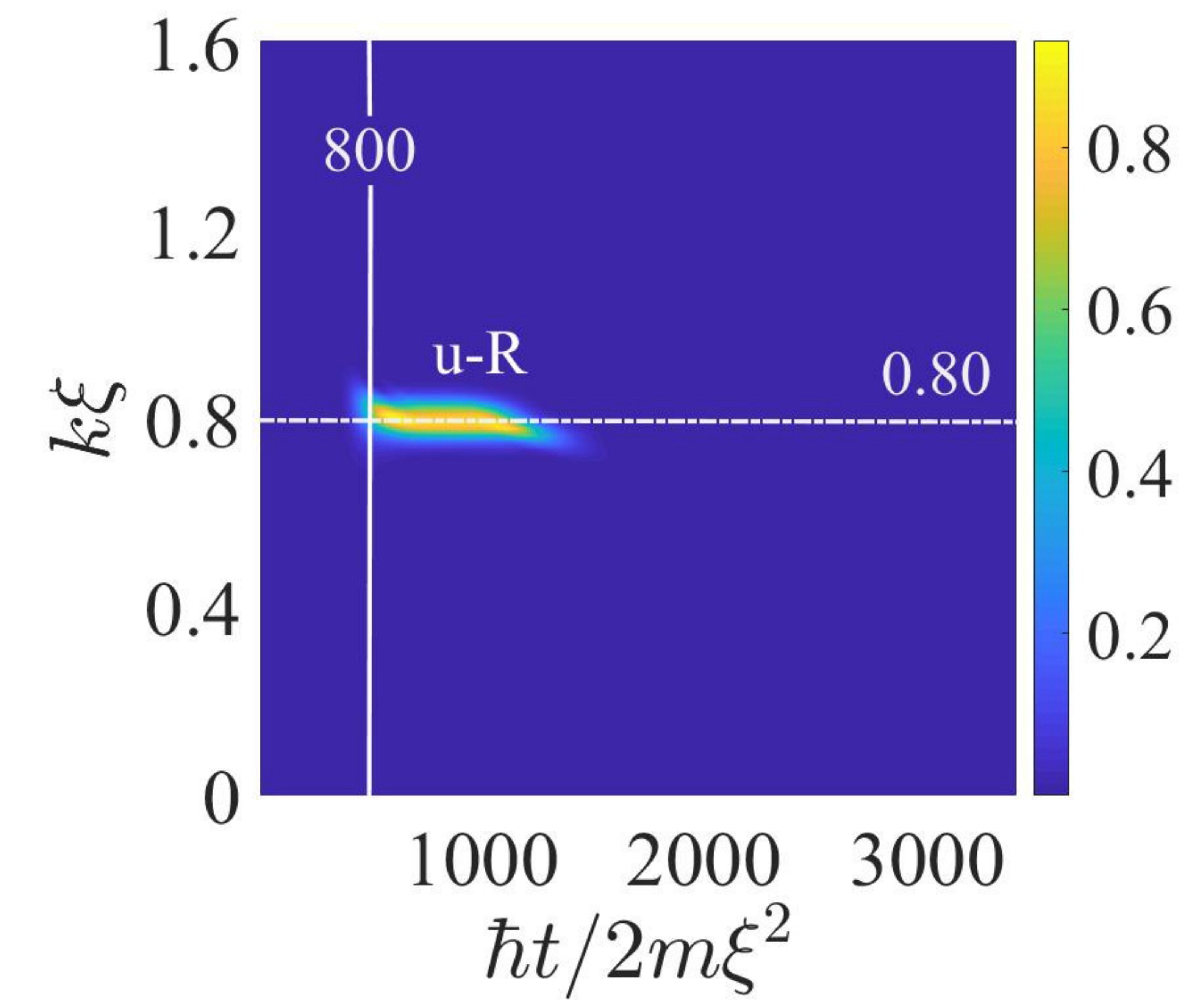}}
\caption{Non-lasing case: evolution in time of the spatial Fourier transform of the spin density of the system in the different regions. a) Downstream region. The wave-vector components of the \emph{d-R} packet given as initial condition for the simulation appear, together with the \emph{d-L} component due to its reflection from the WH. b) Cavity region. The \emph{cav-R} packet transmitted through the WH appears. The \emph{cav-L} component due to the reflection of the \emph{cav-R} packet at the BH is indicated by an arrow but is hardly visible in the figure. c) Upstream region. The \emph{u-R} packet transmitted through the cavity appears. System parameters as in Fig.\ref{fig:F}(b,c).}
\label{fig:L}
\end{figure*}

\begin{figure*}[htbp]
\centering%
\subfigure[\label{fig:M1}]
{\includegraphics[width=0.45\linewidth]{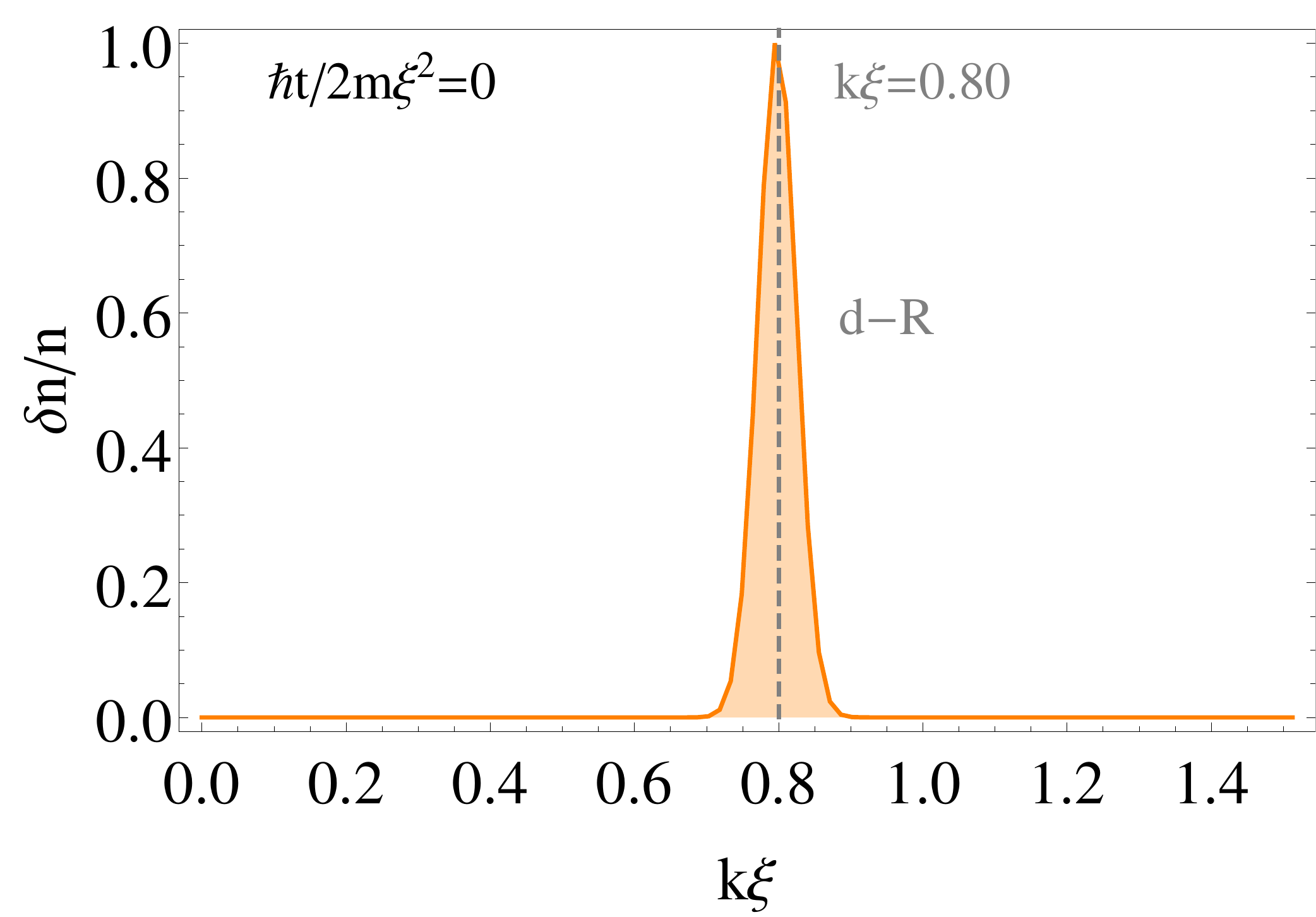}}\quad
\subfigure [\label{fig:M2}]
{\includegraphics[width=0.45\linewidth]{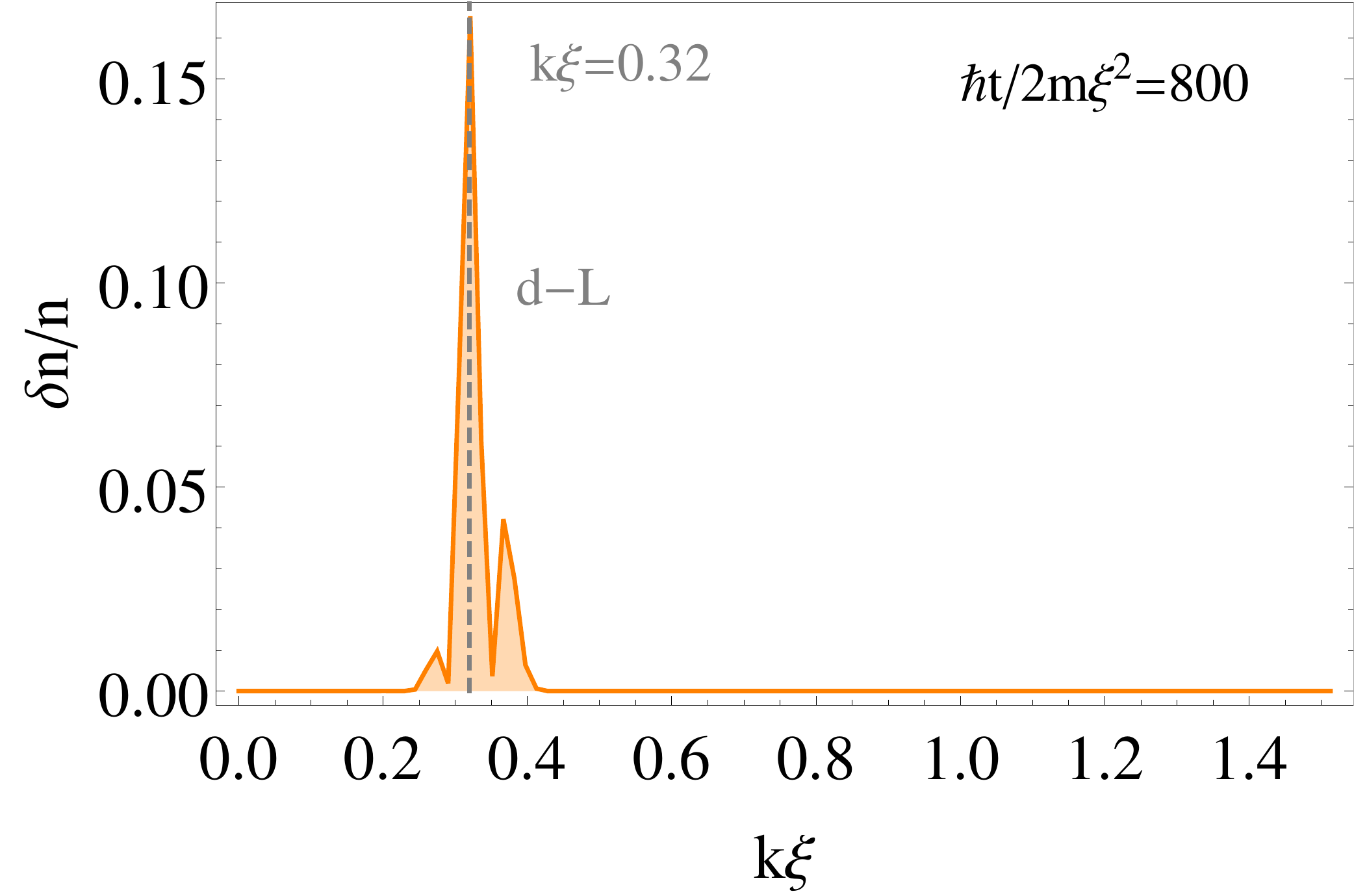}}\\
\subfigure [\label{fig:M3}]
{\includegraphics[width=0.45\linewidth]{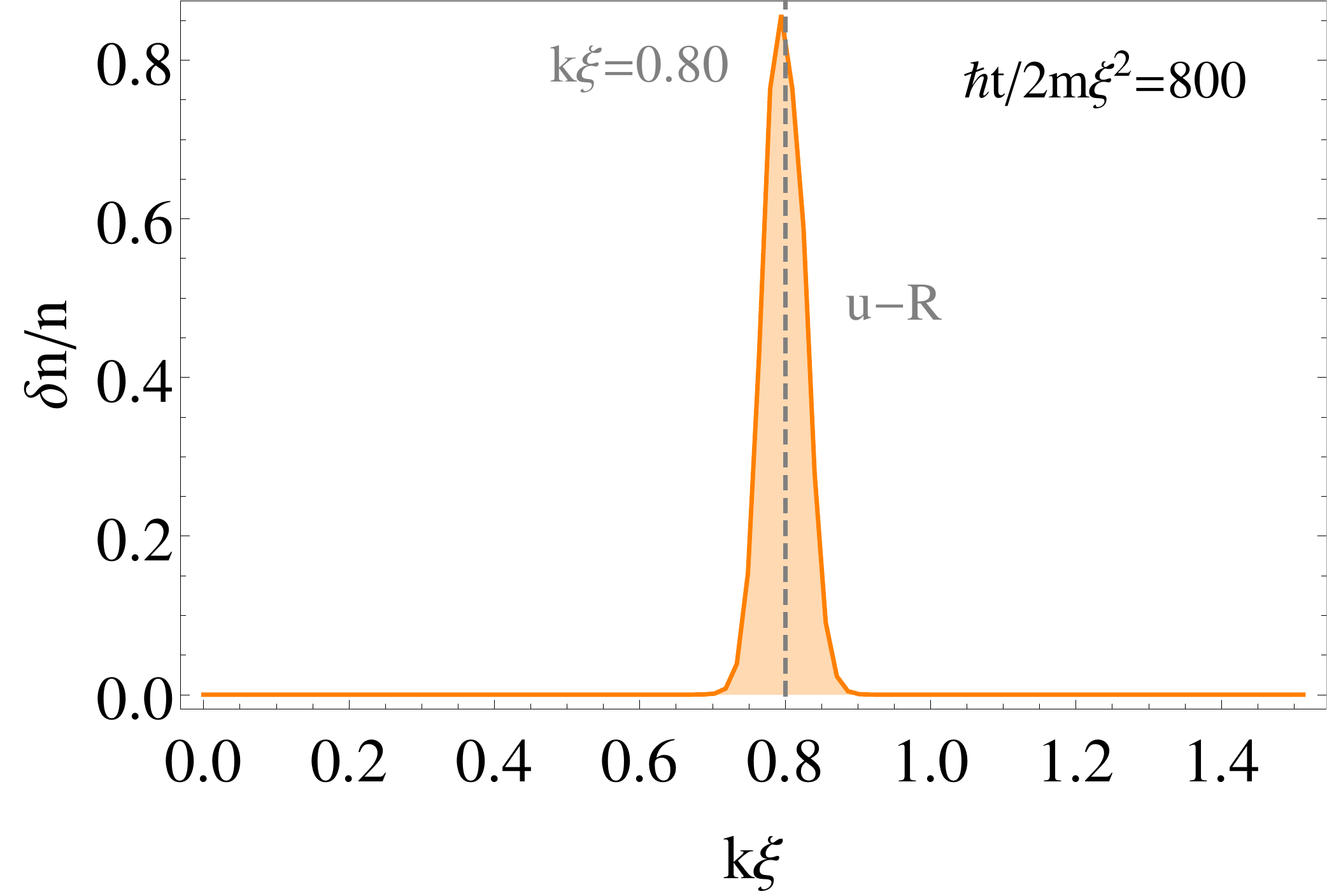}}\quad
\subfigure [\label{fig:M4}]
{\includegraphics[width=0.45\linewidth]{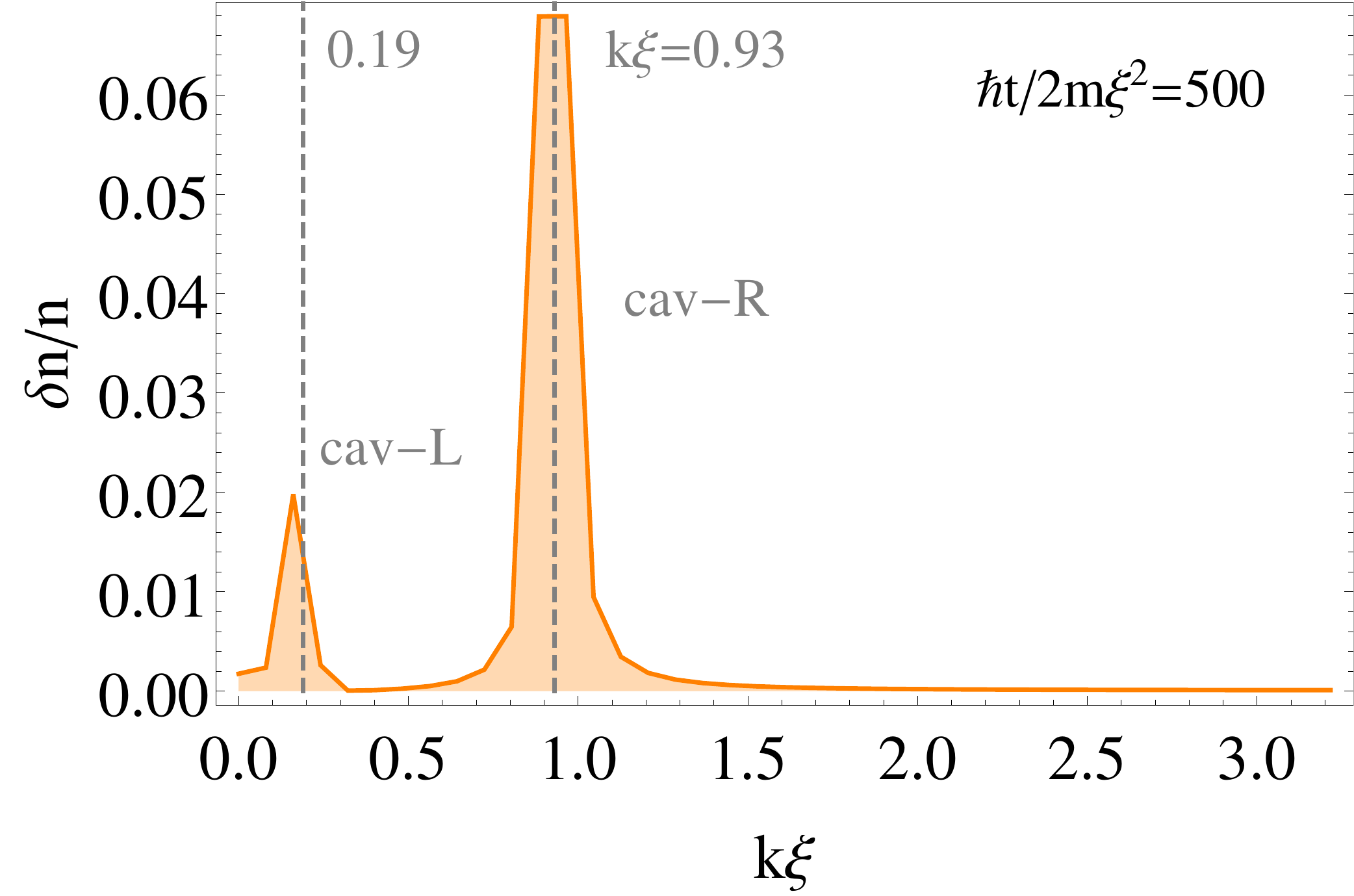}}
\caption{Non-lasing case: spatial Fourier transform of the spin density of the system at different instants of time, showing the spectral content of the transmitted and reflected packets in the downstream, upstream and cavity regions. The dashed lines identify the central wave-vector expected from the dispersion relations shown in Figs. \ref{fig:DispRelOUT_NONlasing} and \ref{fig:DispRelIN_NONlasing}. a) Downstream region, $\hbar t/2m\xi^2=0$: initial condition. b) Downstream region, $\hbar t/2m\xi^2=800$: wave packet reflected by the WH. c) Upstream region, $\hbar t/2m\xi^2=800$: wave packet transmitted through the cavity. d) Cavity region, $\hbar t/2m\xi^2=500$: wave packets transmitted through the WH and reflected by the BH.  System parameters as in Figs.\ref{fig:F}(b,c) and \ref{fig:L}.}
\label{fig:M}
\end{figure*}

\subsubsection{Non-lasing regime}
We start our discussion from a case where the spin excitations are everywhere subsonic and lasing can not occur. Specifically, we choose a value $\Omega'=-0.16$ for the coherent coupling amplitude, except for a region of length $L'_c=80$ (hereinafter referred to as the ``cavity") around $x'=500$, where $\Omega'=-0.12$. With these choices, the spin excitations are everywhere subsonic, and the dispersion relation takes the form shown in Figs. \ref{fig:DispRelOUT_NONlasing} and \ref{fig:DispRelIN_NONlasing}, respectively outside and inside the cavity. The absorbing region is located around the edges of the ring at $x'=0$ and $x'=1000$.

We take as initial condition of the simulation (at $t'=0$) a homogeneous condensate hosting an excitation wave packet as shown in Fig. \ref{fig:F1}. The excitation wave packet is located in the (down-stream) region to the left of the cavity and propagating rightwards, with energy $\hbar\omega'=0.15$ in the laboratory frame. The corresponding wave-vectors are a gaussian distribution around the value $k'\equiv k\xi=0.80$, in the \emph{d-R} branch of the dispersion relation (hereafter we will add the prefixes \emph{d,u,cav} to the mode labels, in order to identify the downstream, upstream and cavity regions of the system to which they refer), with width $\Delta k'=0.04$.

We discretize the space domain with a grid of $1024$ points, and we let the system evolve in time by integrating step-by-step the GP equations \eqref{GP_a} and \eqref{GP_b}, using the time-splitting method, and solving the kinetic terms by the iterated Crank-Nicholson algorithm. In Fig. \ref{fig:F2} and \ref{fig:F3} is reported the state of the system in the real space at the instants $t'=800$ and $t'=3400$ respectively, which in real units correspond to $t=16\,\text{ms}$ $t=68\,\text{ms}$ for the condensate parameters mentioned before.
At the intermediate time (fig.\ref{fig:F2}), a transmitted and a reflected wave packet are visible respectively in the upstream and downstream regions, with a residual component still presents inside the cavity. At the late time (fig.\ref{fig:F3}) instead, all wave packets have been annihilated by the absorbing region and no excitation is left in the system.

More information about the physics involved can be inferred from the time evolution of the system in the wave-vector domain, reported in figs. \ref{fig:L}(a-c). These figures show the time evolution of the spatial Fourier transform of the spin density $\delta n$ respectively in the downstream, cavity and upstream regions. As the density is a real function $\delta n(k)=\delta n^*(-k)$, and we plotted only the positive wave vector domain. As can be seen, the wave-vector content of the observed transmitted and reflected packets matches what is expected by inspection of the dispersion relations in Figs.~\ref{fig:DispRelOUT_NONlasing} and \ref{fig:DispRelIN_NONlasing}. In fig.\ref{fig:L1}, the initial wave packet in the \emph{d-R} mode is visible up to $t'\approx 300$, when it enters into the cavity with the simultaneous emission of a leftwards propagating reflected \emph{d-L} packet. This disappears around $t'\approx 1500$ when it is annihilated by the absorbing region. Fig.\ref{fig:L2} shows the appearance inside the cavity of the wavepacket coming from the downstream region. It propagates as a wavepacket localized in the \emph{cav-R} branch of the dispersion relation until it starts being reflected back by the right edge of the cavity onto the \emph{cav-L} branch around $t'\approx 450$. While this latter component is too weak for being visible in the colorplot, it can be easily recognized in the cut shown in fig.\ref{fig:M4}. Finally, in fig.\ref{fig:L3}, the transmitted wavepacket appears in the upstream region at the instant $t'\approx 450$ and is later on absorbed by the absorbing regions at the late times. In figs.\ref{fig:M}(a-d) plots of the spatial Fourier transform of the spin density at fixed time instants are shown, in order to make clearer the comparison between the observed and the expected components.

\begin{figure*}[h]
\centering%
\subfigure[\label{fig:G1}]
{\includegraphics[width=0.3\linewidth]{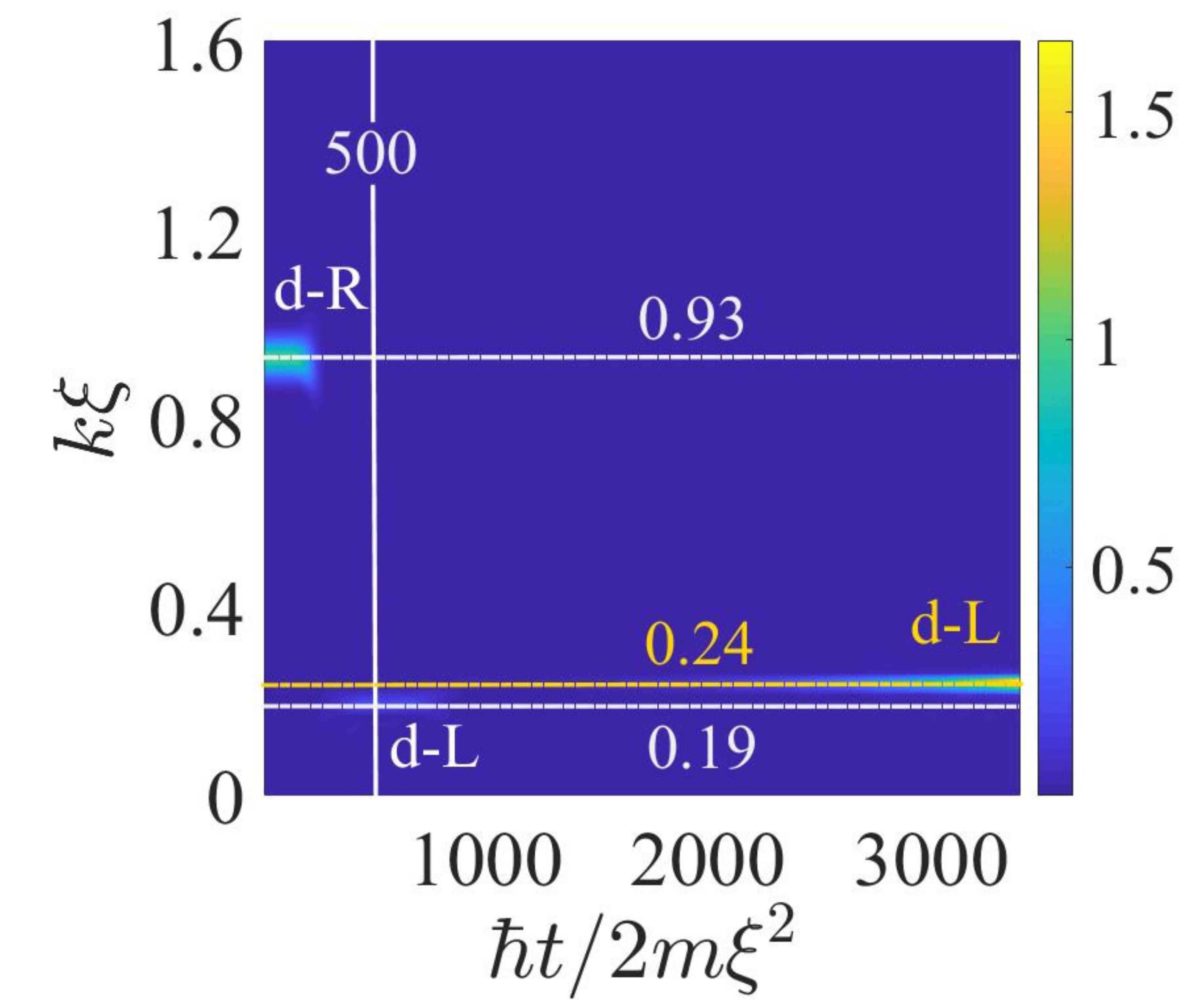}}
\subfigure [\label{fig:G2}]
{\includegraphics[width=0.3\linewidth]{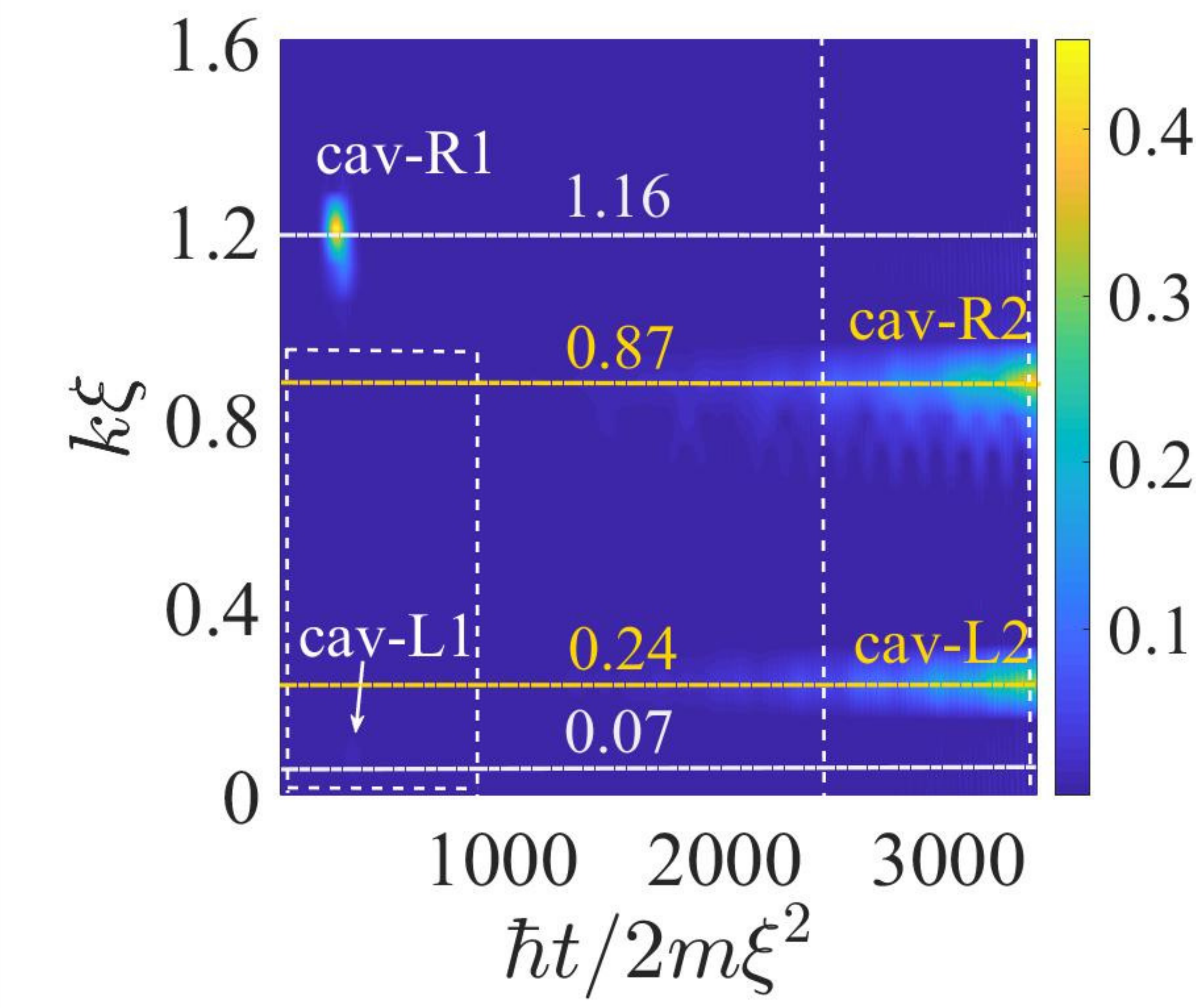}}
\subfigure [\label{fig:G3}]
{\includegraphics[width=0.3\linewidth]{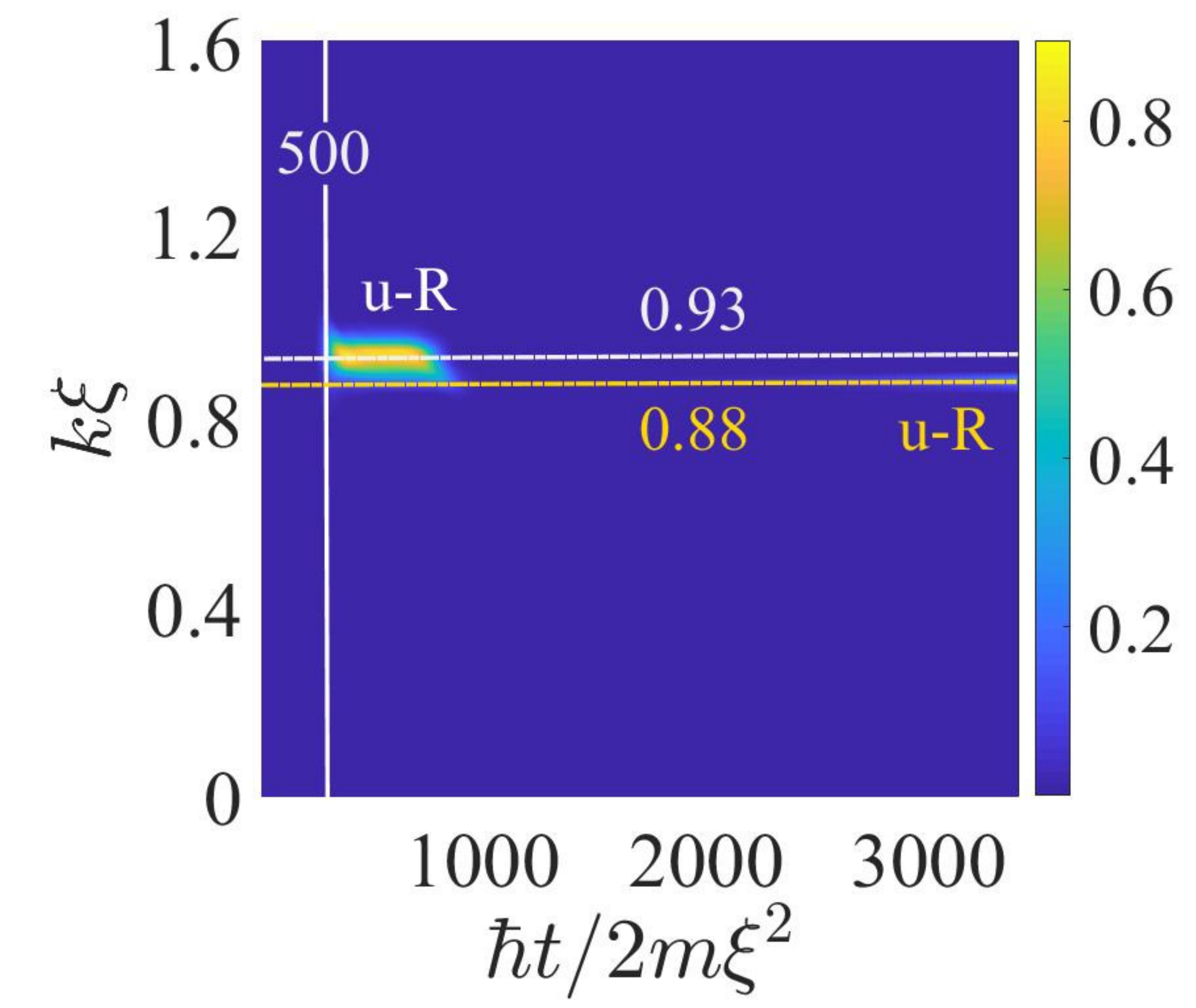}}\\
\subfigure [\label{fig:G4}]
{\includegraphics[width=0.32\linewidth]{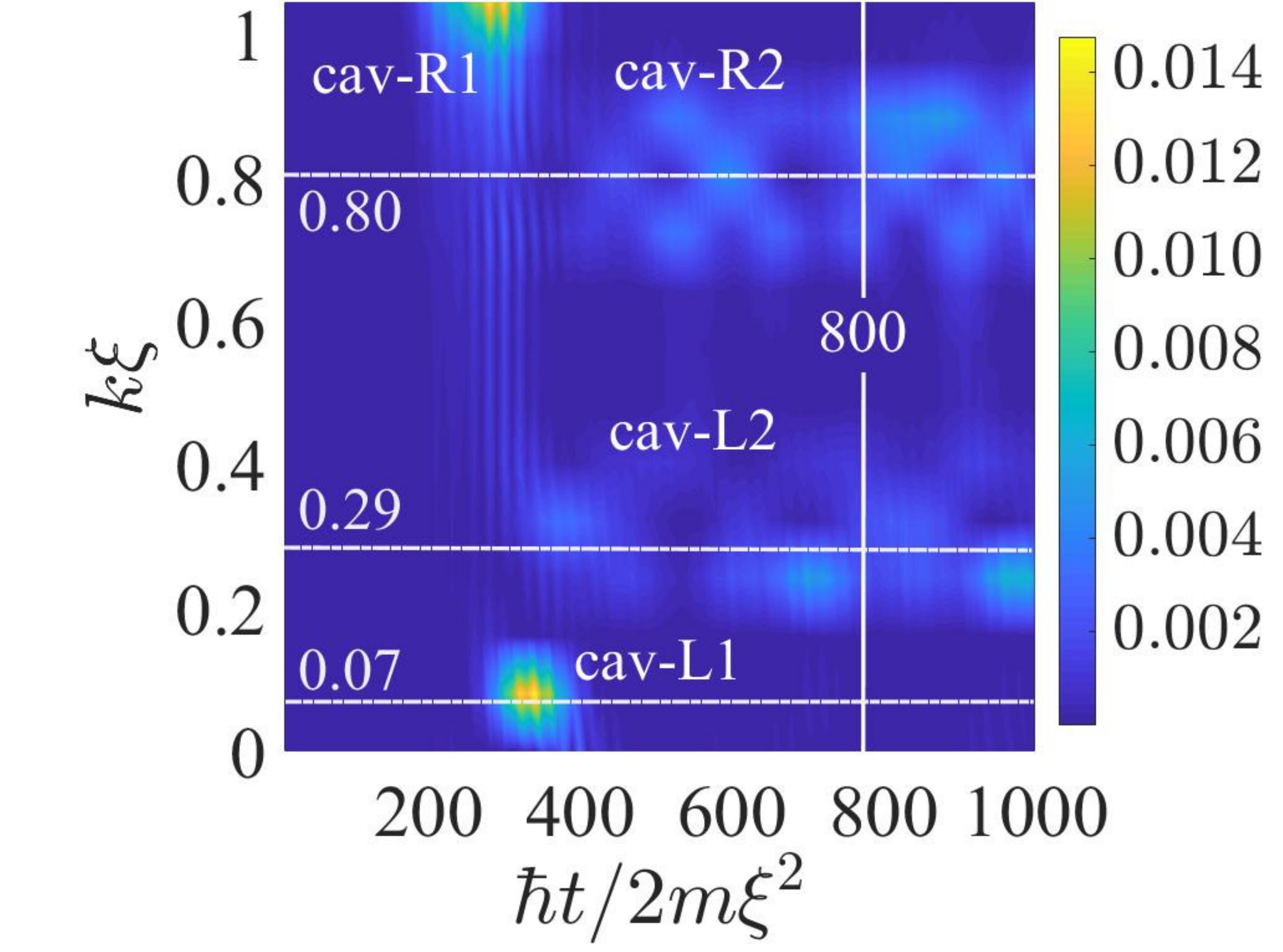}}
\subfigure [\label{fig:G5}]
{\includegraphics[width=0.3\linewidth]{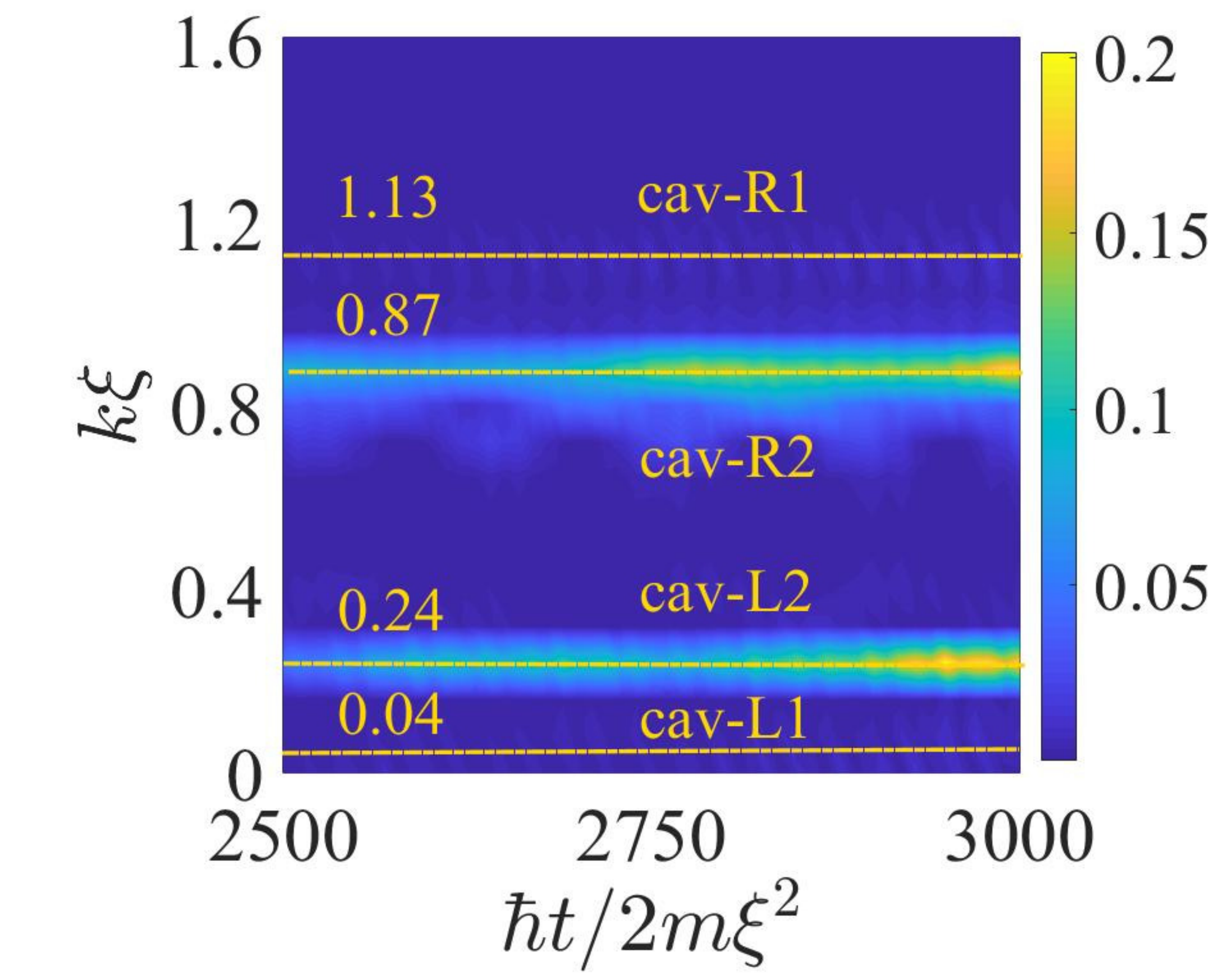}}
\caption{Lasing case: evolution in time of the spatial Fourier transform of the spin density of the system in the different regions. a) Downstream region. The wave-components of the \emph{d-R} packet given as initial condition for the simulation appear, together with the \emph{d-L} contribute dues to its reflection from the WH. At the late times, the leakage of the self-amplified radiation from the cavity appears in the \emph{d-L} modes. b) Cavity region. At the early times, the \emph{cav-R1} packet transmitted through the WH appears, while the \emph{cav-L1} component due to its reflection at the BH is barely visible. At the late times it is evident the onset of the exponential amplification of the unstable modes. c) Upstream region. At the early times, the \emph{u-R} packet transmitted through the cavity appears while, at the late times, it is evident the leakage in the \emph{u-R} mode of the self-amplified radiation from the cavity. d-e) Magnified views of the regions indicated by dashed white rectangles in panel (b). In d), the early times excitation of the \emph{cav-R2} and \emph{cav-L2} modes is now visible. In e) the focus is on the unstable mode. System parameters as in Figs.\ref{fig:F}(d,e).}
\label{fig:G}
\end{figure*}

\begin{figure*}[htbp]
\centering%
\subfigure[\label{fig:H1}]
{\includegraphics[width=0.45\linewidth]{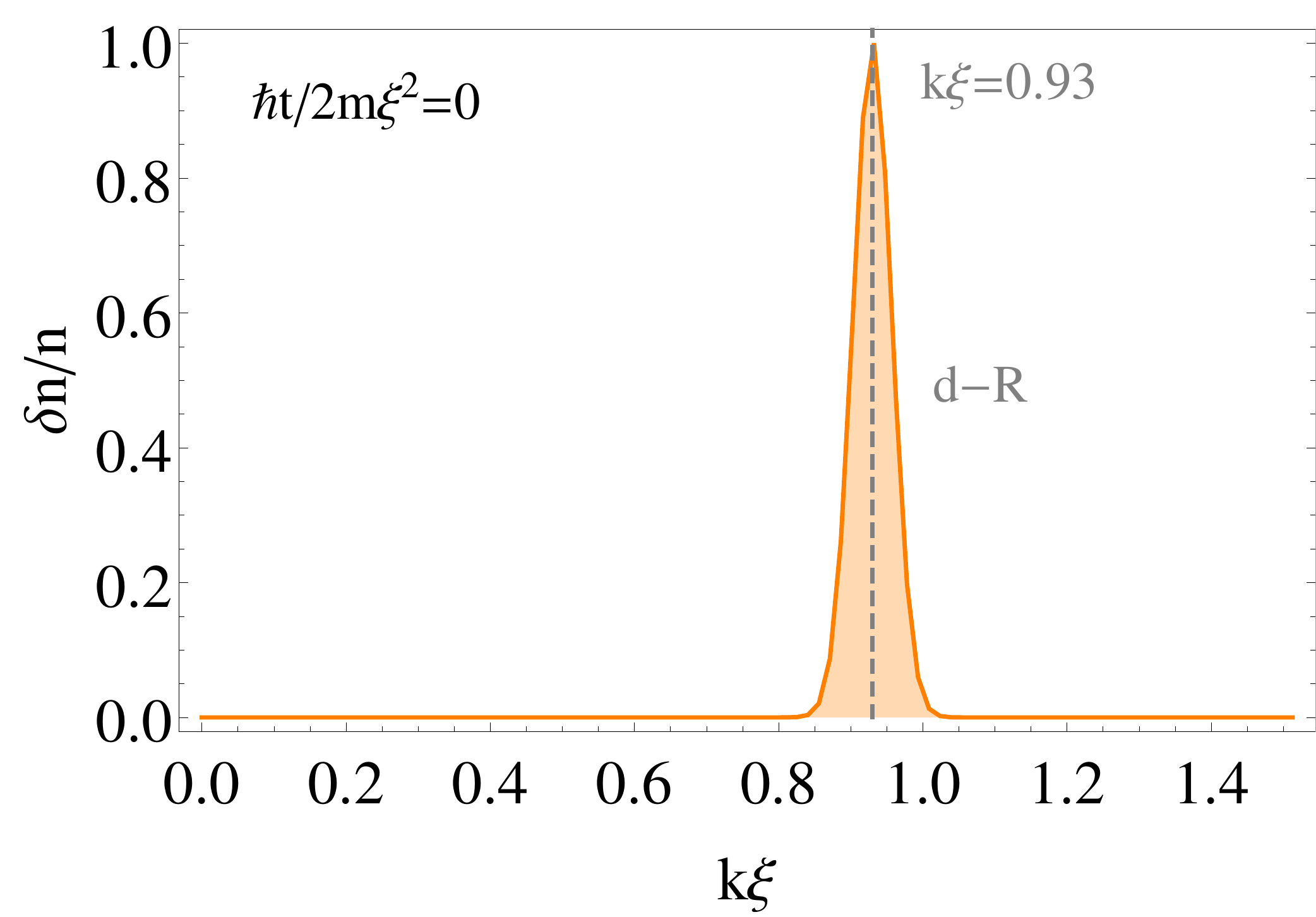}}\quad
\subfigure [\label{fig:H2}]
{\includegraphics[width=0.45\linewidth]{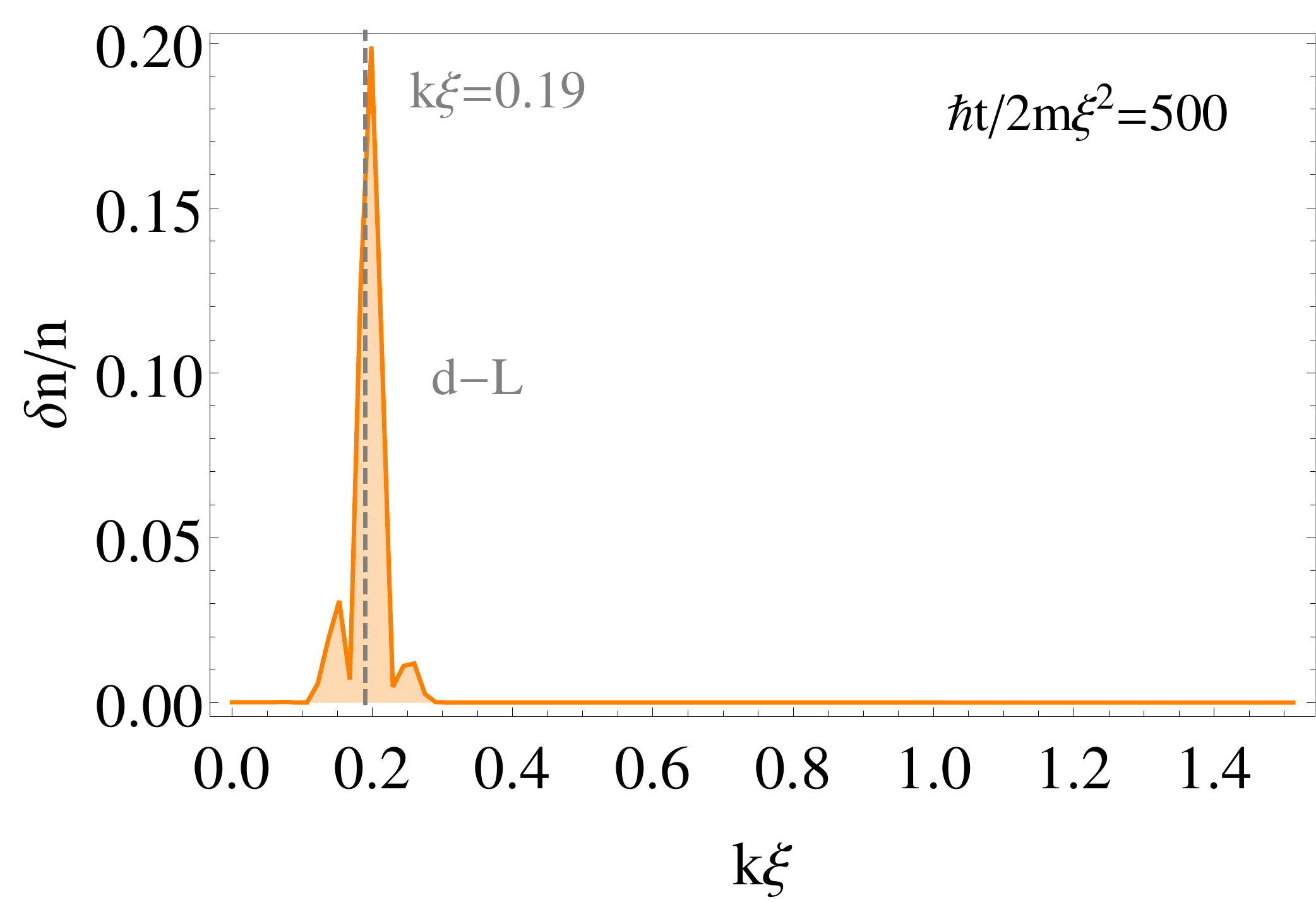}}\\
\subfigure [\label{fig:H3}]
{\includegraphics[width=0.45\linewidth]{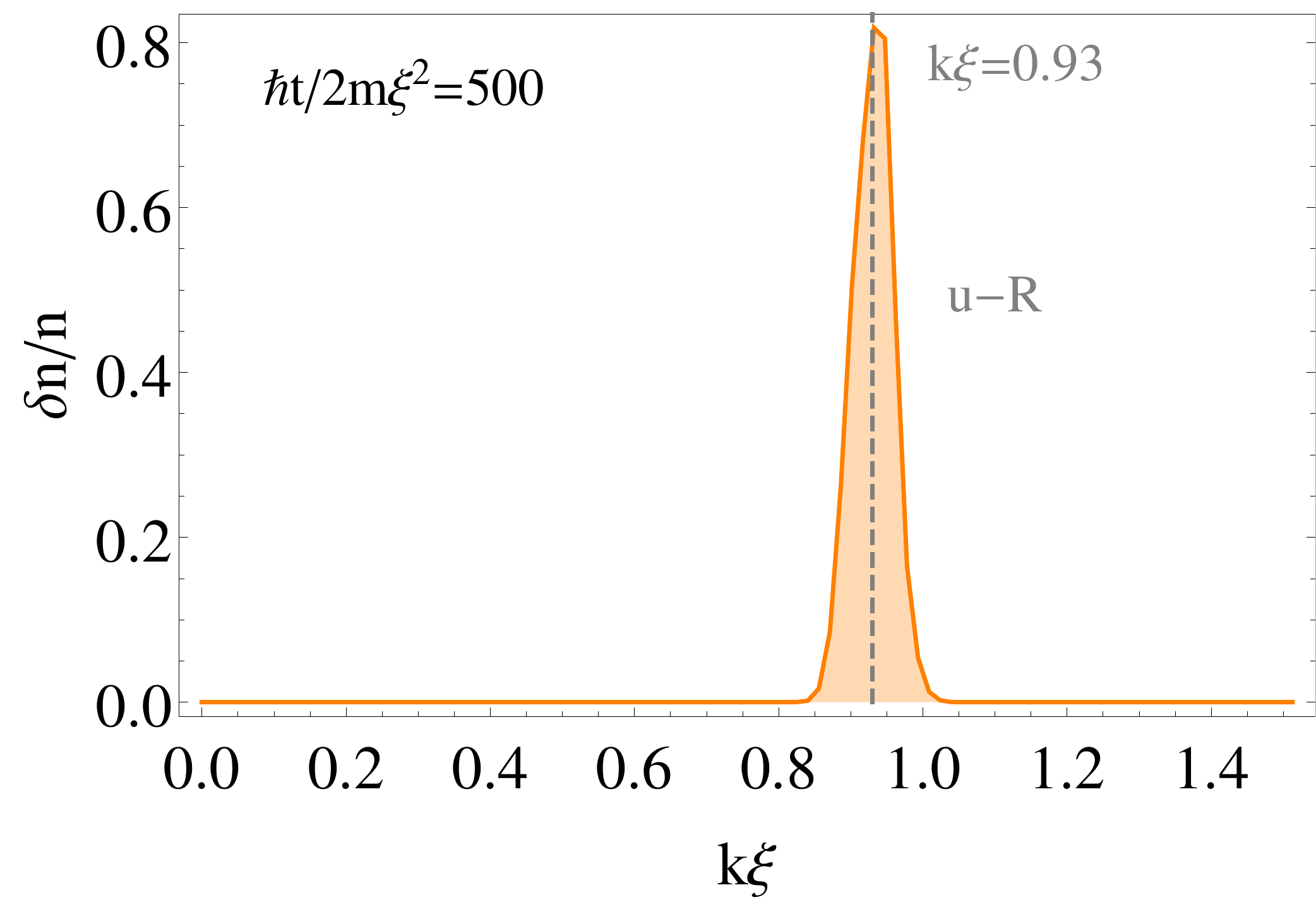}}\quad
\subfigure [\label{fig:H4}]
{\includegraphics[width=0.45\linewidth]{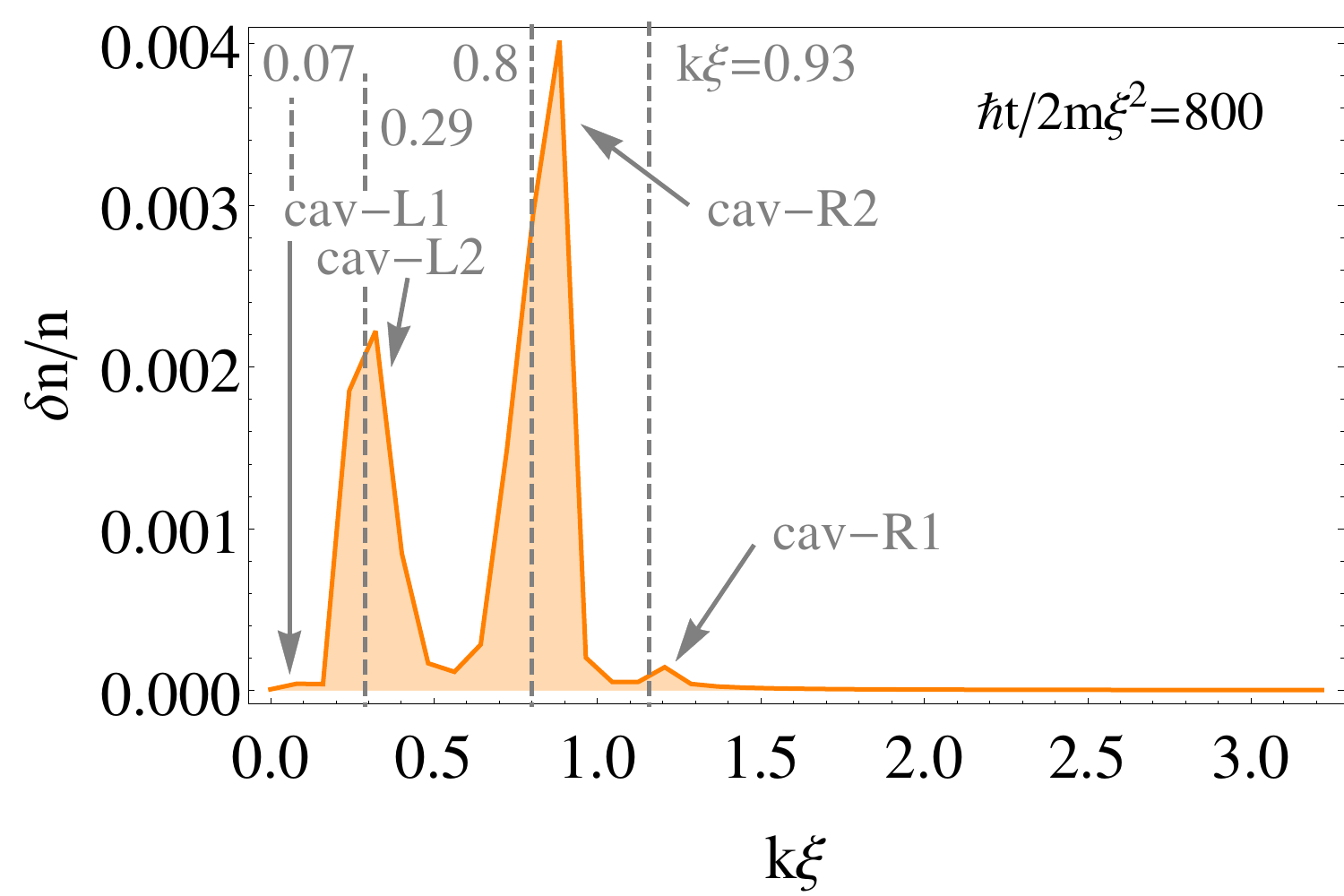}}
\caption{Lasing case at early times: spatial Fourier transform of the spin density of the system at different instants of time and in different regions, showing the
spectral content of the different wave packets. The dashed lines identifies the central wave-vector expected from the analysis of the subsonic and supersonic dispersion relations in Figs. \ref{fig:DispRelOUT_lasing} and \ref{fig:DispRelIN_lasing}. a) Downstream region, $\hbar t/2m\xi^2=0$: initial condition. b) Downstream region, $\hbar t/2m\xi^2=500$: reflected packet by the WH. c) Upstream region, $\hbar t/2m\xi^2=500$: wave packet transmitted through the cavity. d) Cavity region, $\hbar t/2m\xi^2=500$: in contrast to the non-lasing case, the negative norm components are now present and visible. System parameters as in Figs.\ref{fig:F}(d,e) and \ref{fig:G}.}
\label{fig:H}
\end{figure*}

\begin{figure*}[htbp]
\centering%
\subfigure[\label{fig:I1}]
{\includegraphics[width=0.30\linewidth]{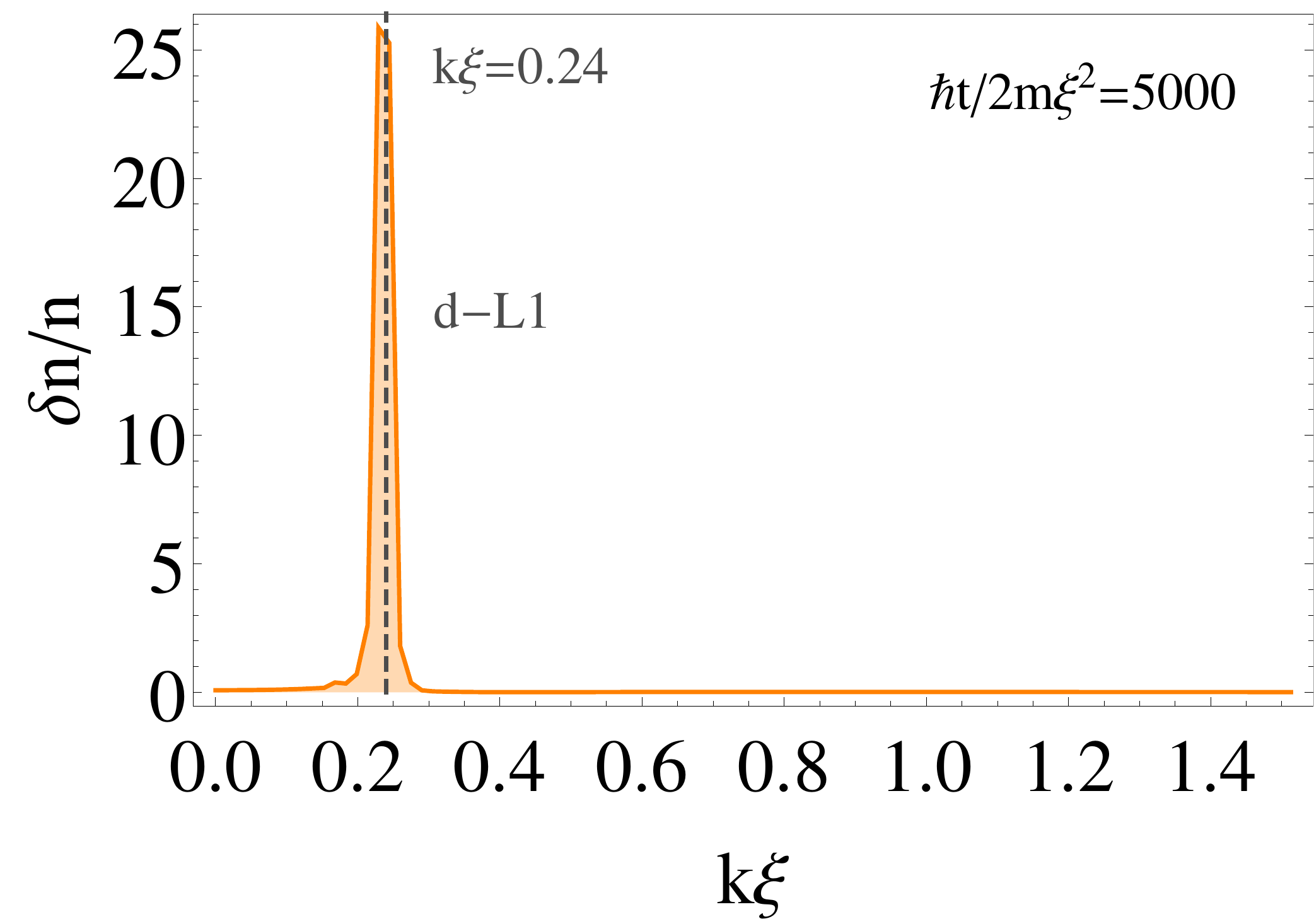}}\quad
\subfigure [\label{fig:I2}]
{\includegraphics[width=0.30\linewidth]{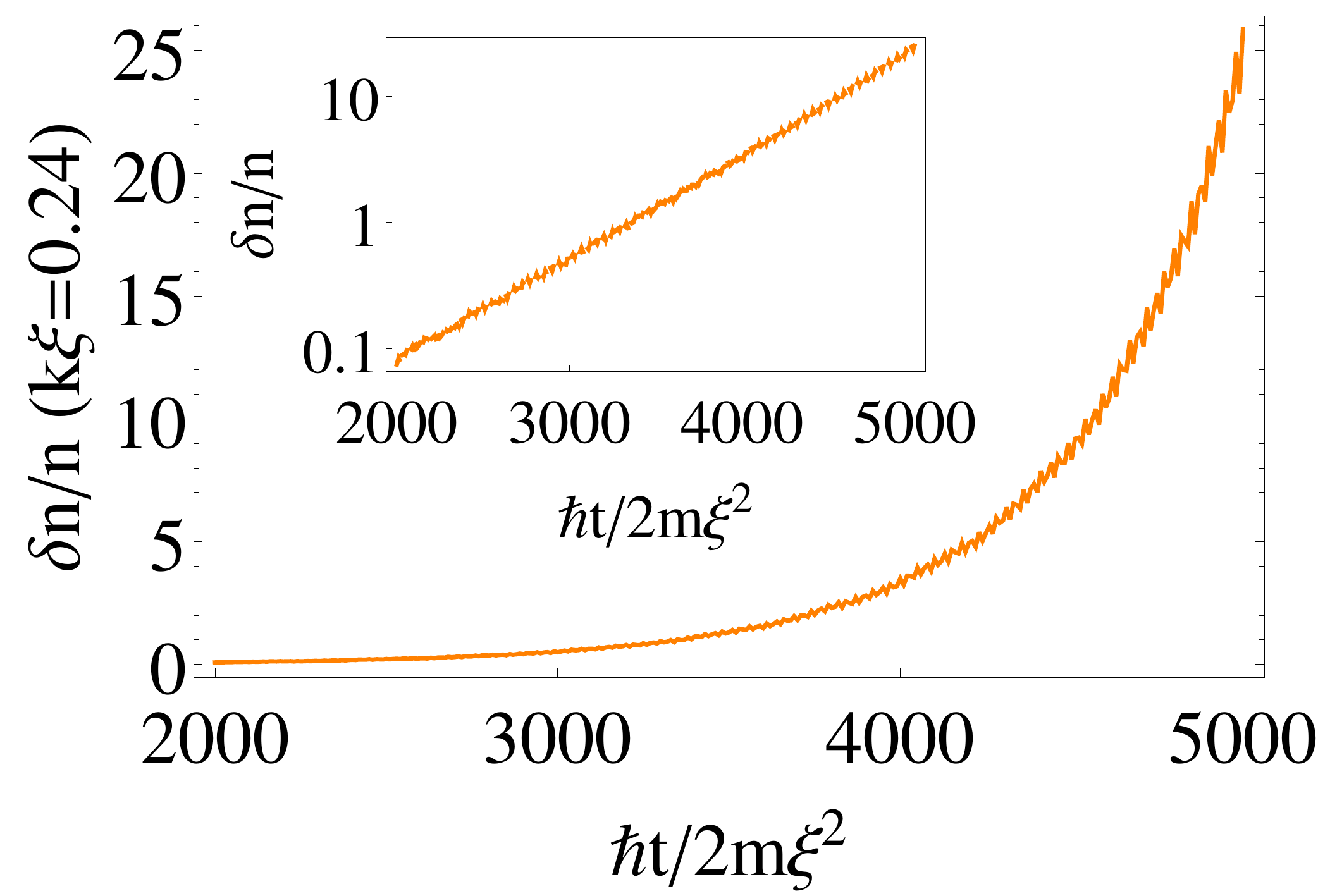}}\\
\subfigure [\label{fig:I3}]
{\includegraphics[width=0.30\linewidth]{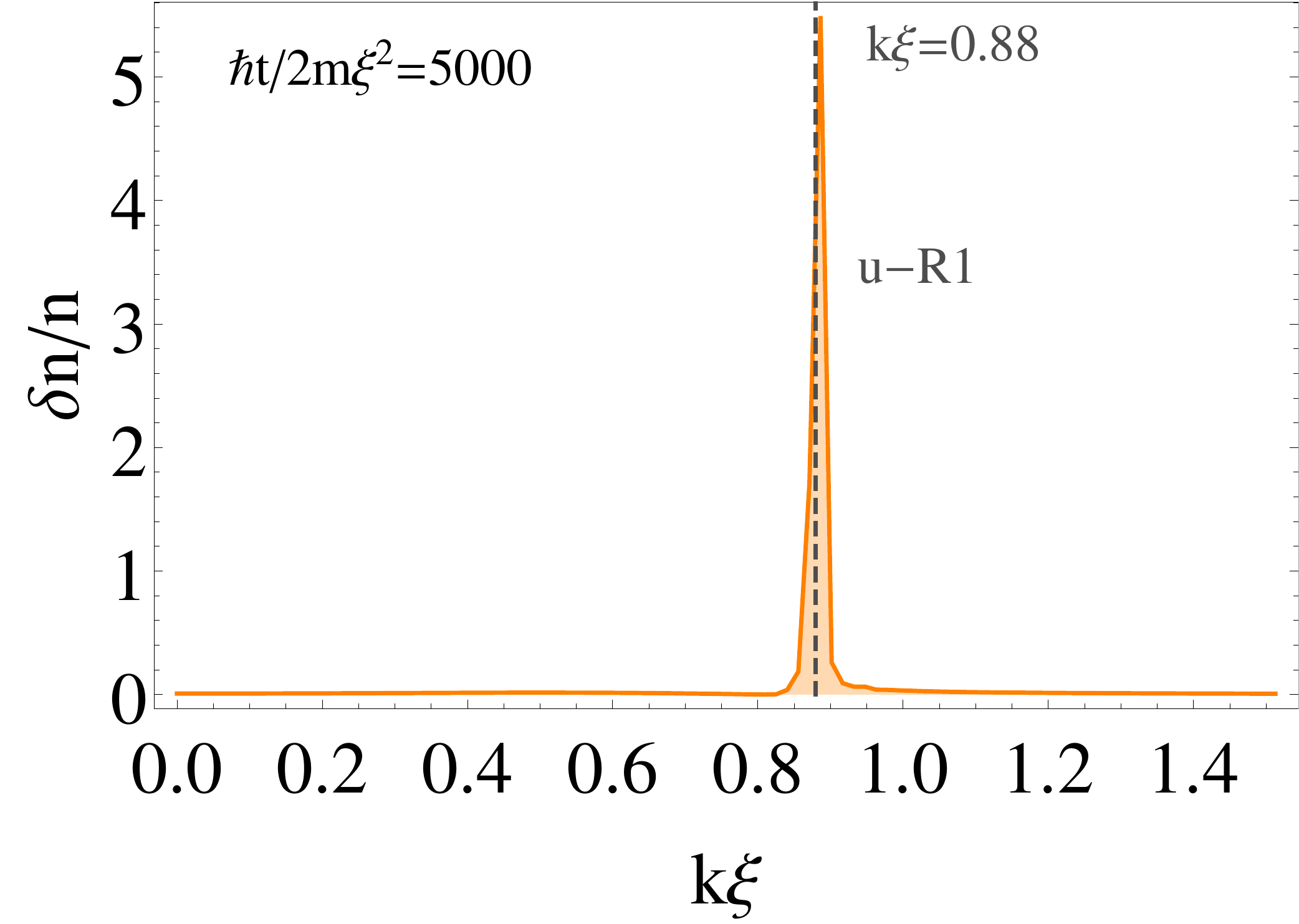}}\quad
\subfigure [\label{fig:I4}]
{\includegraphics[width=0.30\linewidth]{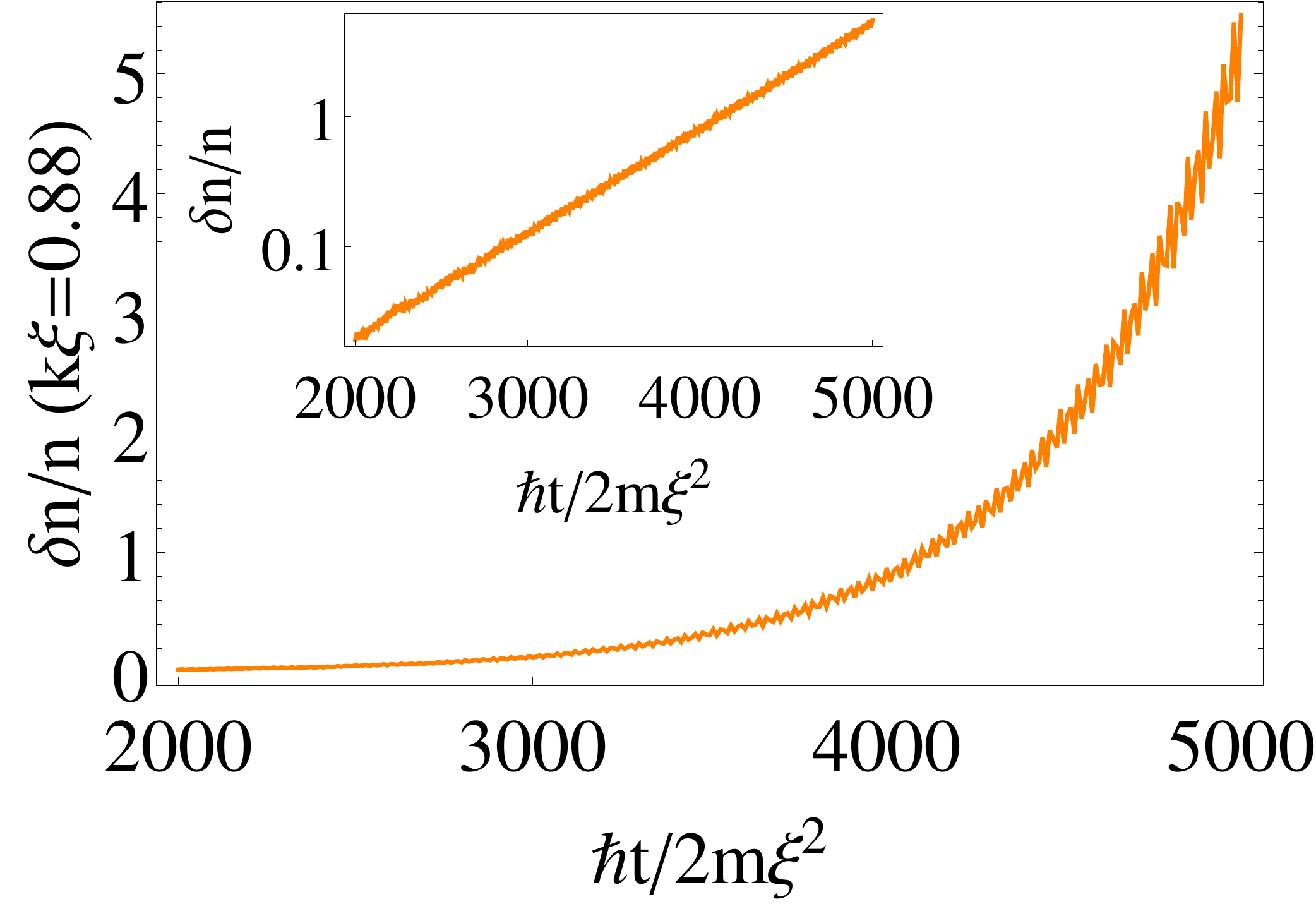}}\\
\subfigure [\label{fig:I5}]
{\includegraphics[width=0.30\linewidth]{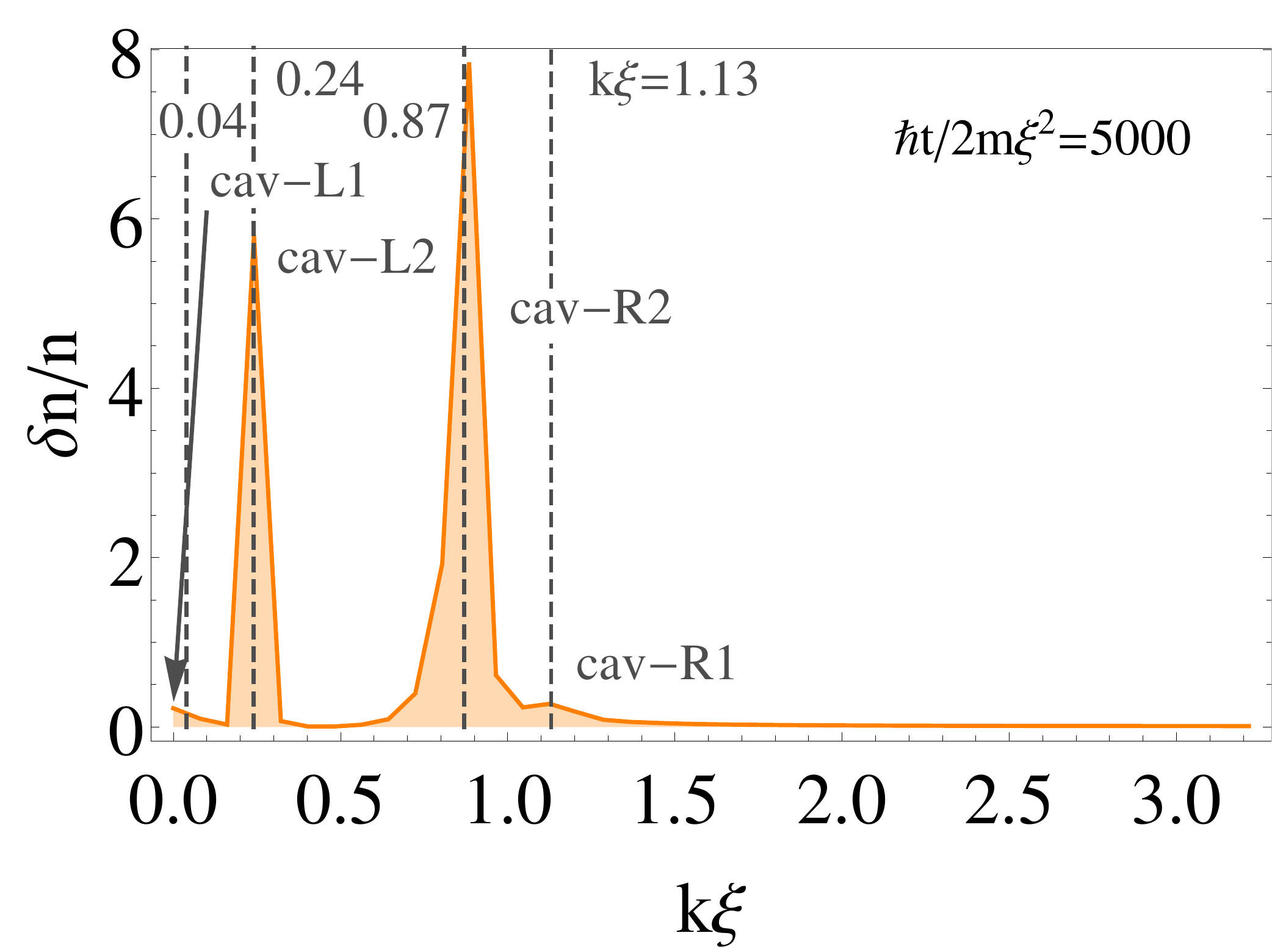}}\quad
\subfigure [\label{fig:I6}]
{\includegraphics[width=0.30\linewidth]{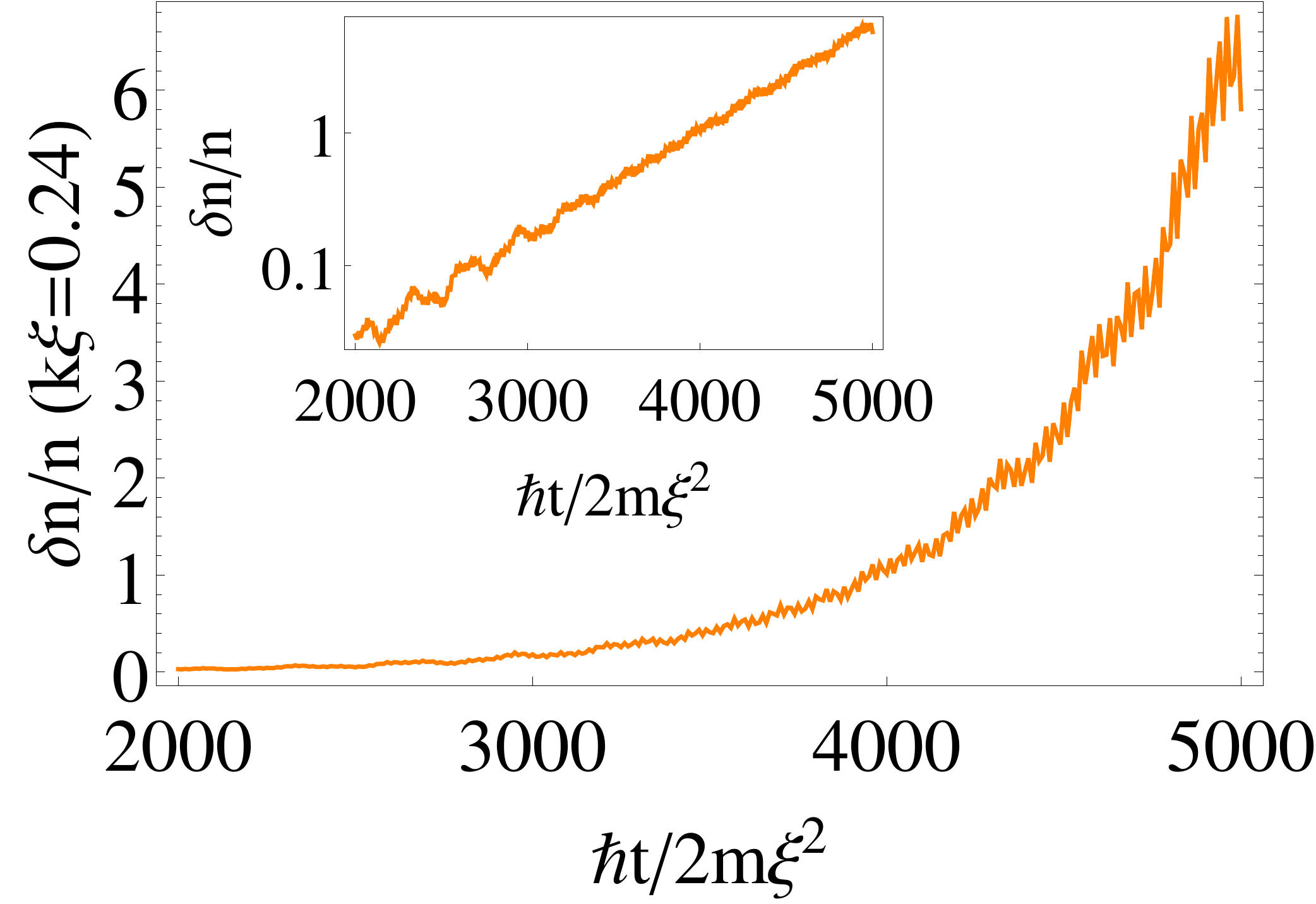}}\\
\subfigure [\label{fig:I7}]
{\includegraphics[width=0.30\linewidth]{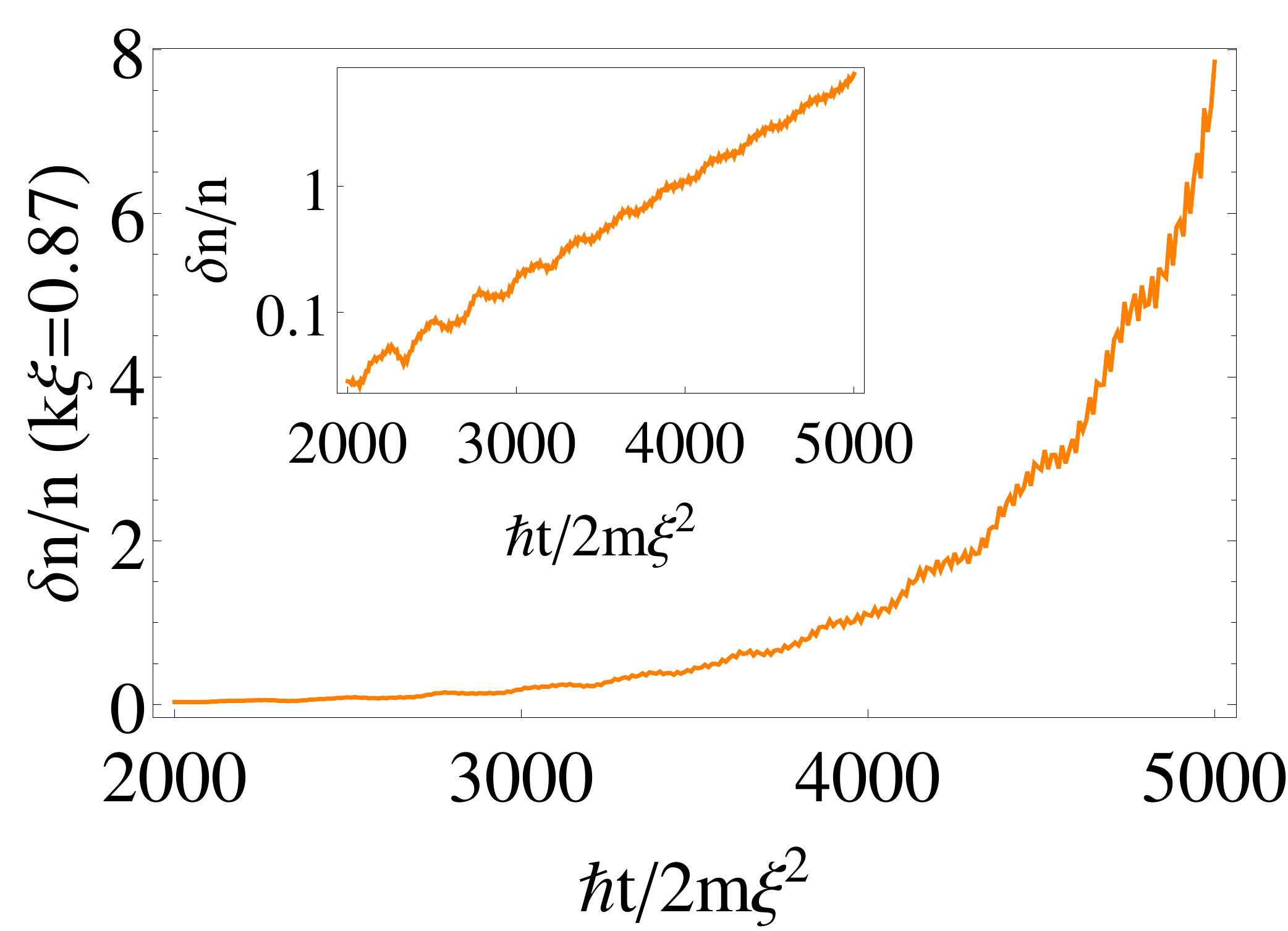}}\quad
\subfigure [\label{fig:I8}]
{\includegraphics[width=0.30\linewidth]{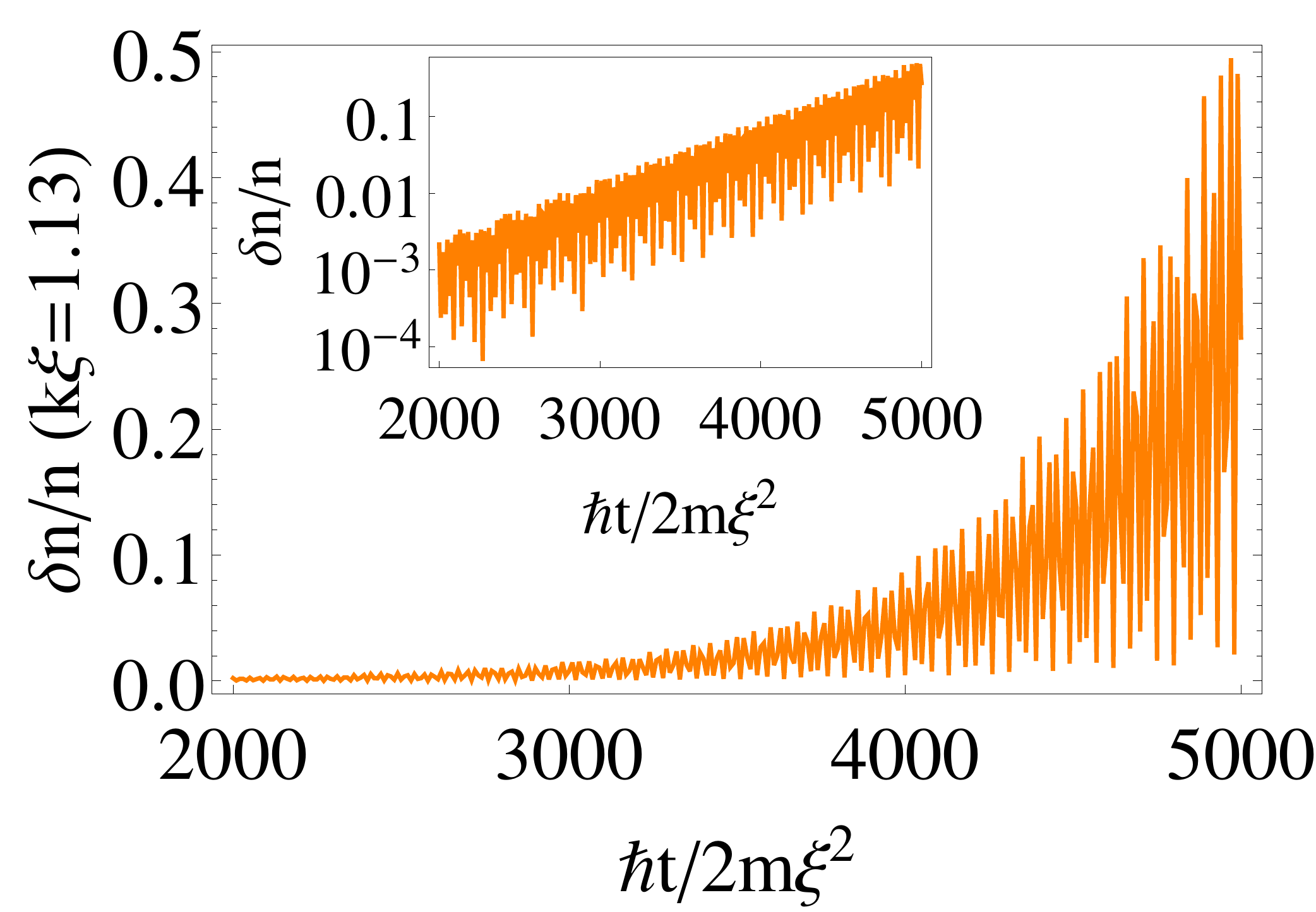}}\quad
\subfigure [\label{fig:I9}]
{\includegraphics[width=0.30\linewidth]{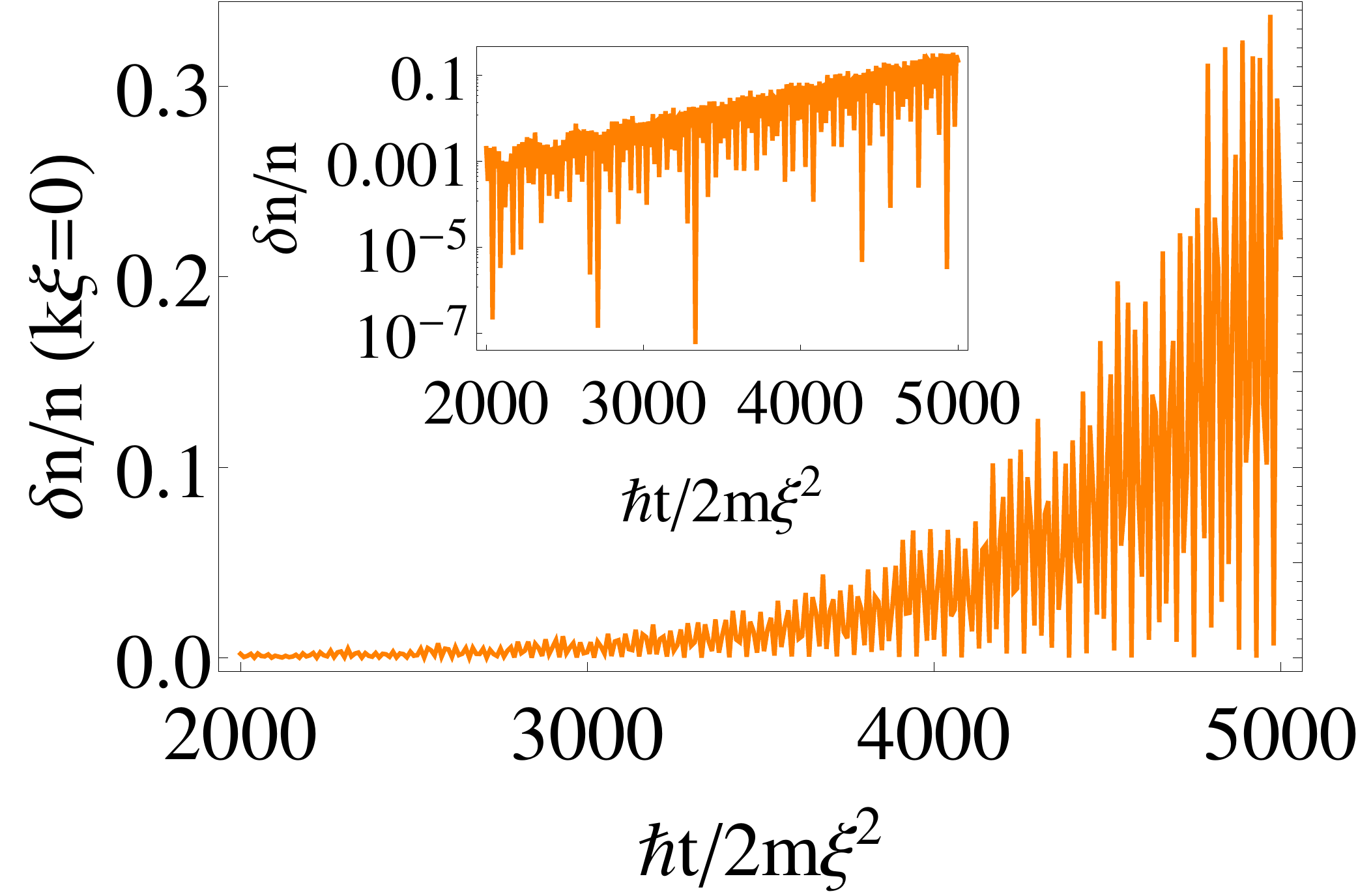}}
\caption{Lasing case at the late times: a,c,e) spatial Fourier transform of the spin density at $\hbar t/2m\xi^2=5000$, in the downstream, upstream and cavity regions, respectively. The pictures show the dominant unstable modes in the cavity and their leakage in the downstream and upstream regions. b,d,f,g,h,i) The time evolution of the unstable modes in the three different regions clearly shows the exponential character of their amplification. Note that fig.(i) shows the $k\xi=0$ component instead of the $k\xi=0.04$ one just because the wavevector resolution in the simulation is equal to $0.08$. System parameters as in Figs.\ref{fig:F}(d,e), \ref{fig:G} and \ref{fig:H}.}
\label{fig:I}
\end{figure*}

\subsubsection{Lasing regime}
To unravel the far richer physics of the lasing regime, we consider the same system as before but with the different values  $\Omega'=-0.01$ and $\Omega'=-0.12$ of the coherent coupling amplitude respectively inside and outside the cavity. In such a case, the cavity region is supersonic and is delimited by a WH-like horizon on its downstream, left-hand side and a BH-like horizon on its upstream, right-hand side. The dispersion relation for the spin excitations takes the form in figs.\ref{fig:DispRelOUT_lasing} and \ref{fig:DispRelIN_lasing} respectively outside and inside the cavity.

We consider again a wave packet, initially located in the downstream region and propagating rightwards towards the cavity. Its energy is again $\hbar\omega'=0.15$, and the wave-vector content is now centred on the value $k'=0.93$ in the \emph{d-R} branch of the subsonic dispersion relation (see Fig. \ref{fig:DispRelOUT_lasing}). As soon as the wave packet hits the WH horizon, new components appear. First of all, standard scattering gives rise to a reflected component which propagates leftwards through the downstream region on the \emph{d-L} branch and a transmitted component propagating rightwards along the cavity on the \emph{cav-R1} branch. At the same time, mode conversion processes at the WH are responsbile for the appearance of an additional negative norm transmitted component on the \emph{cav-R2} branch, which propagates along the cavity rightwards towards the BH. Even further components appear once the two transmitted components hit the BH: in addition to radiation from the cavity into the subsonic upstream region on the \emph{u-R} branch, reflection and mode conversion processes from the BH horizon generate leftward propagating components in the cavity on both \emph{cav-L1} and \emph{cav-L2} branches. This physics is clearly visible in the results of numerical simulations shown in figs.\ref{fig:G}(a-d). As before, in figs.\ref{fig:H}(a-d) is reported the spectral content of the spin density at fixed time instants, with the aim of making clearer the matching between the observed and expected components.

The situation is even more intriguing at much later times, when the amplification of certain wave-vector components in the cavity region is clearly visible in fig.\ref{fig:G2} as a consequence of the presence of unstable modes inside the supersonic cavity. The leakage of this self-amplified radiation into the downstream (fig.\ref{fig:G1}) and upstream (fig.\ref{fig:G3}) regions is also visible. Cuts of the spectral distribution at given instants of time are shown in Fig.\ref{fig:I}(a,c,e), while real-space profiles are reported in figs.\ref{fig:F4} and \ref{fig:F5}. In all these plots, the signature of the unstable mode is clearly recognizable.

By comparing the spectral content of the radiation at the late times and the dispersion relations in Figs. \ref{fig:DispRelOUT_lasing} and \ref{fig:DispRelIN_lasing}, we notice that the unstable cavity modes that are visible in the three regions are the ones close to energy $\hbar\omega'=0.12$. However, by the procedure illustrated in the next section, we verified that the most unstable modes are actually the ones close to $\omega'=0$, in accordance with the results in \cite{Coutant2010}. The reason why in the present case the modes close to $\hbar\omega'=0.12$ are dominant is because they are closest to the triggering wave packets and so are excited first. The exponential character of the amplification of these modes is evident from the Figs. \ref{fig:I}(b,d,f-i) which show their evolution in time.

\begin{figure*}[h]
\centering%
\subfigure[\label{fig:Noise_t1}]
{\includegraphics[width=0.9\linewidth]{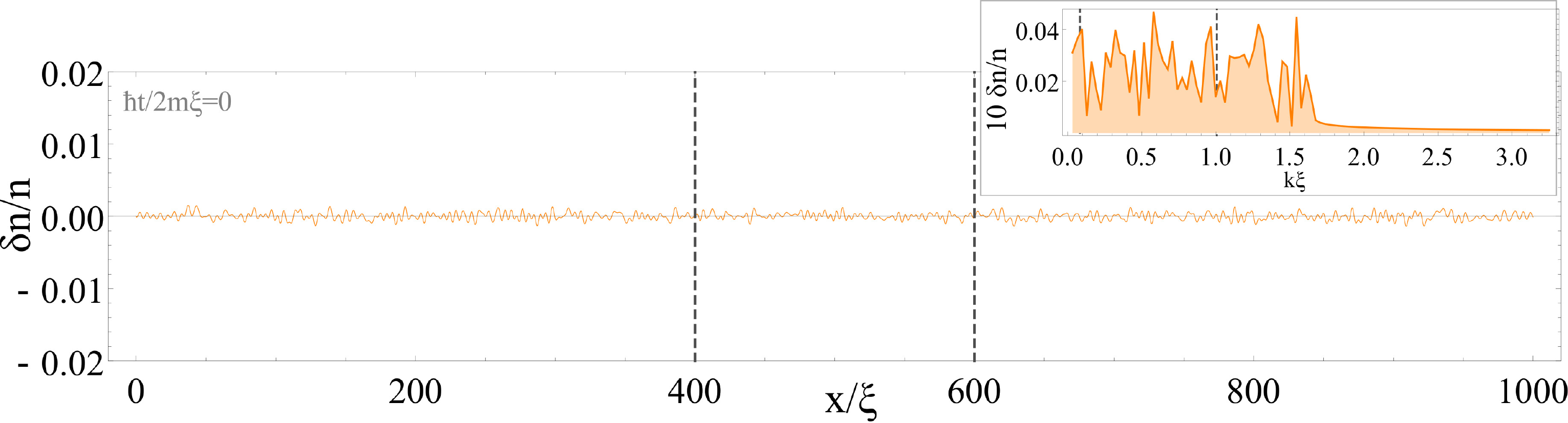}}\\
\subfigure[\label{fig:Noise__t60}]
{\includegraphics[width=0.9\linewidth]{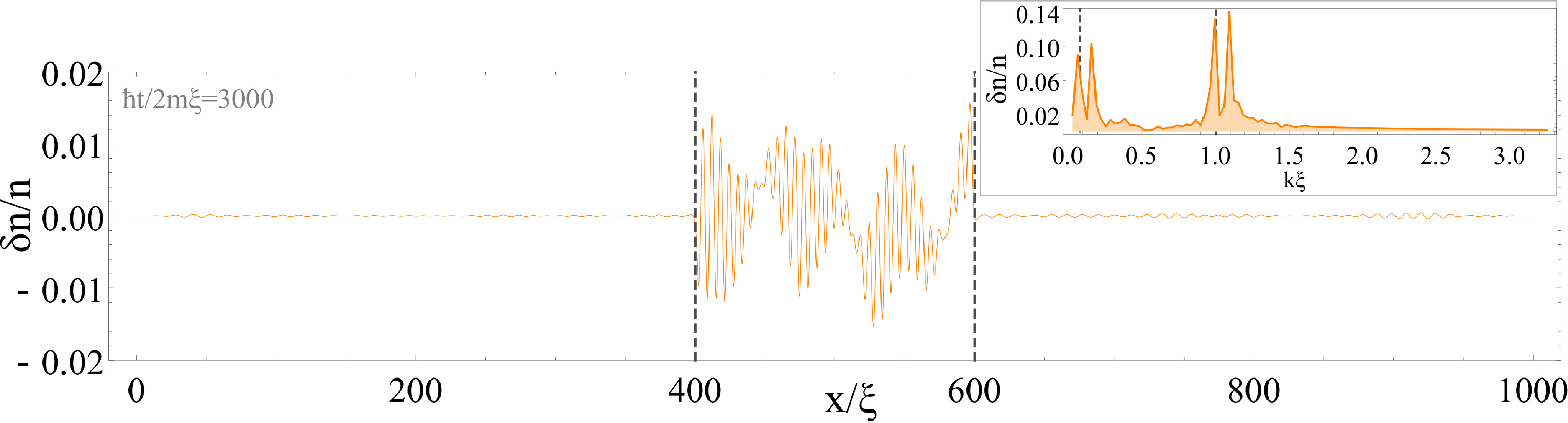}}\\
\subfigure[\label{fig:Noise_t90}]
{\includegraphics[width=0.9\linewidth]{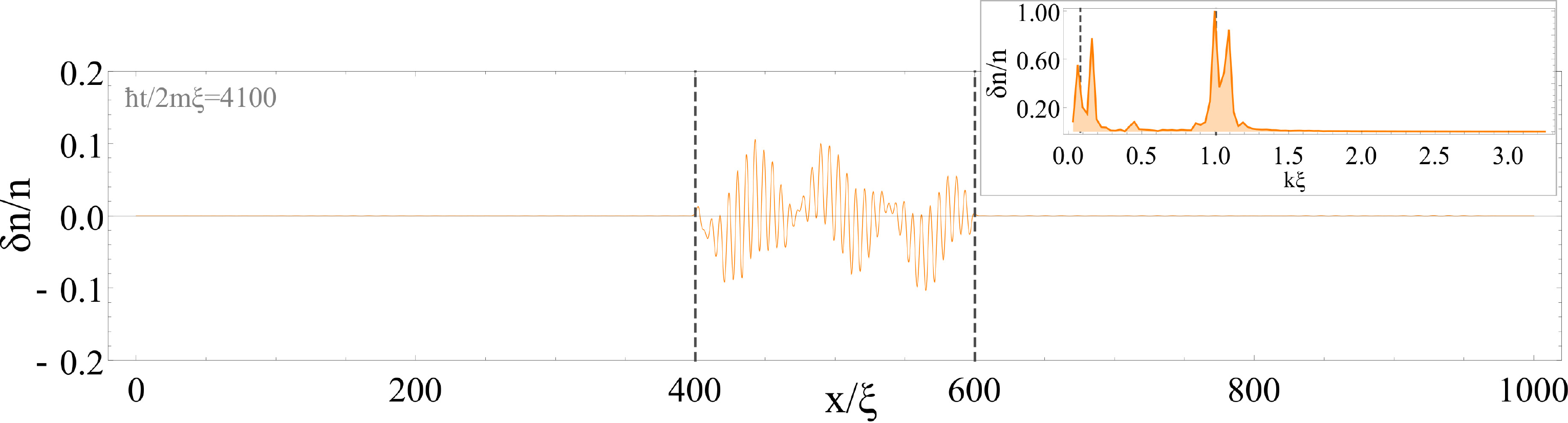}}
\subfigure[\label{fig:DispRel_Noise}]
{\includegraphics[width=0.45\linewidth]{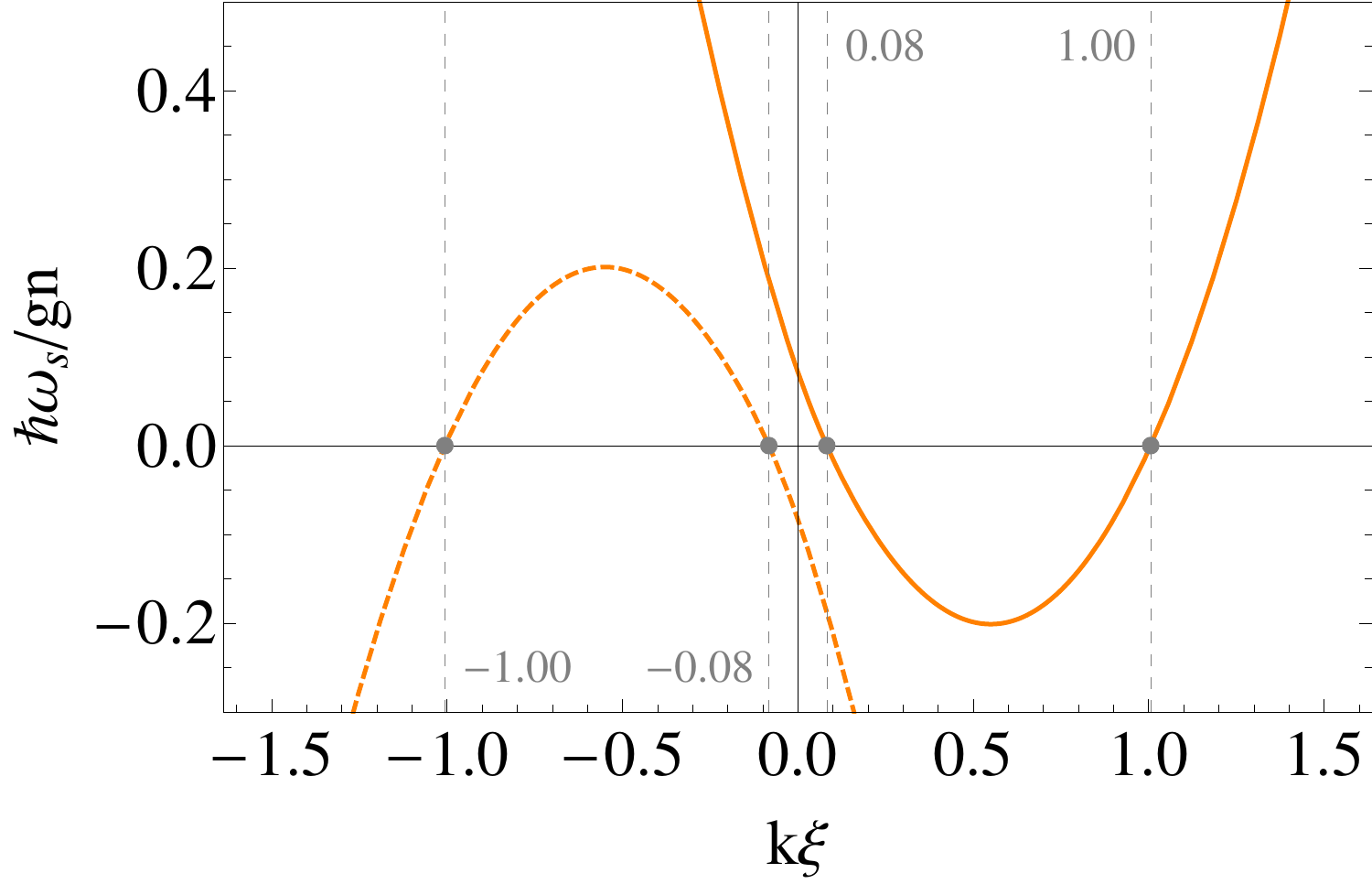}}
\caption{Noise amplification in a lasing configuration with $g=0.8\,g_{ab}$ and $\Omega'=-0.015$ ($\Omega'=-0.5$) inside (outside) the cavity. Spatial profiles of the spin density at different times a) $t'=0$, b) $t'=3000$, c) $t'=4100$ starting from an initially noisy configuration. d) Spin modes dispersion relations inside the cavity for the chosen parameters.} 
\label{fig:Noise}
\end{figure*}

\begin{figure}[h]
\centering%
\includegraphics[width=0.95\linewidth]{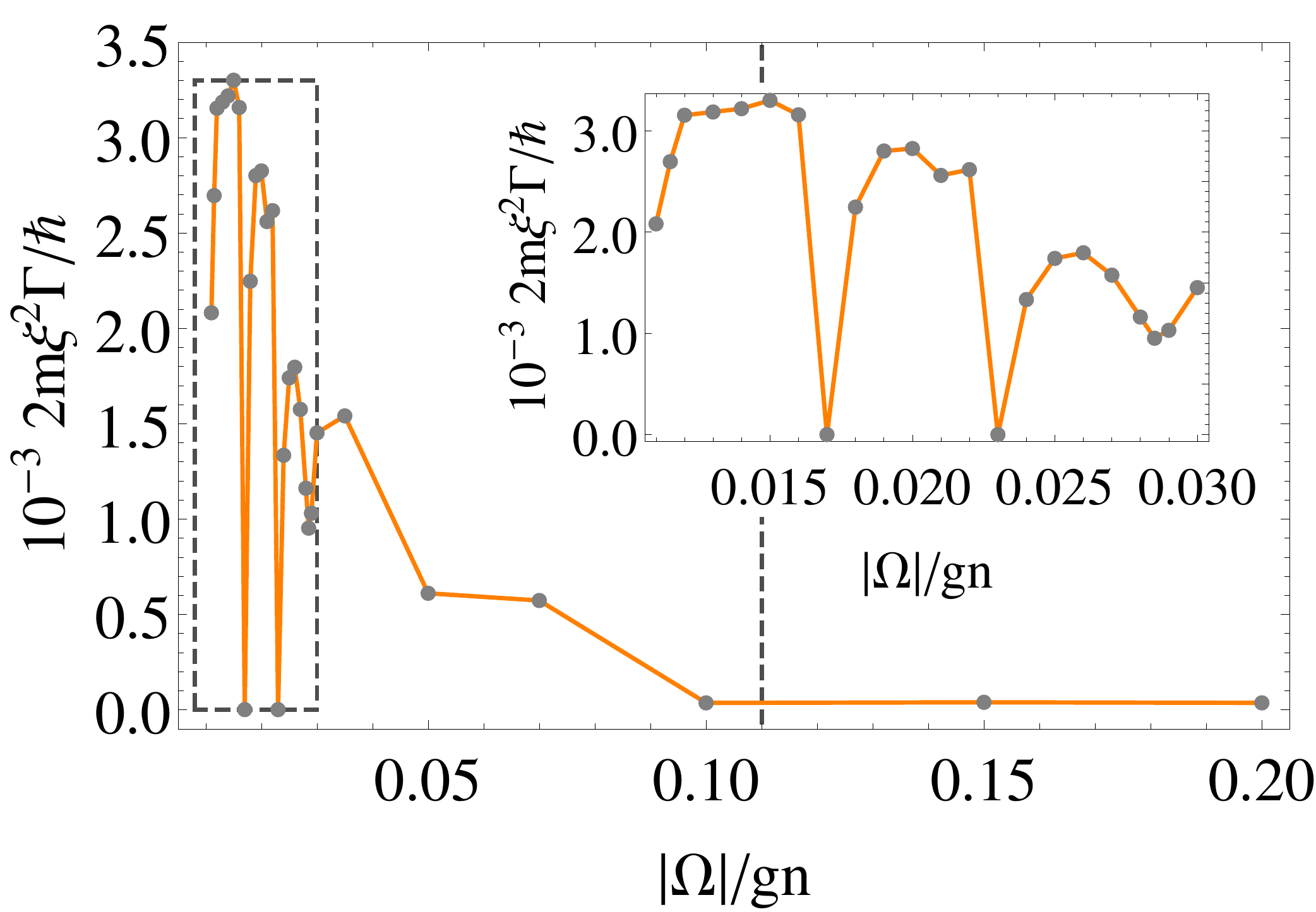}
\caption{The rate $\Gamma'$ of the exponential amplification $\exp(\Gamma' t')$ of the spin excitations inside the cavity. The value of $\Omega/gn$ at which the supersonic-subsonic transition occurs in the cavity is indicated by the dotted line. The inset shows a magnified view of the region within the dashed rectangle.}
\label{fig:Time_Scales}
\end{figure}

\subsection{Noise self-amplification \label{sec:noise}}
The study of the wave packet propagation through the lasing cavity was useful in order to show the appearance of the positive energy, negative norm modes, and the consequent instability of the system once the supersonic regime is attained. The purpose of this section is to further characterize such instability highlighting its dependence on the strength of the supersonic regime attained in the cavity. To this aim, we analyse the time evolution of the same system as before, but imposing this time a random white noise on top of the ground state of the condensate as initial condition for the simulation. In this way we are able to simultaneously seed all potentially unstable modes. 

From an experimental point of view, this initial condition is expected to qualitatively model the evolution starting from an unperturbed system as some fluctuations in the local particle and spin density are always present in an experiment for technical as well as thermal reasons. As in the previous case, experimental information can be retrieved from the spatial spin density profiles of the atomic cloud, as obtained by spin-dependent imaging techniques~\cite{Seo:PRL2015}. If needed, a large number of such spin densities could be combined to get information on the spin-density correlation function. A quantitative study of this quantity goes beyond the present article and will be the subject of future work.

We run the simulation considering different values for the Rabi frequency $\Omega$, while keeping the value of the interaction constants fixed to $g_{ab}/g=0.8$. We use the same grid parameters as in the previous case. In fig.\ref{fig:Noise}(a-c) we show a few examples of spatial profiles at different time steps, with the values $\Omega'=-0.015$ and $\Omega'=-0.5$ inside and outside the cavity respectively. By comparing the Fourier transform of the spin density in the cavity region shown in the insets of fig.\ref{fig:Noise}(a-c) with the corresponding supersonic dispersion relation of the spin modes shown in the lowest panel (d) we see that the most unstable modes are indeed the low-frequency ones close to $\omega'=0$.

Looking at these data in more detail, one can notice that the most unstable modes form doublets of closely spaced peaks in the spatial Fourier transform of the spin density,
centered around each zero-crossing $\omega' = 0$ of the dispersion relation. We interpret these two peaks as corresponding respectively to positive and negative energy components of the unstable mode. Depending on the specific value of $\Omega'$ chosen, the doublets may merge into a single peak corresponding to a zero-frequency instability
\cite{Coutant2010,MichelParentani1,MichelParentani2,deNova2016}.

As a final point, we looked at the dependence of the amplification rate of the most unstable mode on the strength of the coherent coupling $\Omega'$. We evaluated the (dimensionless) time scale $t'_c$ of the exponential self-amplification for different values of $\Omega'$ inside the cavity (keeping $\Omega'=-0.5$ outside). We show the results in Fig. \ref{fig:Time_Scales} in terms of $\Gamma'=1/t'_c$. As expected, the amplification rate shows an overall decreasing trend when the strength of the coherent coupling is increased, until the spin modes become subsonic also inside the cavity, and the instability is completely switched off. 

A close analysis of the dispersion relations reveal that such transition happens close to the value $\Omega'=-0.11$ (given $g_{ab}=0.8$) which is not far from the value $\Omega'=-0.10$ at which it seems to take place in fig.\ref{fig:Time_Scales}. The reason for this small discrepancy is due to the limit of the resolution achievable in determining the amplification rate, which is intrinsic to the time-splitting algorithm we used to solve the time evolution of the system\\
Superimposed on the decreasing trend, one can see in Fig~\ref{fig:Time_Scales} marked oscillations of $\Gamma'$ as a function of $\Omega'$, clearly visible in the rescaled region indicated by the dashed rectangle and rescaled in the inset. Along the lines of \cite{Finazzi2010,Coutant2010,MichelParentani1,MichelParentani2,deNova2016}, these oscillations can be ascribed to the discreteness of the cavity modes between the two horizons: by varying the value of $|\Omega|$, we are in fact effectively changing the amplitude of the gap in the dispersion relations and so the wave-vectors at which it attains the zero energy. As a consequence, new cavity modes enter or leave the supersonic regime. According to this picture, one may interpret the minima in the amplification rate
as relative to conditions in which the unstable mode has just disappeared.

\begin{figure}[htbp]
\centering%
\includegraphics[width=0.9\linewidth]{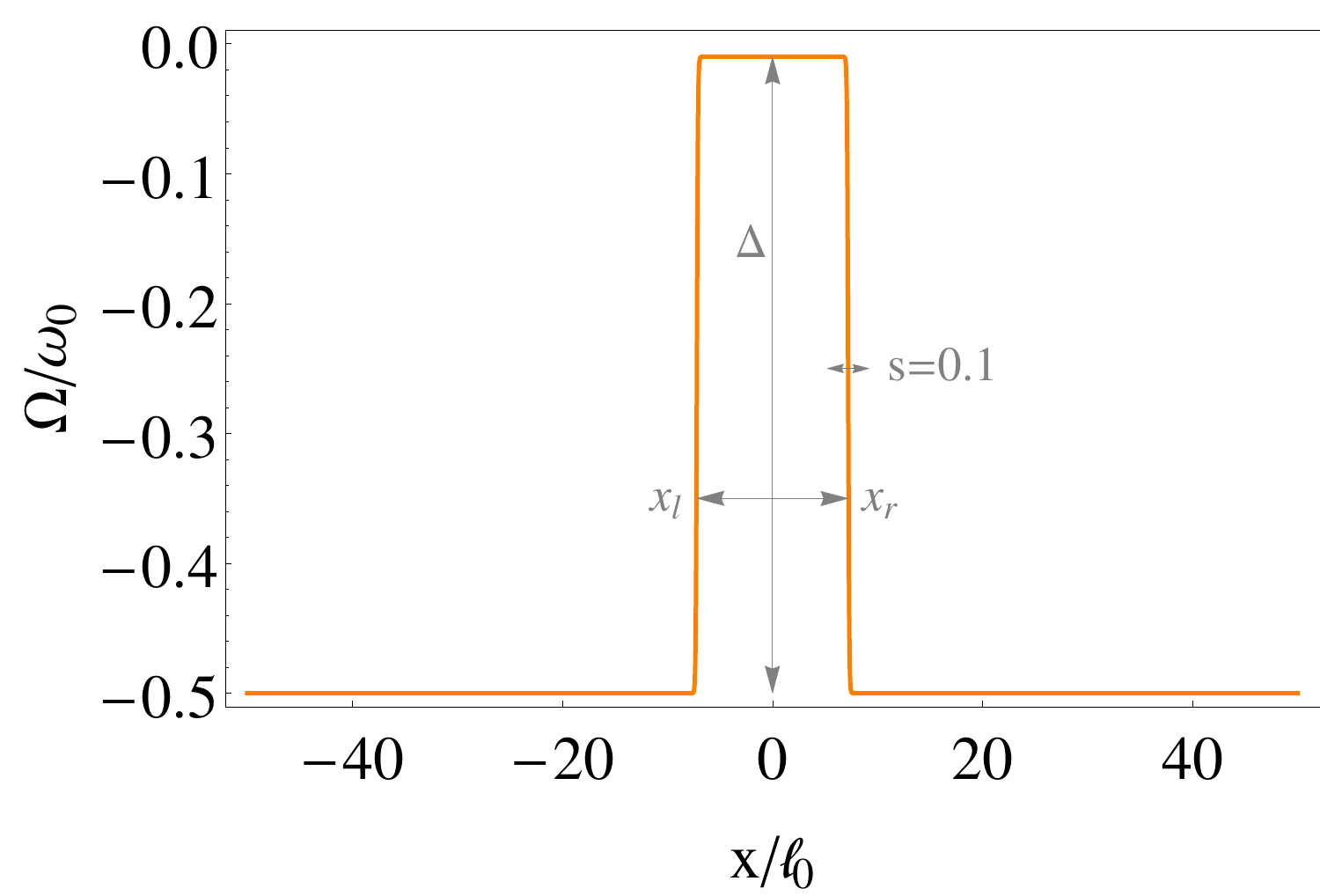}
\caption{Spatial profile of $\Omega/\hbar\omega_0$ (in units of $\chi$) used in the simulations.}
\label{fig:Well_profile}
\end{figure}

\begin{figure*}[htbp]
\centering%
\subfigure[t'=0.\label{fig:HO_t0}]
{\includegraphics[width=0.3\linewidth]{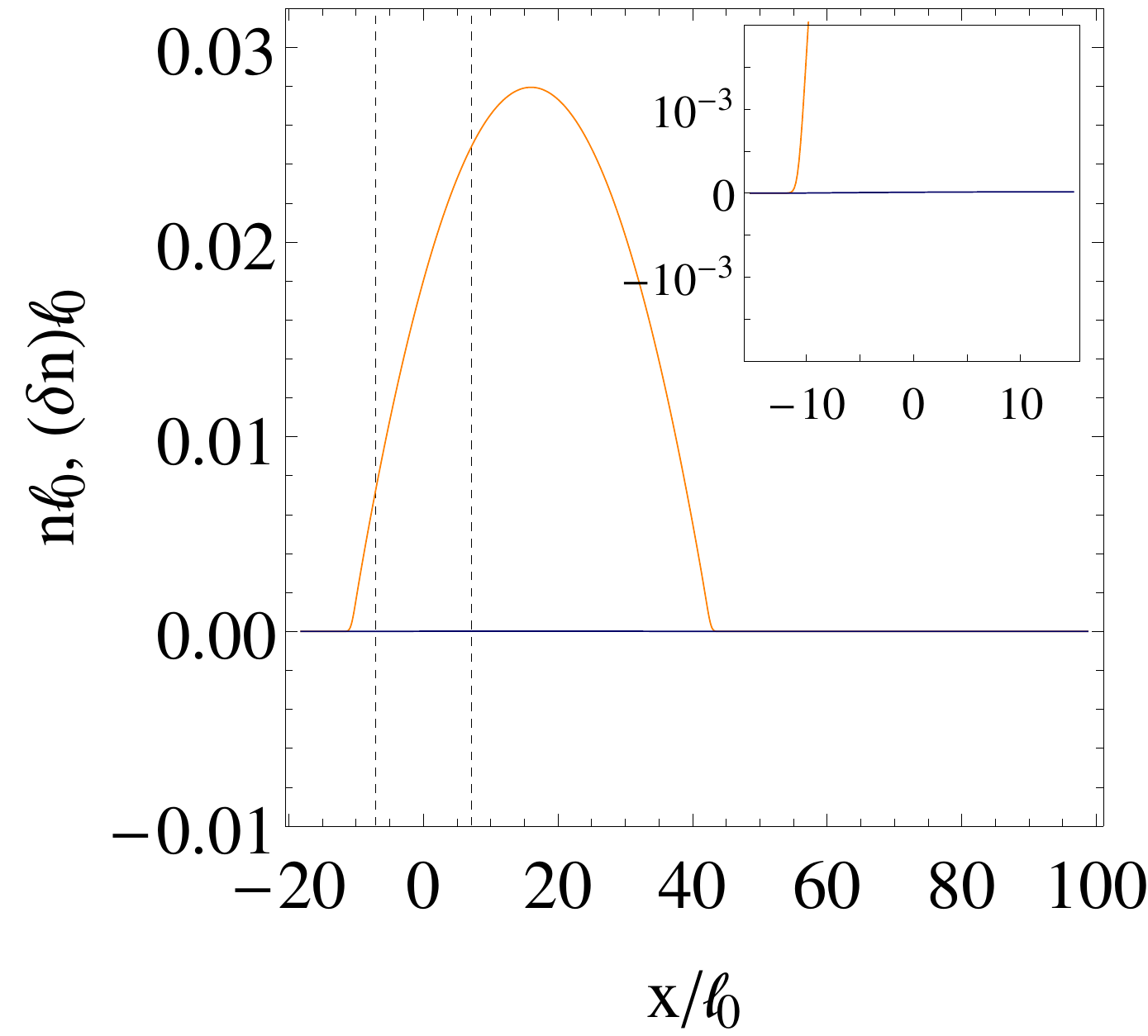}}
\subfigure[t'=40.\label{fig:HO_t40}]
{\includegraphics[width=0.3\linewidth]{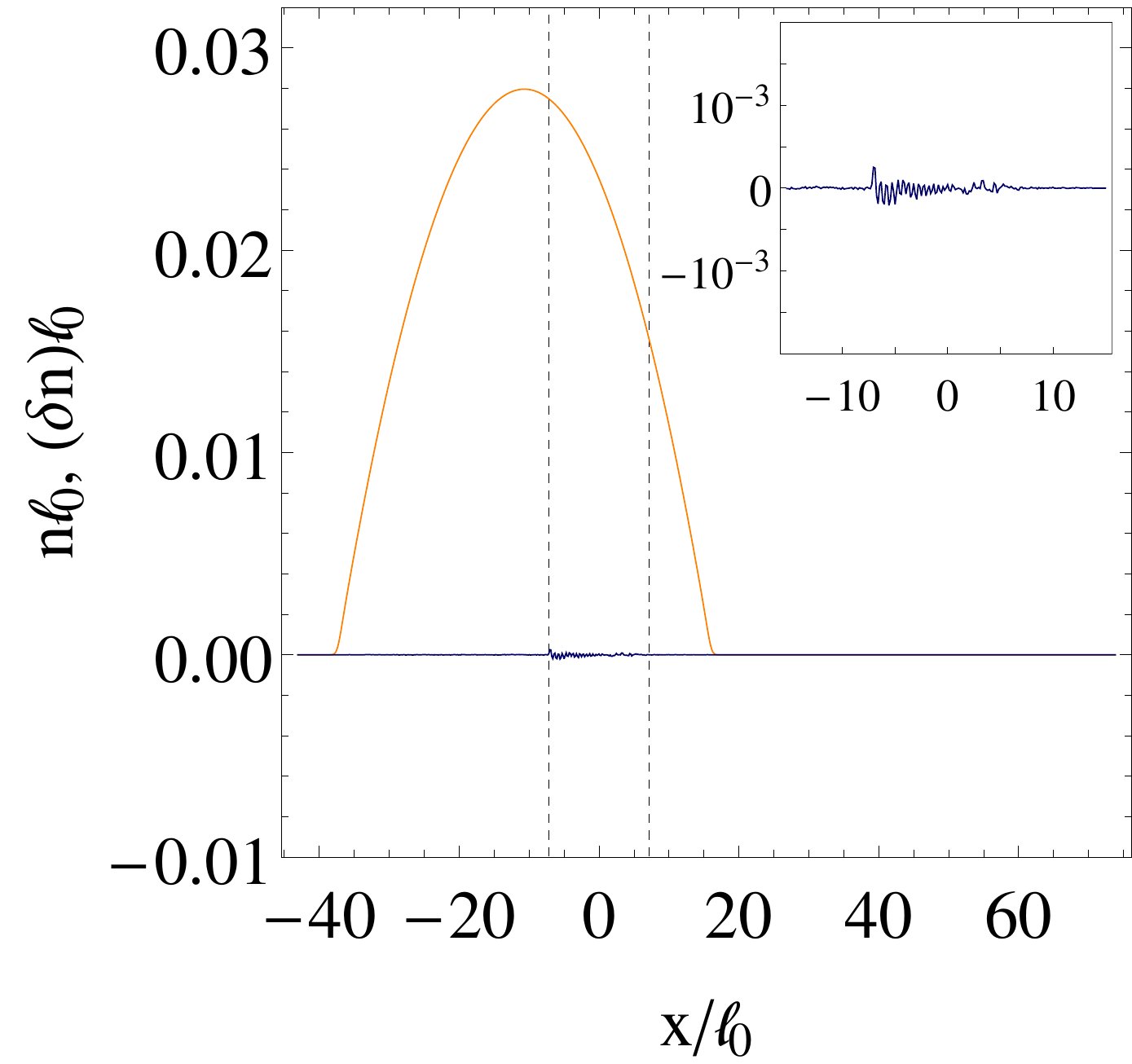}}
\subfigure[t'=45.\label{fig:HO_t45}]
{\includegraphics[width=0.3\linewidth]{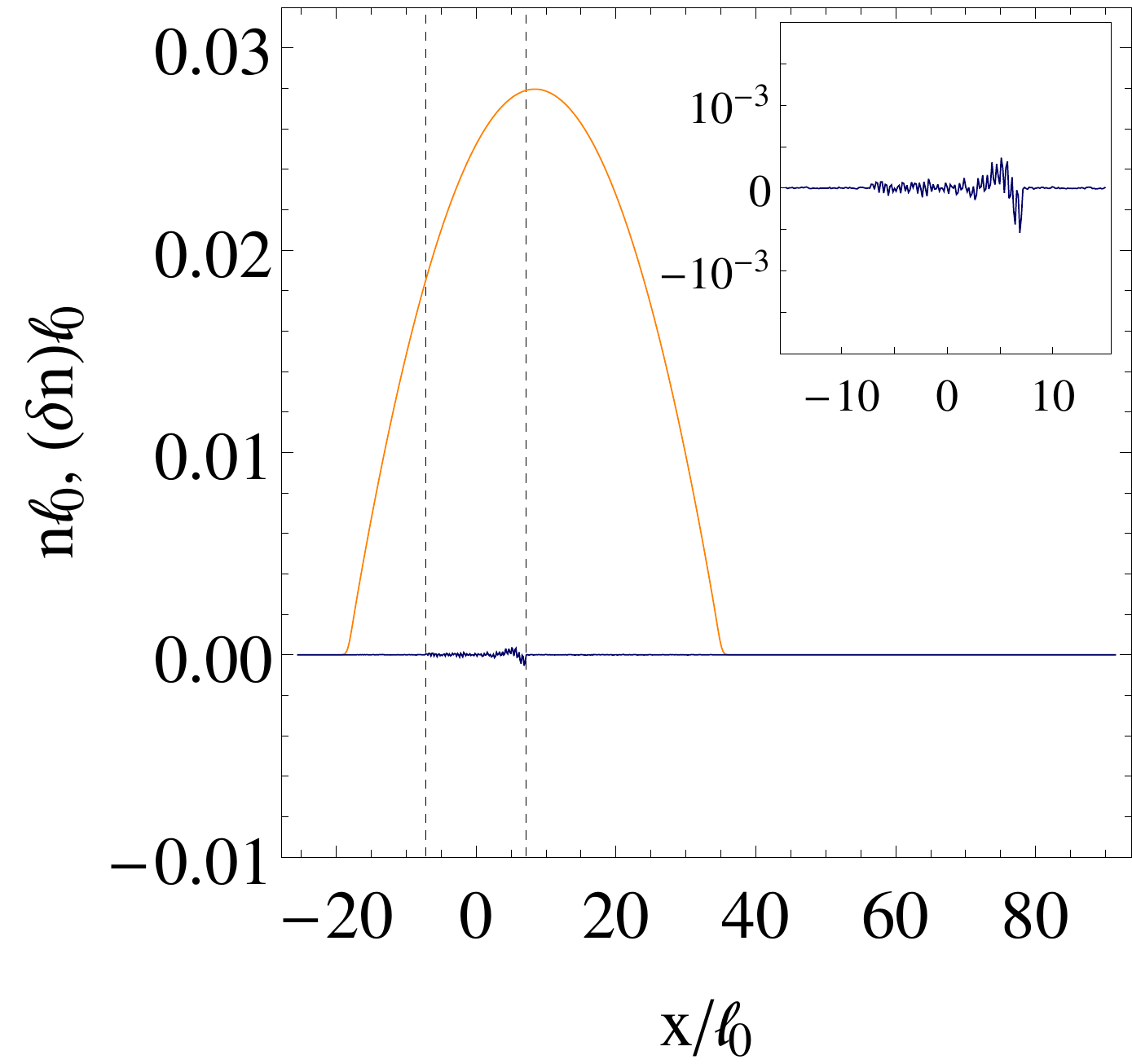}}\\
\subfigure[t'=50.\label{fig:HO_t50}]
{\includegraphics[width=0.3\linewidth]{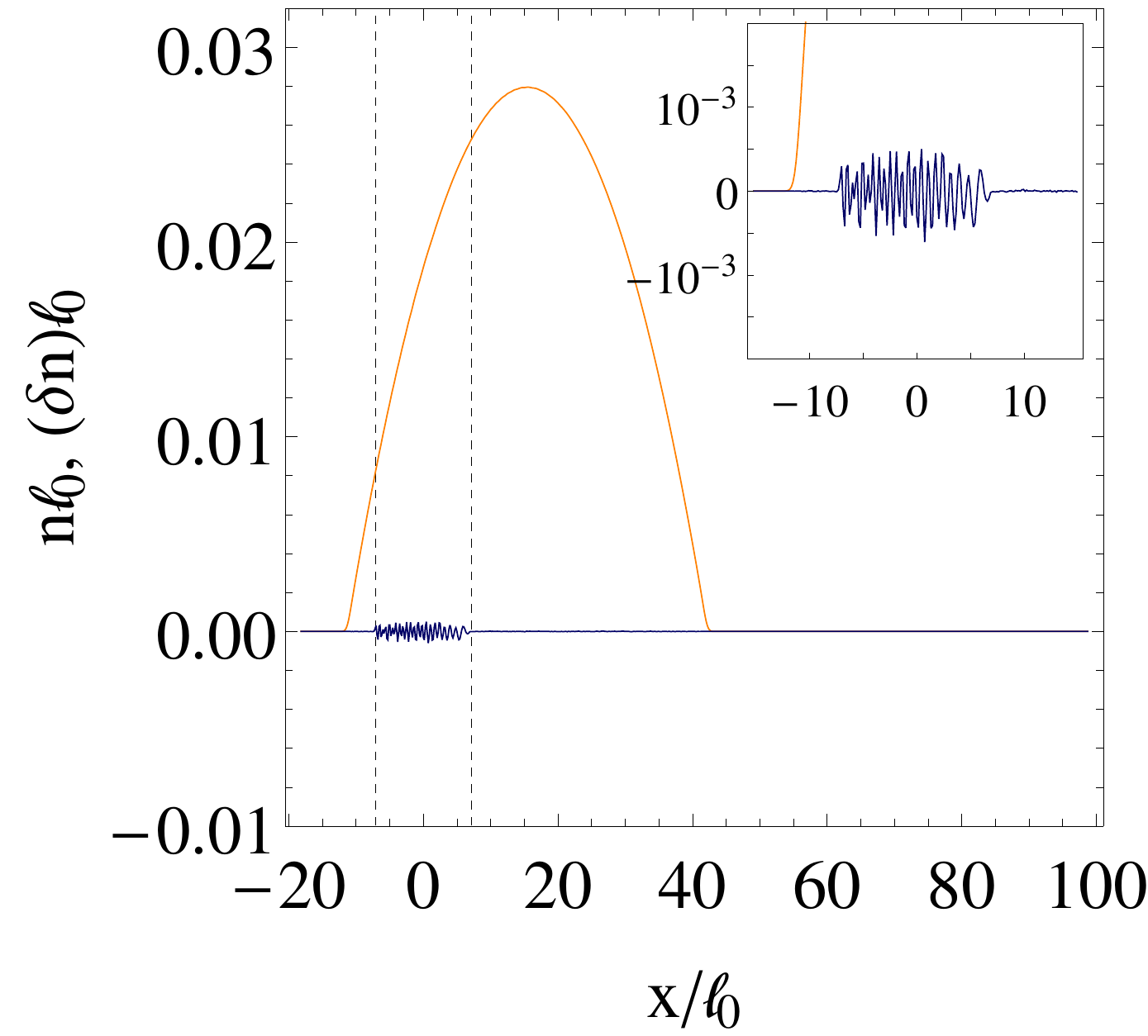}}
\subfigure[t'=55.\label{fig:HO_t55}]
{\includegraphics[width=0.3\linewidth]{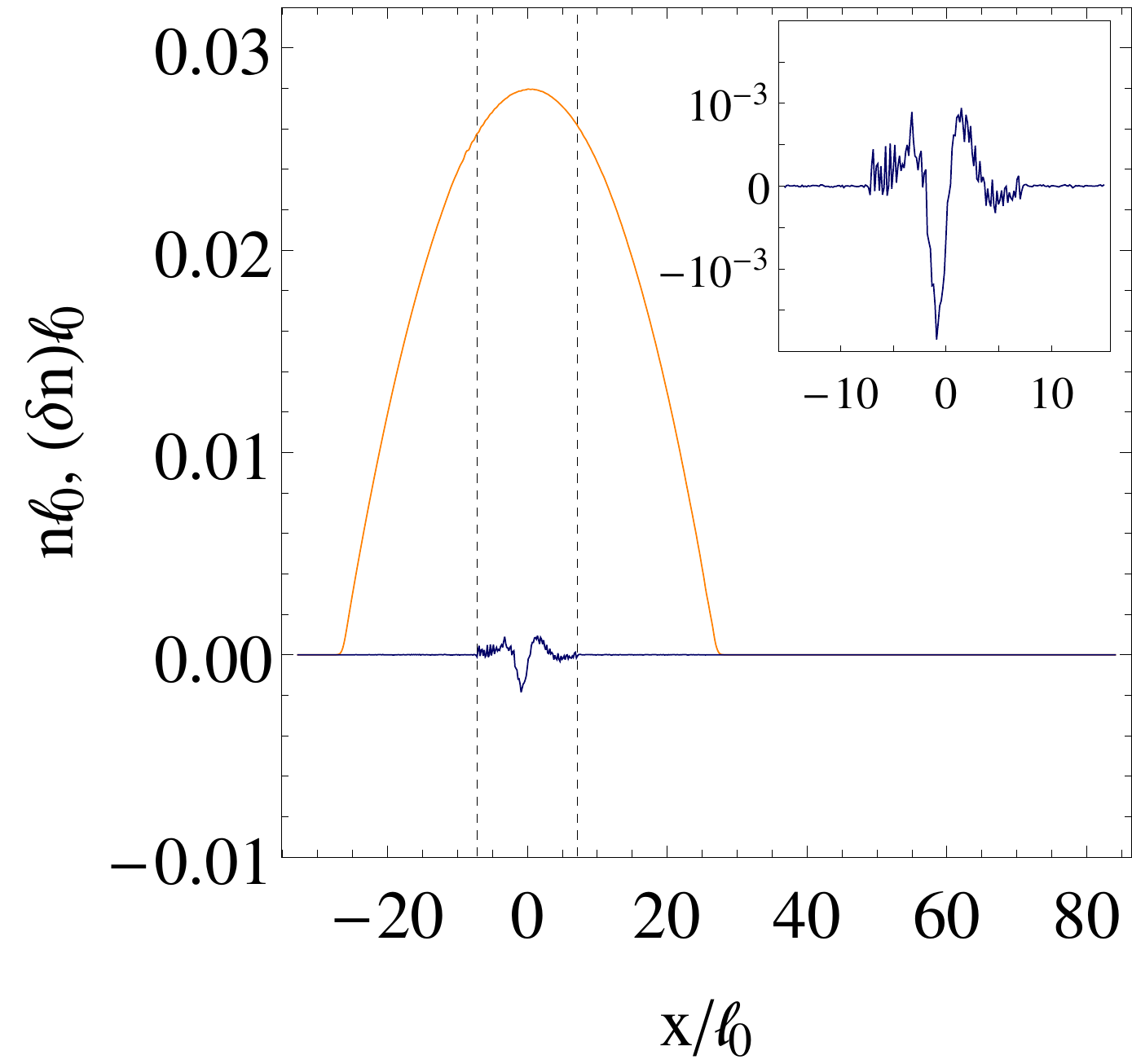}}
\subfigure[t'=60.\label{fig:HO_t60}]
{\includegraphics[width=0.3\linewidth]{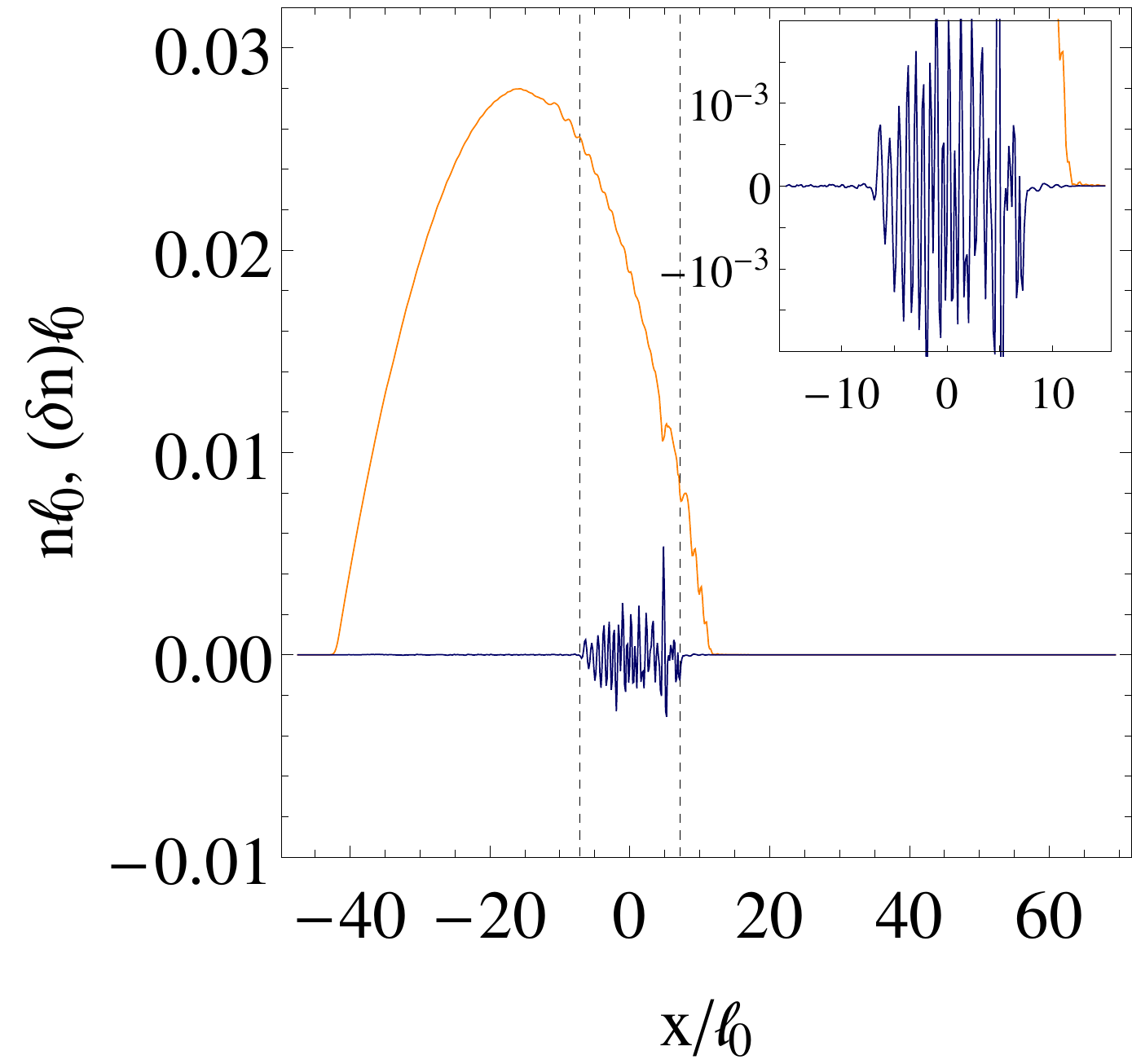}}
\subfigure[t'=64.\label{fig:HO_t64}]
{\includegraphics[width=0.3\linewidth]{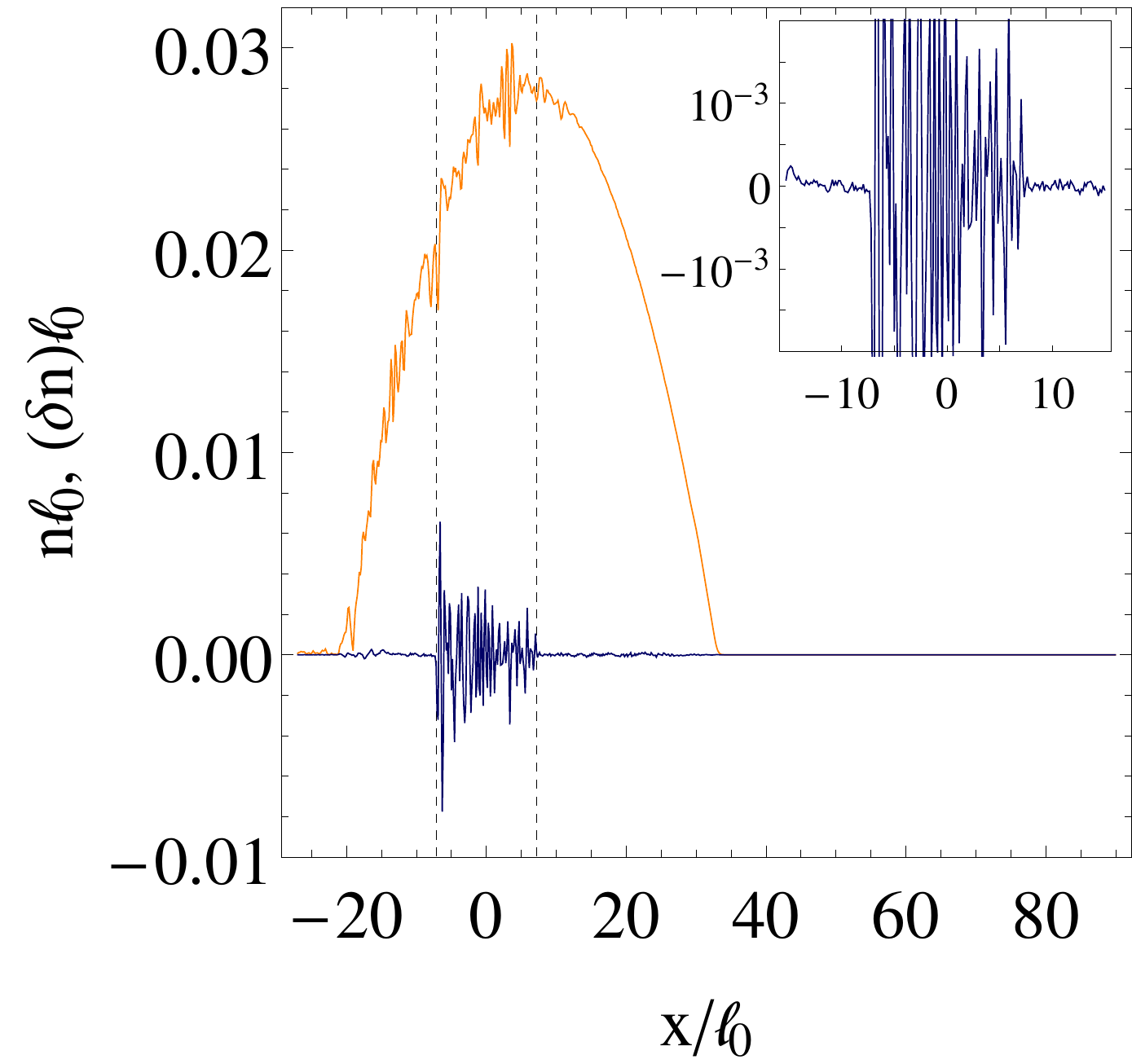}}
\subfigure[\label{fig:StdDev}]
{\includegraphics[width=0.44\linewidth]{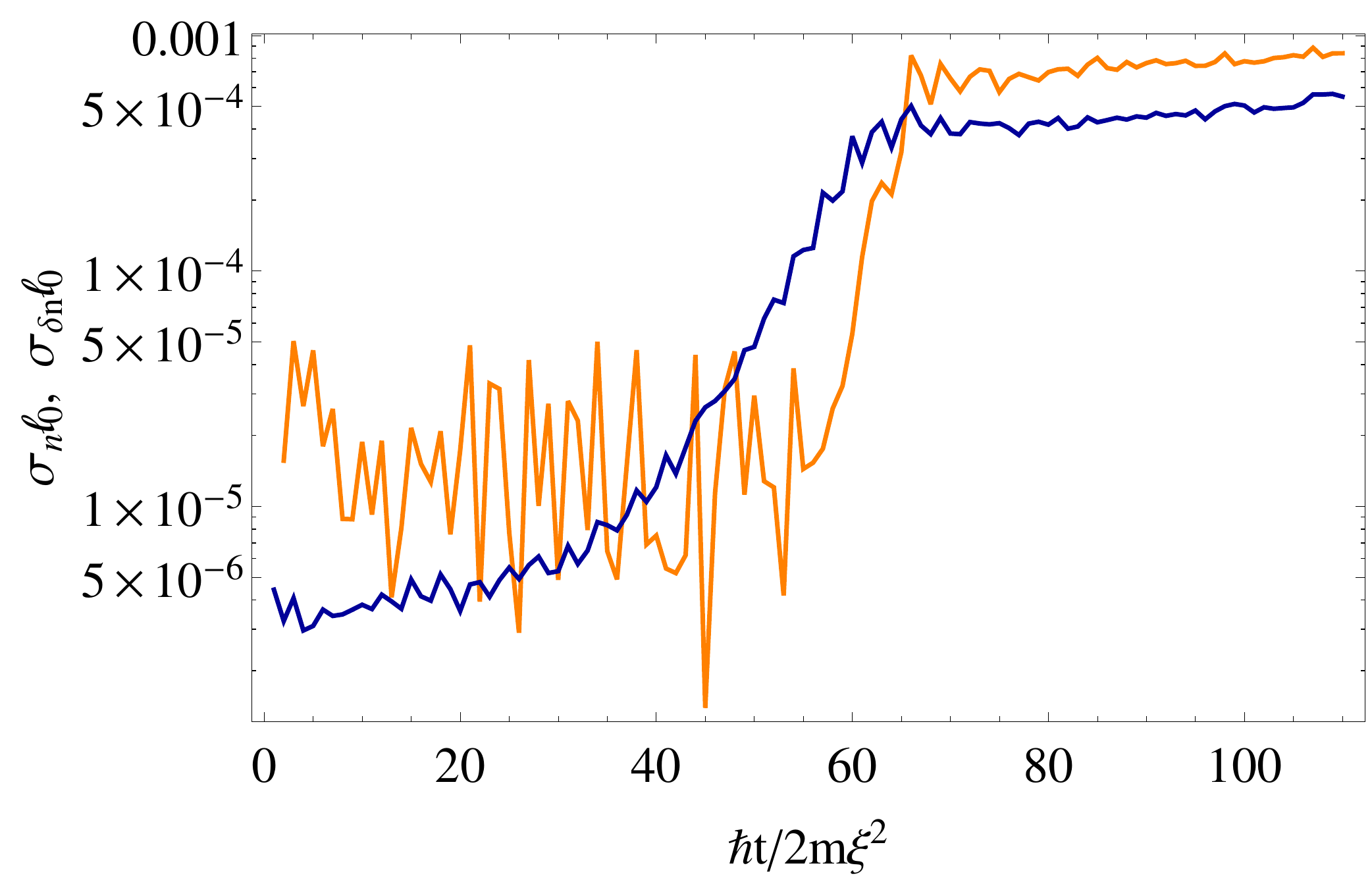}}
\caption{a-g) Different time snapshots of black-hole lasing in a harmonically trapped condensate. The insets show magnified views of the lasing region. The values of the parameters are given in the text. h) Lin-log plot of the time evolution of the standard deviation of the modulation in the density (yellow) and spin (blue) components with respect to their equilibrium values. }
\label{fig:HO}
\end{figure*}

\section{The harmonically trapped condensate\label{sec:HO}}
In all previous sections, we studied the BH-lasing in the spin branch of the excitations in a symmetric, two component, uniform quasi-1D condensate. The relative simplicity of this model, due to the translational invariance of the system, allowed us to identify in a straightforward way the phenomenology of the Hawking physics and of the BH-lasing phenomenon in particular. 

In this section we consider instead the experimentally more feasible case of a harmonically trapped condensate. By using the methods described in the previous sections, the profile of the Rabi frequency can be modulated in space, in order to create a finite supersonic region in the center of the trap, once the condensate is set in motion with a proper velocity. Rather than using the step-like profile for the coupling $\Omega$ as in the previous section for the uniform system, we use here the more realistic shape given by:
\begin{equation}
	\Omega=\frac{\Delta}{2}\left[\tanh\left(\frac{x-x_l}{s}\right)-\tanh\left(\frac{x+x_r}{s}\right)\right] +\Omega_0.
	\label{Omega_profile}
\end{equation}
and shown in Fig. \ref{fig:Well_profile}.
The parameters characterizing this profile are the asymptotic value $\Omega_0$ of the Rabi frequency outside the cavity, the depth $\Delta$ of the well, its left and right delimiting positions $x_l$ and $x_r$, and $s$, which represents the steepness of the walls.

In contrast to the previous sections, normalized dimensionless lengths are defined here in units of the oscillator length $\ell_0=\sqrt{\hbar/2m\omega_0}$, where $\omega_0$ is the characteristic frequency of the harmonic oscillator, while the other dimensionless quantities are defined as: $\Omega'=\Omega/\omega_0$, $t'=\omega_0 t$. In order to show the occurrence of the lasing phenomenon, we consider a set of parameters already used in the previous section for the ring geometry. We use in particular the values $g_{ab}=0.8\,g$ and $\Omega'=-0.01,-0.5$ respectively inside and outside the cavity. Given the profile in Eq. \eqref{Omega_profile}, these values of the Rabi frequency are obtained by choosing $\Omega'_0=-0.5\chi$ and $\Delta'=0.49\chi$. The coefficient $\chi\equiv gn(0)/\hbar\omega_0$, where $n(0)$ is the maximum density in the condensate which, in the Thomas-Fermi (TF) approximation, is equal to $n(0)=2\left(\mu+|\Omega|\right)/\left(g+g_{ab}\right)$,  translates the values of the parameters from one set of units to the other. The chemical potential can be again obtained by the normalization condition $N=\int{dx n(x)}=(2/3\pi)\left[\left(\mu+|\Omega|\right)/\hbar\omega_0\right]^{3/2}\left[S_t/\ell_0(a+a_{ab})\right]$. We choose $s'=0.1$ and  $x'_r=-x'_l=7.2$, so that the length of the cavity is the same as in the cases considered in Sec. \ref{sec:noise}. Given the length of the full system here considered $L'=300$, we choose the number of particles in order to match also the particle density at the center of the trap. The velocity $v'=2k'=90\times4\pi/L$ in the center of the trap $(x'=0)$ is achieved by initially locating the condensate in $x'=16$ at $t'=0$ and then letting it freely oscillate under the effect of the harmonic potential, while the spatial profile of $\Omega'$ is stationary with the trap potential.

The results shown in Fig. \ref{fig:HO} confirm the occurrence of the lasing in the spin modes also for the harmonically trapped condensate. As a further check, the same simulation has been run with the value $\Omega'=-0.2$ inside the cavity, for which the flow is everywhere subsonic. As expected, no instability appears in this case.

While at early times the total density profile remains smooth and unaffected by the instability in the spin degrees of freedom, at later times, when the amplitude of the spin excitation has grown large enough, a significant modulation shows up also in the total particle density profile. This interplay between the spin and density modes is due to the higher-order, nonlinear couplings beyond Bogoliubov theory. From a heuristic point of view, this effect can be associated to the back-reaction of Hawking radiation on the background metric: in this a bit stretched analogy, the spin degrees of freedom correspond to the quantum field theory, while the density ones are the underlying spatio-temporal metric. A more rigorous and comprehensive study along these lines will be the subject of future work.

Independently of this analogy, the nonlinear nature of the effect can be quantitatively assessed in fig.\ref{fig:StdDev}, where we display the time evolution of the standard deviation of the modulation in the density and spin components with respect to their equilibrium values. Exception made for the early times, where the system is in a transient regime, a fast exponential-like amplification appear for both quantities. In particular, it is evident that the modulation in the total density grow faster than the one in the spin component of the system, confirming that the former are indeed an higher-order effect beyond Bogoliubov.

\section{Conclusions\label{sec:conclusions}}
In this article, we showed how a Black-Hole lasing phenomenon can emerge in the spin modes of a flowing one-dimensional, two-component atomic condensate in both spatially homogeneous and harmonically trapped geometries. For the homogeneous case, we studied the propagation of a wave packet of spin excitations through the lasing cavity, and identified the self-amplification of the unstable modes inside the cavity. The rate of such self-amplification was further characterized as a function of the coherent coupling amplitude between the two components by initially imprinting a white random noise into the system: the oscillatory behavior found in the amplification rate reflects the discreteness of the modes sustained by the lasing cavity. We finally confirmed that the effect is preserved also in the experimentally more realistic situation of a harmonically trapped condensate. For this case, an intriguing nonlinear coupling of the spin degrees of freedom to the density ones is highlighted, which may be heuristically associated to the back-reaction of Hawking radiation to the background metric.

Our results demonstrated the promise of multi-component, spinorial atomic Bose-Einstein condensates as analog models of gravitational physics. By simply shaping the amplitude profile of the coherent coupling between the internal states of the atoms, a wide variety of hydrodynamic regimes with different number and shape of the horizons can be achieved. As an example, superradiance from an analogue rotating BH~\cite{Federici2006,Weinfurtner2016} could be investigated within the spin-branch of the system without being disturbed by the microscopic structure of the vortex core. The much larger value of the spin healing length compared to the density one, ensures in fact the existence of a wide ergoregion where radiation enhancement effects can take place on top of a very flat density background far from the vortex core. 

From a more technical experimental point of view the additional flexibility introduced by the richer structure of internal atomic levels and by the possibility of a coherent coupling between them, allows to circumvent a number of issues occurring with single component systems and, in particular, to keep the mean-field hydrodynamics simple and well-controlled while generating arbitrarily complex multi-horizon configurations.

This latter feature appears of paramount interest in view of quantitative experimental studies of the finer details of Hawking radiation, as it will facilitate isolating the quantum effects from the background hydrodynamics. As a most interesting future work, we expect that this will help shining light onto the back reaction effect of quantum fluctuations onto the macroscopic condensate flow, that is the microscopic mechanism underlying black hole evaporation under the effect of the Hawking emission.

\acknowledgments

We are grateful to  Markus Oberthaler, Joerg Schmiedmayer and Alessio Recati for continuous stimulating discussions.
SB acknowledges financial support from EPSRC CM-CDT Grant No. EP/G03673X/1. P\"O acknowledges support from EPSRC EP/M024636/1. IC acknowledges financial support from the EU-FET Proactive grant AQuS, Project No. 640800, and by the Autonomous Province of Trento, partially through the project ``On silicon chip quantum optics for quantum computing and secure communications'' (``SiQuro'').

\appendix
\section{Bogoliubov operator\label{app:BogolOp}}

\begin{equation}
	\mathcal{L}= \begin{pmatrix}
	& d & m_1 & m_2 & o\\
	&-m_1 & -d & -o & -m_2\\
	& m_2 & o & d & m_1\\
	&-o & -m_2 & -m_1 & -d
	\end{pmatrix},
\end{equation}
with
\begin{align*}
	d&=\hbar^2 k^2/2m+g n/2+|\Omega|\\
	o&=g_{ab}n/2\\
	m_1&=g n/2\\
	m_2&=g_{ab} n/2-|\Omega|
\end{align*}

\section{Absorbing boundary conditions\label{app:absBC}}
The absorbing boundary conditions have been implemented in the numerical simulations by adding to the Hamiltonian in the GP equations \eqref{GP_a} and \eqref{GP_b}, terms having the form
\begin{multline}
	H_{\text{abs}}(x,t)=\\
	if(x)\sum_{j=a,b}\left[C_1^j e^{i k_j x}\frac{d^2}{d x^2} e^{-i k_j x}-C_2^j\left(\left|\psi_j\right|^2-\left|\psi_{0j}\right|^2)\right)\right],
\label{H_abs}
\end{multline}
where we indicated by $\psi_{0j}=\left|\psi_{0j}\right| e^{i k_j x}$ $(i=a,b)$ the order parameters describing the unperturbed state of the system. These terms vanish when acting on the unperturbed state $\boldsymbol{\psi}_0\equiv\left(\psi_{0a},\psi_{0b}\right)$. In eq.\eqref{H_abs}, $f(x)$ is a function which specifies the location of the absorbing region, sufficiently smooth in order to avoid spurious reflections. In our calculations, we have taken it to have a Gaussian shape. The first term between brackets is a diffusive term which damps wave-vector components of the signal other than $k_j$, while the second one suppresses fluctuations of the modulus of the order parameter.

\bibliography{BH_lasing}

\end{document}